# Trust and Lottery-Related Anomalies

Reza Bradrania [1], Ya Gao [2]

**Abstract**

We examine whether social trust shapes the pricing of lottery-like stocks by influencing investors' beliefs regarding optimistic information associated with extreme future payoffs. We argue that higher social trust reduces investors' skepticism toward optimistic information associated with extreme future payoffs, increasing demand for lottery-type stocks and leading to overpricing and lower subsequent returns. Using a sample of 8,875 firms across 49 US states over the period 1996–2018 and MAX as a proxy for lottery-like characteristics, we show that the MAX effect is concentrated in regions with relatively high trust and is weak or absent in regions with relatively low trust. We further find that this trust conditioning is stronger in regions with lower education or income levels, consistent with social trust playing a relatively more important role when investors have more limited access to private information. The results are robust to alternative lottery proxies and a range of robustness checks. Our findings identify social trust as a novel determinant of the pricing of lottery-like stocks and provide evidence that social capital may amplify belief-driven mispricing in assets with highly skewed payoffs.

[1] Corresponding author. Adelaide University, College of Business and Law, Adelaide, Australia. E-mail: reza.bradrania@Adelaide.edu.au

[2] Henan Finance University, Zhengzhou, Henan, China. E-mail: gaoya@hafu.edu.cn

*"People are not accustomed to thinking hard, and are often content to trust a plausible judgment that quickly comes to mind."*

— Daniel Kahneman (2003)

## 1. Introduction

Lottery-related anomalies are widely documented in the literature.[3] That is, stocks with lottery-like payoffs (i.e., stocks with skewed distributions that have a small probability of large payoffs) tend to significantly underperform compared to stocks with non-lottery-like payoffs. Using the maximum daily return (MAX) over a month as the proxy for lottery-like stocks[4], Bali et al. (2011) find that in the US stock market, stocks with high MAX in the previous month generate low returns in the subsequent month, suggesting that lottery-like stocks underperform compared to non-lottery-like stocks. This negative relation is usually referred to as the MAX effect or MAX anomaly. Following Bali et al. (2011), a growing body of studies has explored the existence of this effect in other stock markets[5], as well as the plausible driving factors of the MAX effect; these include investor sentiment (Fong and Toh, 2014), levels of mispricing (Zhong and Gray, 2016; Seif, Docherty, and Shamsuddin, 2018), reference-dependent preference (An, Wang, Wang, and Yu, 2020), an upper bound for price named the 52-week high (Byun, Goh, and Kim, 2020), individualism related to overconfidence and overoptimism (Cheon and Lee, 2018), individual investors' stock preference (Lin and Liu, 2018), earnings announcements (Nguyen and Truong, 2018; Baars and Mohrschladt, 2021), attention and social interaction (Bali, Hirshleifer, Peng, Tang, and Wang, 2024), property crime (Gao and Bradrania, 2024), and investors' weather-induced mood (Bradrania and Gao, 2024).

In this paper, we add to this strand of literature by examining the impact of social trust on lottery demand. Specifically, we study the role of social trust on the gambling behavior of investors and on the resulting underperformance of lottery-type stocks. Social trust, also referred to as generalized trust in the literature, is defined as the trust in anonymous others or the probability individuals attribute to the possibility of being cheated (Guiso, Sapienza, and Zingales, 2004a, 2008; Pevzner, Xie, and Xin, 2015; Wei and Zhang, 2020; Limbach, Rau and Schürmann, 2021; Drobetz, Mönkemeyer, Requejo, and Schröder, 2023).[6] More formally, it is considered as the subjective probability that another individual will perform an action that is beneficial, or at least not harmful, to an individual, and the probability is high enough for that individual to consider cooperating with another individual in a society (Gambetta, 1988).

[3] For example, see Kumar (2009), Boyer, Mitton, and Vorkink (2010), Bali, Cakici, and Whitelaw (2011), and Conrad, Kapadia, and Xing (2014).

[4] Lottery-like stocks, or equivalently, lottery-type stocks refer to stocks with lottery-like payoffs. We use lottery-like stocks and lottery-type stocks interchangeably hereafter.

[5] See, for example, the European market (Annaert, De Ceuster, and Verstegen, 2013; Walkshäusl, 2014), the South Korean market (Nartea, Wu, and Liu, 2014), the Australian market (Zhong and Gray, 2016), the Canadian market (Aboulamer and Kryzanowski, 2016), the Chinese market (Nartea, Kong, and Wu, 2017), the Taiwanese market (Hung and Yang, 2018), the Brazilian market (Berggrun, Cardona, and Lizarzaburu, 2019), the global market (Cheon and Lee, 2018), and nine advanced emerging markets (Seif, Docherty, and Shamsuddin, 2018).

[6] We use trust, generalized trust and social trust interchangeably hereafter. Unless otherwise stated, these terms refer to generalized interpersonal trust rather than individuals' trust or confidence in financial markets, which may operate through different channels.

Prior research shows that social trust enhances participation in trust-intensive economic activities, including financial market investment (Guiso et al., 2004a, 2008; El-Attar and Poschke, 2011; Barbi, Febo, and Giudici, 2023). For example, Guiso et al. (2004a) document that individuals with higher levels of trust are more likely to invest in stocks, which are the most trust-intensive form of investment, and less likely to hold cash. Guiso et al. (2008) further show that high-trust individuals are more likely to invest in stocks and allocate a larger share of their wealth to financial assets. These studies suggest that social trust primarily influences investors' beliefs about the credibility and reliability of information and counterparties, rather than their attitudes toward risk or low-probability outcomes.

Building on this insight, we argue that social trust affects the pricing of lottery-like stocks through a belief-based channel. Lottery-like stocks are characterized by highly skewed return distributions and salient extreme positive payoff realizations, whose extreme upside narratives and payoff potential are inherently difficult to verify. In high-trust environments, investors may apply a smaller skepticism discount to optimistic narratives and signals associated with these extreme upside outcomes, thereby assigning greater credibility to claims regarding their extreme future payoff potential. This trust-induced belief distortion contributes to stronger demand for lottery-like stocks, leading to overpricing and subsequently lower expected returns. Consequently, we hypothesize that the negative lottery premium, as captured by the MAX effect, is stronger in regions with higher levels of social trust. Unlike probability-weighting or preference-based explanations, our proposed mechanism operates through investors' beliefs about the credibility of optimistic but difficult-to-verify information. To our knowledge, this study is the first to examine the role of social trust in shaping lottery demand and the pricing of lottery-like stocks. Our findings identify social trust as a source of cross-sectional heterogeneity in investors' beliefs and the pricing consequences of lottery demand, with important implications for asset pricing.

The importance of trust in economic settings has been widely documented in the literature. Prior studies show that trust promotes investment and economic growth (e.g., Knack and Keefer, 1997; Zak and Knack, 2001; Algan and Cahuc, 2010), enhances the performance of large organizations (e.g., Fukuyama, 1995; La Porta et al., 1997), and improves economic efficiency and financial development (Guiso et al., 2004a). In financial markets, trust has been shown to encourage stock market participation (e.g., Guiso et al., 2008; El-Attar and Poschke, 2011), shape investors' local and foreign investment biases (Wei and Zhang, 2020; Shao and Wang, 2021; Drobetz et al., 2023), impact cross-border mergers and acquisitions (Ahern, Daminelli, and Fracassi, 2015), influence investors' reactions to corporate earnings announcements (Pevzner et al., 2015), and affect firms' financing costs, including the cost of equity and debt (Gupta et al., 2018; Meng and Yin, 2019; Brockman et al., 2020). While these studies demonstrate that social trust has important implications for firms' financing policies, financing conditions, and other corporate outcomes, relatively little is known about whether social trust may also have unintended asset-pricing consequences when investors evaluate optimistic but difficult-to-verify information. This study contributes to the literature by showing that although social trust generally enhances economic and financial outcomes, it may also amplify belief-driven mispricing in settings where valuations depend on optimistic but difficult-to-verify narratives, such as lottery-like stocks. Unlike prior studies that focus on corporate financing outcomes, our study examines the asset-pricing implications of investor beliefs about lottery-like stocks. By linking social trust to the pricing of lottery-like stocks, our study identifies a novel belief-based channel through which social trust can influence asset prices.[7]

To examine our hypotheses, we focus on all common stocks traded on Nasdaq and headquartered in the US from January 1996 to December 2018. We restrict our sample to Nasdaq firms because prior studies suggest that trading in Nasdaq stocks is more localized, allowing local investor characteristics to play a more prominent role in asset pricing (Schultz, 2003; Loughran and Schultz, 2004; Anand et al., 2011). Following Bali et al. (2011), we proxy for lottery-like stocks using the maximum daily return (MAX) over the previous month in our main analysis.[8] We measure social trust using responses to the World Values Survey (WVS) question: "Generally speaking, would you say that most people can be trusted or that you need to be very careful in dealing with people?", which is widely employed in the trust literature (e.g., La Porta et al., 1997; Guiso et al., 2008; Aghion, Algan, Cahuc, and Shleifer, 2010; Pevzner et al., 2015; Wei and Zhang, 2020). Following Wei and Zhang (2020) and Pevzner et al. (2015), we construct a regional social trust index (TRUST) within the US. This approach differs from earlier studies that primarily use WVS data to examine cross-country differences in trust. Focusing on the US setting offers several advantages. First, there is substantial cross-sectional and time-series variation in social trust across US regions, which allows us to study how trust relates to lottery demand and asset pricing within a unified institutional environment. Second, the common regulatory, legal, and accounting standards faced by US firms mitigate concerns that international differences such as insider trading regulations, disclosure requirements, or accounting rules drive our results. Our final sample consists of 8,875 firms across 10 US geographical regions that cover 49 states.

We investigate our hypothesis in several ways. First, we confirm the presence of a MAX effect in our sample by sorting stocks into deciles based on their maximum daily return in the previous month. We then examine whether social trust relates to the strength of the MAX effect by conducting univariate portfolio analyses across regions that are classified as having relatively high and relatively low trust. We find that the underperformance of lottery-like stocks is concentrated in regions with relatively high trust, while the MAX effect is weak or absent in regions with relatively low trust. For example, using the median of TRUST to classify regions, the value-weighted high-minus-low MAX portfolio earns −1.34% per month in regions with relatively high trust, compared with an insignificant −0.87% in regions with relatively low trust, and its CAPM alpha is −1.85% per month, compared with −1.40%.

[7] Our trust channel is also distinct from a financing channel, under which social trust improves firms' financing conditions (e.g., Gupta et al., 2018; Meng and Yin, 2019; Brockman et al., 2020), which might be relevant for lottery-like firms that tend to be younger and rely more on external financing. The financing channel concerns how trust affects firms' access to capital and their subsequent issuance and investment, whereas our mechanism concerns how credible investors find optimistic but difficult-to-verify narratives about extreme payoffs. We show in Section IA-13 that our findings are not explained by firms' financing activity, investment, or age.

[8] In the robustness tests, we also use a set of alternative proxies for lottery-type characteristics such as alternative MAX measures and idiosyncratic skewness to examine the hypothesis.

We further examine this relation using Fama and MacBeth (1973) cross-sectional regressions. Consistent with the portfolio evidence, the negative relation between MAX and future returns is significantly stronger for stocks headquartered in regions with relatively high trust than for those in relatively low trust regions. When we include social trust and its interaction with MAX in the regressions, the results indicate that the MAX effect exists only in regions with relatively high trust. Economically, the coefficient on the interaction term implies that moving MAX from the $25^{th}$ to the $75^{th}$ percentile of its distribution (3.83 to 10.89) is associated with approximately 18 basis points lower next-month returns in high-trust regions relative to low-trust regions, equivalent to about 2.1% on an annualized basis.

These findings support the view that social trust influences investors' beliefs by reducing skepticism toward optimistic information associated with extreme upside payoffs. This belief-based mechanism contributes to higher demand for lottery-like stocks, greater overpricing, and subsequently lower returns. Our results are robust to controlling for regional characteristics that may be correlated with both trust and lottery demand, including religion, education, income levels, lottery exposure, economic development, unemployment, population density, and local financial development (Kumar, 2009; Kumar, Page, and Spalt, 2011; Guiso et al., 2003, 2004a, 2004b, 2008).

Social trust reflects both institutional characteristics of the financial system, such as the perceived likelihood of fraud or opportunistic behavior, and individuals' subjective beliefs formed from prior experiences and information (Guiso et al., 2004a, 2008). When trust primarily reflects such subjective priors, its influence on financial decision-making should depend on the availability and quality of private information. More educated or wealthier individuals are more likely to possess superior financial knowledge, greater access to information, and a better understanding of how financial markets operate, reducing their reliance on social trust when forming beliefs about optimistic but difficult-to-verify information associated with lottery-like stocks (Guiso et al., 2004a, 2008; El-Attar and Poschke, 2011). Accordingly, we conjecture that the effect of social trust on lottery demand is weaker in regions with higher levels of education or income, where investors rely more on private information and less on social trust in forming beliefs about optimistic but difficult-to-verify extreme upside narratives.

To examine this conjecture, we split the sample by regional levels of education and income and re-estimate the Fama and MacBeth (1973) cross-sectional regressions. We find that the influence of social trust on the MAX effect is stronger in regions with lower education or income levels, consistent with the notion that investors with weaker information environments rely more heavily on trust when forming beliefs about lottery-like stocks. Importantly, this effect is asymmetric across trust environments. Among firms headquartered in regions with relatively high trust, the MAX effect is substantially stronger in areas with lower education or income levels than in those with higher education or income levels. In contrast, within regions with relatively low trust, the MAX effect in low education or income areas is weak. These patterns are consistent with prior evidence that less educated or less wealthy investors exhibit a stronger propensity to invest in lottery-type stocks (Kumar, 2009). However, our results highlight a novel interaction: such lottery demand translates into pronounced mispricing only in regions with relatively high trust, where trust-based belief distortions lead investors to assign greater credibility to optimistic but difficult-to-verify narratives regarding extreme future payoffs.

We conduct several robustness tests to support our main findings. First, we construct an orthogonalized MAX measure by removing the component of MAX that is correlated with the social trust index (TRUST). When this orthogonalized MAX is used as the sorting variable in the univariate portfolio analysis, the MAX effect disappears. This result suggests that the underperformance of lottery-like stocks is closely linked to the trust-related component of MAX, consistent with our proposed belief-based mechanism. Second, we replace MAX with alternative proxies for lottery-like payoffs, including alternative constructions of MAX and idiosyncratic skewness, and find that our results remain robust. Third, we estimate pooled panel regressions with regional and monthly fixed effects to address concerns that our findings are driven by persistent regional heterogeneity or common time shocks. Our main results remain robust after controlling for these regional and time fixed effects. Fourth, we further control for political ideology, racial diversity, immigrant share, and other dimensions of social capital, and find that the effect of social trust on the MAX effect remains robust. Fifth, we split the sample by firms' institutional ownership and re-estimate the cross-sectional Fama and MacBeth (1973) regressions. We find a stronger trust effect among firms with lower institutional ownership. This finding provides additional evidence relevant to the transmission of regional social trust to stock prices, although institutional ownership does not identify whether non-institutional investors reside in the firm's headquarters region. Sixth, we conduct a geographic permutation placebo test in which we randomly reassign regional trust to firms and re-estimate the baseline regression. The observed coefficient on the interaction between MAX and trust lies in the extreme left tail of the placebo distribution, indicating that our findings depend on the actual correspondence between firms' headquarters locations and regional trust. Seventh, we examine whether our findings reflect a financing channel. Because lottery-like firms tend to be younger, have greater growth opportunities, and rely more heavily on external financing, an improvement in financing conditions associated with social trust could be particularly relevant for these firms. We find that high-MAX firms in regions with relatively high trust do not exhibit greater share issuance, external financing, or investment, and are not younger than those in regions with relatively low trust. Moreover, controlling for these characteristics and their interactions with MAX does not affect our main findings. Finally, we show that the impact of social trust on the MAX effect is distinct from social transmission bias (Hirshleifer, 2020). Using population density, social capital, and Facebook-based social connectedness as proxies for the intensity of local social transmission, we find that these proxies do not amplify the MAX effect, while the interaction between MAX and trust remains negative. Details of these analyses and additional robustness tests are reported in the Robustness Tests Section and the Internet Appendix.

This study contributes to the literature in several ways. First, this study contributes to the literature on lottery-related anomalies by identifying social trust as a novel determinant of the MAX effect and the pricing of lottery-like stocks. While prior studies document the existence of these anomalies and explore their underlying drivers, we show that social trust conditions the pricing of lottery-like stocks through a belief-based mechanism. More broadly, our findings suggest that although social trust generally improves economic and financial outcomes, it may also amplify belief-driven mispricing when valuations rely on optimistic but difficult-to-verify information. Second, the paper extends the literature on social trust and financial decision-making by demonstrating that the pricing consequences of lottery-like stocks vary systematically with social trust, with this effect varying systematically across regions with different education and income levels. Importantly, our proposition is not that social trust is inherently detrimental to financial markets. Rather, we argue that in the specific context of lottery-like stocks, where valuations depend on highly uncertain and positively skewed payoff distributions, social trust may reduce investors' skepticism toward optimistic narratives, thereby contributing to stronger belief-driven overpricing and subsequent underperformance. In this sense, our findings identify a context-specific asset-pricing implication of social trust rather than a general negative consequence of trust. Third, we show that the effect of education and income on the pricing of lottery-like stocks is conditional on social trust. Specifically, the tendency of less educated or less wealthy investors to invest more heavily in lottery-type stocks (Kumar, 2009) manifests only in regions with relatively high trust. In low-trust regions, differences in education or income do not translate into a stronger MAX effect. This asymmetric pattern highlights social trust as a necessary condition through which lower education or income levels amplify the belief-based mechanism underlying the pricing of lottery-like stocks, while more educated or wealthier investors rely less on trust when forming financial decisions. Finally, since trust is a central dimension of social capital and culture (Putnam, 1993; Fukuyama, 1995; Knack and Keefer, 1997; Guiso et al., 2004a, 2006, 2011; Bloom, Sadun, and Van Reenen, 2012; Ahern et al., 2015; Wei and Zhang, 2020), our study contributes more broadly to the growing literature on the role of social capital and culture in financial markets.

Our results highlight that social trust plays an important role in shaping the pricing of lottery-like stocks through its influence on investors' beliefs, particularly for assets whose valuations depend on optimistic but difficult-to-verify information. When trust reduces investors' skepticism toward such information, investors may assign greater credibility to optimistic narratives regarding extreme future payoffs, leading to belief-driven mispricing. This mechanism is especially relevant in regions with lower education or income levels, where investors rely more heavily on social trust when evaluating such information. Our findings also suggest that improving access to reliable information and financial literacy may help reduce pricing distortions arising from excessive reliance on social trust in these settings. Importantly, our findings should not be interpreted as implying that social trust is detrimental to financial markets. Rather, they suggest that its asset-pricing implications depend on the informational environment and may contribute to mispricing when optimistic investment narratives are inherently difficult to verify.

The remainder of the paper is organized as follows. Section 2 describes the relevant data and variable definitions. In Section 3, we show the underperformance of lottery-type stocks. In Section 4, we present our empirical results on the relationship between MAX and trust, as well as how the level of education or income can impact the effect of trust on the MAX effect. Section 5 provides further robustness checks. Section 6 concludes.

## 2. Data and main variables

In the present study, our measure of trust is calculated based on the responses to the Word Values Survey (WVS) 1981–2020 Integrated Questionnaire[9] (Inglehart et al., 2020), which is widely used in the prior research on social trust. Since we focus on social trust across different regions of the US, we collect our region-level social trust data from five waves (waves 3, 4, 5, 6, and 7) of the WVS from 1995 to 2020. These five waves of surveys were conducted approximately once every five years by WVS in the US, in 1995, 1999, 2006, 2011 and 2017. Therefore, our sample period is from January 1996 to December 2018. The surveys report the location (state) of survey respondents and group them into 10 US geographical regions[10], which comprise New England, Middle Atlantic, South Atlantic, East South Central, West South Central, East North Central, West North Central, Rocky Mountain, Northwest, and California. In total, there are approximately 8,479 survey respondents in the five waves of the WVS.

Following prior studies on trust (e.g., La Porta et al., 1997; Guiso et al., 2008; Aghion et al., 2010; Pevzner et al., 2015; Wei and Zhang, 2020), we measure social trust based on the respondents' answers to the following WVS question: "Generally speaking, would you say that most people can be trusted or that you need to be very careful in dealing with people?" Similar to Pevzner et al. (2015), in each region for each survey year, we recode the response to this question to 1 if a respondent answers that most people can be trusted and 0 otherwise, and then we calculate the mean of the responses as our measure of social trust index (labeled as TRUST) in the year. Since TRUST is calculated for each survey year, we follow Wei and Zhang (2020) and use linear interpolation to estimate the value of TRUST in the years between two consecutive surveys for each region.[11]

In this study, we include all common stocks (CRSP share codes 10 and 11) traded on the Nasdaq from January 1996 to December 2018.[12] We focus on Nasdaq stocks since previous research suggests that trading in Nasdaq stocks is localized (Loughran and Schultz, 2004; Schultz, 2003; Anand et al., 2011). We follow Gupta et al. (2018) and Wei and Zhang (2020) and define a firm's location as the location of its headquarters. We collect firms' annual headquarters-state data from Compustat and merge them with the corresponding regional TRUST measure to construct our sample. We use headquarters state because the literature on local culture and social trust typically assigns regional characteristics based on firms' headquarters. Headquarters location reflects a firm's primary local environment including its managerial, social, media, and investor visibility, and therefore provides a reasonable proxy for the regional social environment that is most closely associated with the firm. Although many Nasdaq firms operate across multiple states, headquarters-state assignment is the standard approach in the literature and facilitates comparisons with prior studies examining the effects of local social and cultural characteristics on corporate and market outcomes (Gupta et al., 2018; Wei and Zhang, 2020).

[9] http://www.worldvaluessurvey.org/WVSDocumentationWVL.jsp.

[10] We follow Wei and Zhang (2020) and identify these geographical regions from the WVS. The states in each region are as follows. New England (Maine, New Hampshire, Vermont, Massachusetts, Rhode Island and Connecticut), Middle Atlantic (New York, Pennsylvania and New Jersey), South Atlantic (Delaware, Maryland, Washington, D.C., Virginia, West Virginia, North Carolina, South Carolina, Georgia and Florida), East South Central (Kentucky, Tennessee, Mississippi and Alabama), West South Central (Oklahoma, Texas, Arkansas and Louisiana), East North Central (Wisconsin, Michigan, Illinois, Indiana and Ohio), West North Central (Missouri, North Dakota, South Dakota, Nebraska, Kansas, Minnesota and Iowa), Rocky Mountain (Montana, Wyoming, Nevada, Utah, Colorado, Arizona, New Mexico and Idaho), Northwest (Oregon and Washington), and California.

[11] Generalized interpersonal trust is widely viewed as a slow-moving regional characteristic rather than a high-frequency state variable. Accordingly, interpolation is intended to approximate the gradual evolution of trust between survey waves rather than capture short-term fluctuations. The interpolated series should therefore be interpreted as an approximation of the latent evolution of regional trust between survey observations.

[12] Firm-months with missing returns, including delisting months, are excluded from our main analysis. In Section 4.3, footnote 22, we show that our results are robust to incorporating delisting returns.

Table 1, Panel A, presents the details of the 10 US geographical regions in our study, the values of trust index (TRUST) for each region during the five survey years, the number of firms headquartered in each region during our sample period, as well as the total number of firms in our sample.

<Insert Table 1 about here>

In total, there are 49 US states and 8,875 firms during our sample period. Table 1 shows that the related rankings of social trust index (TRUST) in each region are relatively stable for each of the five survey waves (1995, 1999, 2006, 2011 and 2017). Higher index (TRUST) values correspond to higher trust; the Northwest region has the highest average TRUST, with values from 0.429 to 0.573, whereas the East South Central region has the lowest average TRUST, with values from 0.214 to 0.291.

TRUST also varies over time across these 10 different US geographical regions. For example, TRUST in the New England region increases from 0.331 in 1995 to 0.433 in 1999, then decreases slightly and remains stable between 0.425 and 0.415 from 2006 to 2011, but increases to 0.487 in 2017. However, the index in the Rocky Mountain region first increases from 0.275 in 1995 to 0.367 in 1999 and continues to increase to 0.434 in 2006, but its value decreases to 0.394 in 2011 and then to 0.382 in 2017.

We collect regional demographic variables, such as the religious belief (CATH) at the state level from the Association of Religion Data Archives, the education level (EDU) in each state from the Economic Research Service in the US Department of Agriculture, and the state-level median household income (MHI) and the state-level per capita personal income (PCI) from the US Census Bureau. CATH is the percentage of Catholic adherents in each state. The state-level CATH data are available for 1990, 2000, 2010 and 2020. We obtain CATH for each survey year and linearly interpolate CATH for interim years. EDU is the percentage of the population over 25 years old in each state that holds a bachelor's degree or higher. The data for EDU at the state level are available for 1990, 2000, 2010 and 2019; therefore, we linearly interpolate EDU for interim years. Similarly, data for state-level median household income (MHI) and per capita personal income (PCI) are available for 1990, 2000 and 2010. We linearly interpolate MHI and PCI for intermediate years and then apply their values in 2010 to the period after 2010 (the period 2010–2018)[13]. Since both median household income and per capita personal income have been used as the state income in prior studies (see, e.g., Wei and Zhang, 2020; Shu et al., 2012), we use MHI as the proxy for income in our main analysis, and PCI as an alternative

[13] We follow Hasan, Hoi, Wu, and Zhang (2017) as well as Shu et al. (2012) to backfill the data for the missing years using the available data from the most recent preceding year. Also, interpolation is a standard approach for estimating regional characteristics for non-available data points (Wei and Zhang, 2020; Hilary and Hui, 2009; Kumar et al., 2011; Shu et al., 2012).

proxy for income in the robustness analysis.

We further collect six additional state-level regional variables that include the state lottery revenue from the US Census Bureau Annual Survey of State and Local Government Finances, compiled by the Urban-Brookings Tax Policy Center,[14] the state gross domestic product (GDP) and population from the US Bureau of Economic Analysis (BEA), the state unemployment rate (UNEMP) from the Local Area Unemployment Statistics (LAUS) of the US Bureau of Labor Statistics, the state land area from the US Census Bureau 2010 Census, the religious adherents (REL) at the state level from the Association of Religion Data Archives, and the number of bank branches in each state from the summary of deposits of the Federal Deposit Insurance Corporation (FDIC). LOTTERY is the natural log of one plus net state lottery revenue (ticket sales minus prizes) per capita in dollars, and it captures the local propensity to gamble. The state lottery revenue data are available annually, and we set lottery revenue to zero for states and years without a state lottery. GDPPC is the natural log of state GDP per capita in dollars. UNEMP is the annual unemployment rate in each state in percent, computed as the total number of unemployed persons divided by the total labor force across all counties in each state. POPDEN is the natural log of state population per square mile of land area. Since the population data are available annually and land area is time-invariant, POPDEN varies annually. REL is the percentage of the state population who are adherents of any religious congregation. Similar to CATH, the state-level REL data are available for 1990, 2000, 2010 and 2020, and we linearly interpolate REL for interim years. BRANCH is the number of bank branches per 10,000 residents in each state, and it captures local financial development (Guiso et al., 2004b; Beck, Demirgüç-Kunt, and Martinez Peria, 2007). The Summary of Deposits data are available annually as of June 30 of each year. For each state in month *t*, these regional variables take their values in the corresponding year (year *t*).

Panel B of Table 1 shows the summary statistics of regional variables in our sample. These include the mean, standard deviation, 25th percentile, median, and 75th percentile of TRUST, CATH, EDU, MHI, PCI, REL, LOTTERY, GDPPC, UNEMP, POPDEN, and BRANCH. For the entire sample, the mean (median) value of TRUST is 0.38 (0.38), the mean (median) value of CATH is 23.93% (26.38%), the mean (median) value of EDU is 27.59% (27.17%), the mean (median) value for MHI is 48726.19 (47517.00) dollars, the mean (median) of PCI is 24741.23 (24260.50) dollars, the mean (median) of REL is 50.18% (49.75%), the mean (median) of LOTTERY is 3.88 (4.08), which corresponds to a median net lottery revenue of about 58 dollars per capita, the mean (median) of GDPPC is 10.68 (10.66), which corresponds to a median GDP per capita of about 42,700 dollars, the mean (median) of UNEMP is 5.66% (5.27%), the mean (median) of POPDEN is 5.31 (5.41), which corresponds to a median population density of about 224 residents per square mile, and the mean (median) of BRANCH is 2.88 (2.92) bank branches per 10,000 residents.

We obtain the firm-level transaction data (price, returns, volume and number of shares outstanding)

---

[14] The state lottery revenue data are obtained from the Urban-Brookings Tax Policy Center (2026), "Lottery Revenue From 1977 to 2022", available at https://taxpolicycenter.org/statistics/lottery-revenue. The Tax Policy Center compiles these data from the US Census Bureau's Annual Survey of State and Local Government Finances. Lottery revenue is measured in thousands of dollars on a state fiscal-year basis and is net of prizes, but not of administrative expenses. We scale lottery revenue by state population from the US Bureau of Economic Analysis to obtain lottery revenue per capita.

from the Center for Research in Security Prices (CRSP). We use daily stock returns to compute the maximum daily stock returns (MAX) for each stock in each month and other variables of interest including the market beta (BETA) and idiosyncratic skewness (ISKEW). We also use daily stock returns and volume data to calculate a measure of illiquidity (ILLIQ) following Amihud (2002). The CRSP daily value-weighted index is used as the proxy for market return. We use monthly returns to compute proxies for intermediate-term momentum (MOM) and short-term reversals (REV). Monthly share prices and shares outstanding are used to calculate market capitalization (MKTCAP). Additionally, we use Compustat to obtain equity book value of firms to calculate the book-to-market ratios (BTM) for each firm.

Panel C of Table 1 presents the summary statistics for stock variables of interest in our sample. These include the mean, standard deviation, 25th percentile, median, and 75th percentile of the stock variables. SIZE is market capitalization in millions of dollars (MKTCAP) after taking the natural logarithm.

Panel D reports the correlation matrix of the regional variables used in our analysis. The correlations are modest, suggesting that TRUST captures a distinct regional characteristic. TRUST is positively correlated with Catholic adherence (CATH: 0.32), education (EDU: 0.42), income (MHI: 0.41; PCI: 0.38), economic development (GDPPC: 0.32), lottery revenue (LOTTERY: 0.27), and local financial development (BRANCH: 0.19), and weakly negatively correlated with REL (−0.18), UNEMP (−0.12), and POPDEN (−0.16).

Detailed information on all variables constructed in this section is provided in the Appendix.

## 3. Lottery demand and the cross-section of expected stock returns

In this section, we investigate the negative relation between extreme positive daily return (MAX) in the previous month and the expected stock returns, namely the MAX effect, in our sample. Every month from January 1996 to December 2018 we sort all sample stocks into 10 decile portfolios based on the maximum daily return (MAX) within the previous month. Table 2 reports both value-weighted and equal-weighted average monthly raw returns, and alphas relative to the CAPM, Fama and French (1993) three-factor model (FF3), Fama and French (1993) and Carhart (1997) four-factor model (FFC4), Fama and French (2015) five-factor model (FF5), and Fama and French (2018) six-factor model, which augments FF5 with the momentum factor (FF6)[15], for the decile portfolios and the zero-cost portfolio[16] that is long the high MAX (decile 10) portfolio and short the low MAX (decile 1) portfolio. The *t*-statistics, adjusted following Newey-West (1987) using six lags, are presented in parentheses.

[15] Following Fama and French (1993, 2015, 2018) and Carhart (1997), the FF6 alpha is the estimated intercept from the regression of the excess portfolio return on the excess return of the market portfolio (MKTRF), the size factor (SMB), the value factor (HML), the profitability factor (RMW), the investment factor (CMA), and the momentum factor (UMD). We follow the same approach to calculate the FF5, FFC4, FF3 and CAPM alphas by using five factors (MKTRF, SMB and HML, RMW, and CMA), four factors (MKTRF, SMB and HML, and UMD), three factors (MKTRF, SMB and HML) and the market factor (MKTRF) as risk factors in regressions, respectively. MKTRF, SMB (small minus big), HML (high minus low), RMW (robust minus weak), CMA (conservative minus aggressive), and UMD (winner minus loser) are described in and obtained from Kenneth French's data library (http://mba.tuck.dartmouth.edu/pages/faculty/ken.french/).

[16] In other words, the zero-cost portfolio is the difference between high MAX (decile 10) and low MAX (decile 1) portfolios, also called the 'high-low MAX' portfolio.

<Insert Table 2 about here>

Consistent with the negative MAX premium between the high MAX (decile 10) portfolio and the low MAX (decile 1) portfolio in Bali et al. (2011), the results in Table 2 show that there is a negative relation between MAX and the cross-section of expected returns in our sample. The value-weighted average raw return for the zero-cost portfolio is −1.03% per month and marginally significant at the 10% level (*t*-statistic = −1.88). This high-low MAX spread becomes stronger, and both economically and statistically significant, at all conventional levels after adjusting for risk factors. The alphas with respect to CAPM, FF3, FFC4, FF5, and FF6 models are −1.57% (*t*-statistic = −3.34), −1.42% (*t*-statistic = −4.70), −1.16% (*t*-statistic = −3.87), −0.65% (*t*-statistic = −2.62), and −0.49% (*t*-statistic = −1.95) per month, respectively. Taking a closer look, the value-weighted average raw returns of the MAX-sorted decile portfolios tend to decrease, albeit not monotonically, from 1.28% per month for the low MAX (decile 1) portfolio to 0.25% per month for the high MAX (decile 10) portfolio. The corresponding alphas exhibit a similar pattern across the decile portfolios of MAX. The FF5 and FF6 alphas are smaller in magnitude than the other risk-adjusted returns, which is not surprising. High-MAX stocks are typically small and unprofitable firms, and the profitability factor (RMW) in these models is built on precisely this characteristic. Consistent with this, the zero-cost MAX portfolio loads heavily and negatively on RMW[17], so the profitability factor absorbs part of the unconditional MAX premium.

The raw returns and alphas of equal-weighted portfolios show a similar pattern to those of value-weighted portfolios, though they are less significant. For instance, we find that the equal-weighted average raw return of the zero-cost portfolio is −0.59% per month (*t*-statistic = −1.05). This is consistent with the results of Bali et al. (2011), who demonstrate that the negative MAX premium is stronger for value-weighted compared to equal-weighted portfolios. Similar to the value-weighted portfolios, the difference in returns between high-low MAX portfolios becomes stronger after adjusting for CAPM, FF3 and FFC4 risk factors, with alphas of −1.11% (*t*-statistic = −2.30), −0.97% (*t*-statistic = −2.79) and −0.64% (*t*-statistic = −1.81) per month, respectively. However, the FF5 and FF6 alphas are statistically insignificant, which is expected given that the profitability factor absorbs much of the MAX effect, as explained above.

In summary, the results in Table 2 show that the lottery-like portfolio of stocks (decile 10 with high MAX) significantly underperforms the non-lottery-like portfolio (decile 1 with low MAX), which confirms that there is a MAX effect in our sample of stocks.

## 4. Trust and the lottery demand anomaly

### 4.1. Univariate portfolio analysis

In this section, we examine whether social trust can influence the MAX anomaly. We begin our analysis by using univariate portfolio analysis to test if social trust has an impact on the MAX effect.

[17] In untabulated results, the zero-cost MAX portfolio has a loading of −1.58 (*t*-statistic = −10.7) on RMW in the FF5 model.

We first create a ranking (or threshold) of TRUST over time for each region where firms are headquartered, and assign each firm-month to the high-TRUST or low-TRUST group depending on whether the one-month lagged value of TRUST for the firm's region is above or below the threshold of TRUST, respectively.[18] We consider the median as well as mean of TRUST as alternative thresholds for TRUST. Then, for each group in each month, we sort our stocks into deciles based on MAX and calculate the average monthly raw returns and several risk-adjusted returns for decile portfolios and for the difference between high (decile 10) and low (decile 1) MAX portfolio for the high-TRUST and low-TRUST groups. The results are presented in Table 3. Panel A and Panel B of Table 3 report the results of value-weighted returns and equal-weighted returns for the high-TRUST and low-TRUST groups, respectively.[19] The Newey-West (1987) adjusted *t*-statistics are presented in parentheses using six lags.

<Insert Table 3 about here>

The results in Panel A of Table 3 show that for the value-weighted portfolios, the MAX effect is stronger, and both economically and statistically significant, for the high-TRUST group compared to the low-TRUST group. When we use the median of TRUST as the threshold, the raw return of the high-low MAX portfolio in the low-TRUST group is −0.87% per month with a *t*-statistic of −1.54. However, the corresponding raw return in the high-TRUST group is −1.34% per month with a *t*-statistic of −2.49. The magnitude of the high-low MAX spread is approximately 1.3 times larger than that generated by the unconditional portfolio analysis in Table 2 (−1.03% per month with a *t*-statistic of −1.88). We find a similar pattern for the risk-adjusted returns. For example, the CAPM alpha of the high-low MAX portfolio in the high-TRUST group is −1.85% per month with a *t*-statistic of −3.91, which is statistically significantly more negative compared with the corresponding return of −1.40% per month with a *t*-statistic of −2.82 in the low-TRUST group. The results are qualitatively similar when we use the mean of TRUST as the threshold.

We repeat the portfolio analysis in Panel A of Table 3 with the equal-weighted portfolios and report the results in Panel B of Table 3. The results for equal-weighted portfolios in Panel B show a similar pattern of raw returns and risk-adjusted returns to those for value-weighted portfolios in Panel A, albeit weaker. For instance, for the median of TRUST as the threshold, the FFC4 alpha of the high-low MAX portfolio in the high-TRUST group is −0.97% per month (*t*-statistic = −2.53), which is statistically significantly more negative compared with the corresponding return of −0.45% per month (*t*-statistic = −1.21) in the low-TRUST group. The FF5 and FF6 alphas of the high-low MAX portfolio are statistically insignificant in both trust groups. This is consistent with Table 2, where these models absorb the unconditional MAX premium for equal-weighted portfolios. Even in these cases, the high-low MAX alpha is more negative in the high-TRUST group than in the low-TRUST group in every specification. For example, for the median of TRUST as the threshold, the FF5 alpha is −0.53% per month in the high-TRUST group compared with 0.00% in the low-TRUST group. The results in Panel B again provide evidence that the MAX effect is stronger in regions with relatively high trust than in regions with relatively low trust.

[18] We use time-series variation in the trust index (TRUST) for each region because TRUST varies over time across different US geographical regions. Both the cross-sectional and time-series variation in TRUST across regions could be used to identify the link between social trust and the MAX effect. In the Internet Appendix (Table IA2), we also create a ranking of TRUST using region-level TRUST in each month to group our sample stocks into subsamples based on the highest and lowest TRUST; the results follow a similar pattern.

[19] For brevity, we do not report the results of the risk-adjusted returns for decile portfolios of MAX in Table 3.

Overall, the results in Table 3 support our hypothesis that social trust plays a significant role in shaping the MAX effect and the pricing of lottery-like stocks. Consistent with our belief-based framework, the evidence suggests that higher levels of social trust amplify the pricing of lottery-like stocks by reducing investors' skepticism toward optimistic information associated with extreme future payoffs. In this sense, our findings complement prior studies showing that trust facilitates participation in trust-intensive financial activities (e.g., Guiso et al., 2004a, 2008; El-Attar and Poschke, 2011; Barbi et al., 2023). More broadly, our results suggest that the role of social trust may differ across informational environments. Whereas prior studies show that trust enhances market efficiency through the incorporation of verifiable information (e.g., Pevzner et al., 2015), our findings are consistent with a belief-based mechanism through which social trust may contribute to stronger overpricing of lottery-like stocks.

### 4.2. Cross-sectional regression analysis

To investigate the impact of social trust on the MAX anomaly while controlling for other variables that have asset-pricing implications in the cross-section, we continue our analysis by conducting Fama-MacBeth (1973) cross-sectional regressions across firms headquartered in different regions. Consistent with the previous portfolio-level analysis, to differentiate the effect of different levels of social trust on the MAX anomaly, we run the cross-sectional regressions under regions with high and low trust, respectively.

Similarly, we first divide our sample into two groups (high-TRUST and low-TRUST) based on the ranking of TRUST over time for each region where firms are headquartered. If the firm's region's one-month lagged value of TRUST is above or below the thresholds of TRUST of the region over time, then the firm-month is placed into the high-TRUST or low-TRUST group, respectively. The thresholds are median and mean of TRUST of the region of the firm's headquarters over time. For ease of exposition, throughout the remainder of the paper we refer to observations with TRUST above (below) a region's historical threshold as high-TRUST (low-TRUST) regions. These terms should be understood as relative to each region's own historical average rather than absolute trust levels across regions.

Then, we run the monthly cross-sectional regressions for each high-TRUST and low-TRUST group. The full regression specification is as follows:

$$R_{i,t} = \lambda_{0,t-1} + \lambda_{1,t-1} MAX_{i,t-1} + \Lambda_{t-1} Z_{i,t-1} + \mathcal{E}_{i,t} \quad (1)$$

where $R_{i,t}$ is the realized return (in percent) on stock $i$ in month $t$, $MAX_{i,t-1}$ is the maximum daily return for stock $i$ in month $t-1$ and $\boldsymbol{Z}_{i,t-1}$ is a vector containing the control variables, which include market beta (BETA), natural log of market capitalization (SIZE), natural log of book-to-market ratio (BM), momentum (MOM), short-term reversal (REV) and illiquidity (ILLIQ). All the control variables are widely considered as determinants of stock returns in the cross-section, and their detailed definitions and constructions can be found in the Appendix. All independent variables are winsorized on a monthly basis at the 1st and 99th percentiles to alleviate concerns with outliers.

Table 4 reports the time-series averages of the slope coefficients, along with the Newey-West (1987) adjusted $t$-statistics using six lags (in parentheses) from the monthly cross-sectional regressions separately for the high-TRUST and low-TRUST groups.

<Insert Table 4 about here>

We find that the underperformance of lottery stocks that is proxied by MAX is more pronounced among the regions with high levels of trust than among the regions with low levels of trust. More specifically, for both median and mean thresholds, when the regression specification does not include control variables, the average coefficients on MAX in the high-TRUST group are more negative, and both economically and statistically significant, with values of −0.054 ($t$-statistic = −2.87) and −0.051 ($t$-statistic = −2.78), respectively. Conversely, the average coefficients on MAX in the low-TRUST group are insignificant, with values of −0.028 ($t$-statistic = −1.41) and −0.025 ($t$-statistic = −1.12), respectively, which are approximately half the values of those coefficients in the high-TRUST group. These results indicate that the MAX effect is roughly doubled across the regions with high trust.

When we add control variables to the regression specification, the results follow a similar pattern to the previous regression specification. For both median and mean thresholds, the average coefficients on MAX in the high-TRUST group decrease slightly in absolute magnitude, with values of −0.050 ($t$-statistic = −3.00) and −0.045 ($t$-statistic = −2.76), respectively, but these remain negative and statistically significant. However, the coefficients on MAX in the low-TRUST group become statistically insignificant, with values of −0.019 ($t$-statistic = −1.10) and −0.020 ($t$-statistic = −0.97), respectively. Similarly, we find that the coefficients on MAX for the high-TRUST group are approximately 2.3 to 2.6 times larger in absolute magnitude than those for the low-TRUST group.

In summary, the results in Table 4 provide direct cross-sectional evidence that social trust conditions the MAX effect and the pricing of lottery-like stocks. The negative relation between MAX and future returns is economically large and statistically significant for firms headquartered in high-trust regions, while it is weak and statistically insignificant in low-trust regions. This pattern indicates that social trust plays a central role in determining whether extreme positive return realizations translate into stronger overpricing and subsequent underperformance. Consistent with our hypothesis, the results suggest that social trust influences the pricing of lottery-like stocks by shaping investors' beliefs about optimistic information associated with extreme future payoffs, rather than through differences in risk tolerance (e.g., Guiso et al., 2004a, 2008; El-Attar and Poschke, 2011; Barbi et al., 2023).

### 4.3. Cross-sectional regression analysis with an interaction term

To examine whether our results of the cross-sectional regression across the full sample are robust and further examine the impact of social trust on the MAX effect, we re-run the Fama-MacBeth (1973) cross-sectional regressions after controlling for TRUST and an interaction term between MAX and a corresponding dummy variable that differentiates between firms that are headquartered in different regions with low and high TRUST values, without and with control variables. With this method, we can investigate how the MAX effect varies across firms' regions with different levels of social trust. We run the following monthly cross-sectional regression:

$$R_{i,t} = \lambda_{0,t-1} + \lambda_{1,t-1}MAX_{i,t-1} + \lambda_{2,t-1}TRUST_{i,t-1} + \lambda_{3,t-1}MAX_{i,t-1} \times DTRUST_{i,t-1} + \boldsymbol{\Lambda}_{t-1}\boldsymbol{Z}_{i,t-1} + \mathcal{E}_{i,t} \quad (2)$$

where $R_{i,t}$ is the realized return (in percent) on stock $i$ in month $t$, $MAX_{i,t-1}$ is the maximum daily return for stock $i$ in month $t-1$, and $TRUST_{i,t-1}$ is the social trust index in month $t-1$ for the region where firm $i$ (i.e., stock $i$'s firm) is headquartered. $DTRUST_{i,t-1}$ is a dummy variable that equals one or zero, which is defined based on TRUST of the region where firm $i$ is headquartered. If the firm's region's TRUST in month $t-1$ is above the median of TRUST of the firm's headquarters region over time (high level of trust), then $DTRUST_{i,t-1}$ is one and zero otherwise.[20] $\boldsymbol{Z}_{i,t-1}$ is a vector containing the control variables. Detailed information of the variables is provided in the Appendix. All independent variables are winsorized on a monthly basis at the 1st and 99th percentiles to eliminate the potential impact of the outliers.[21]

Table 5 presents the time-series averages of the slope coefficients for our full sample stocks from the outlined monthly cross-sectional regressions. The Newey-West (1987) adjusted $t$-statistics using six lags are presented in parentheses.

<Insert Table 5 about here>

Models (1) and (2) indicate that there is a significantly negative relation between the maximum daily return (MAX) over the previous month and the cross-sectional expected stock returns, which is consistent with the previous findings of the MAX effect in the US (e.g., Bali et al., 2011, among others). We find that the average coefficients on MAX are negative, and both economically and statistically significant, in the univariate model (1) and the multivariate model (2) that includes all control variables, with values of −0.038 ($t$-statistic = −2.08) and −0.030 ($t$-statistic = −1.93), respectively.

---

[20] In Table IA3 of the Internet Appendix, we also use the mean of TRUST as the threshold to define the dummy variable for TRUST (DTRUST). The results are robust.

[21] TRUST and DTRUST serve different purposes in the regression specification. TRUST captures the continuous association between social trust and expected stock returns, whereas DTRUST is used only to classify observations into relatively high- and low-trust environments for the interaction analysis. Accordingly, the interaction term MAX × DTRUST directly tests our hypothesis that the MAX effect differs across relatively high- and low-trust environments. In Table IA4 of the Internet Appendix, we also run a test that includes the main effect of DTRUST; the results remain qualitatively unchanged.

We then only include MAX and TRUST, without and with the control variables in the regression specification. The results are shown in models (3) and (4). In both models, the slope coefficients on MAX are statistically significant and they remain similar in absolute magnitude compared to those in models (1) and (2). This implies that MAX has information beyond social trust in explaining the cross-section of stock returns.

To explore the relation between the MAX effect and social trust, we consider the interaction term between MAX and the dummy variable for TRUST ($MAX \times DTRUST$). In models (5) and (6), we include MAX, TRUST, and $MAX \times DTRUST$ before and after considering control variables. The coefficient estimate on the interaction term can provide the economic difference of the MAX effect between firms headquartered in regions with low and high levels of social trust.

The results from model (5) show that the estimated coefficient on MAX becomes insignificant, with a value of $-0.030$ ($t$-statistic = $-1.51$), compared to the coefficients on MAX in models (1) and (3), whereas the estimated coefficient on the interaction term ($MAX \times DTRUST$) is significantly negative at the 5% significance level, with a value of $-0.022$ ($t$-statistic = $-2.22$). The results indicate that the MAX effect is weaker and statistically insignificant (coefficient estimate = $-0.030$) among firms' regions with low trust; however, it becomes stronger among firms' regions with high trust, with the estimated coefficient of $-0.022$ (formally, ($-0.022$) × 1, $t$-statistic = $-2.22$).

When we include all control variables in model (6), the results are consistent with those shown in model (5). The estimated coefficient on MAX tends to be less significant with a value of $-0.022$ ($t$-statistic = $-1.32$). However, the estimated coefficient on the interaction term ($MAX \times DTRUST$) is $-0.025$ ($t$-statistic = $-2.71$), which is significantly negative at the 1% significance level.[22] Similarly, the results imply that the MAX effect tends to be weaker and statistically insignificant (coefficient estimate = $-0.022$) among firms' regions with low trust, whereas it is stronger among firms' regions with high trust, with the estimated coefficient of $-0.025$ ($t$-statistic = $-2.71$). Thus, the results in models (5) and (6) show that the MAX effect among firms headquartered in high-trust regions is stronger than for firms headquartered in low-trust regions. Additionally, the slope coefficients on TRUST in both models (5) and (6) are marginally significant, with values of 1.321 ($t$-statistic = 1.87) and 1.070 ($t$-statistic = 1.70), respectively. The positive coefficients on TRUST suggest that there seems to be a positive relation between trust and expected stock returns, which is consistent with the finding of previous literature that lack of trust (the perceived risk of being cheated) decreases the expected return on the stock investment, making participation less attractive (e.g., Guiso et al., 2008; El-Attar and Poschke, 2011).

One potential concern regarding our findings of the effect of social trust on the MAX effect is that our social trust measure may be correlated with the regional characteristics that can also impact investor preference for lottery stocks. Kumar (2009) finds that Catholics and investors with lower levels of education and income have a stronger propensity to invest in lottery-like stocks. He also shows that investors who live in regions with greater participation in state lotteries and in urban areas are more inclined to hold lottery-type stocks, and that demand for such stocks increases when economic conditions deteriorate. Kumar et al. (2011) further show that local religious beliefs shape gambling attitudes and the demand for lottery-like stocks. Guiso et al. (2003) show that Catholics are also more trusting. Also, the effect of trust can vary with the level of education or wealth, because less educated or less wealthy people possess less reliable information and rely more on trust in making economic or financial decisions (e.g., Guiso et al., 2004a, 2008; El-Attar and Poschke, 2011; Pevzner et al., 2015). Moreover, local financial development affects households' access to financial services and their participation in financial markets (Guiso et al., 2004a).

[22] Thus, the additional decline in next-month returns associated with moving MAX from the 25th to the 75th percentile is approximately 18 basis points ($-0.025 \times (10.89 - 3.83) = -0.1765\%$), equivalent to approximately 2.1% on an annualized basis (0.1765% × 12).

To address this concern and ensure that the impact of trust on the lottery demand is not driven by these regional characteristics, we follow Bali et al. (2024) and control for a firm's headquarters characteristics at the state level. The regional factors include the percentage of the Catholic population (CATH); the percentage of the population aged 25 and older who attained a bachelor's degree or higher (EDU); the median household income (MHI) measured in thousands of dollars in each state;[23] the local propensity to gamble, measured by the natural log of one plus net state lottery revenue per capita (LOTTERY); local economic development, measured by natural log of state GDP per capita (GDPPC); local economic conditions, measured by the state unemployment rate (UNEMP); urbanization, measured by the natural log of state population density (POPDEN); overall religiosity, measured by the percentage of the state population who are adherents of any religious congregation (REL); and local financial development, measured by the number of bank branches per 10,000 residents (BRANCH). All regional variables are measured for the state in which a firm is headquartered. The detailed information of the variables can be found in the Appendix. We run monthly cross-sectional regressions for the following specification:

$$R_{i,t} = \lambda_{0,t-1} + \lambda_{1,t-1}MAX_{i,t-1} + \lambda_{2,t-1}TRUST_{i,t-1} + \lambda_{3,t-1}MAX_{i,t-1} \times DTRUST_{i,t-1} + \boldsymbol{\Lambda}_{t-1}\boldsymbol{Z}_{i,t-1} + \boldsymbol{\Gamma}_{t-1}\boldsymbol{REG}_{i,t-1} + \varepsilon_{i,t} \quad (3)$$

where $Z_{i,t-1}$ is the vector of stock-level control variables defined in Eq. (2) and $REG_{i,t-1}$ is the vector of regional control variables. The results are reported in Table 6. The Newey-West (1987) adjusted *t*-statistics using six lags are presented in parentheses.

<Insert Table 6 about here>

Models (2) to (11) of Table 6 show that after controlling for regional factors such as CATH, EDU, MHI, LOTTERY, GDPPC, UNEMP, POPDEN, REL, and BRANCH, the results for the estimated coefficients on MAX and the interaction term are qualitatively similar to those in model (6) of Table 5. For example, in model (11) of Table 6, consistent with the pattern in model (6) of Table 5, the estimated coefficient on MAX becomes less significant with a value of −0.025 ($t$-statistic = −1.53), whereas the estimated coefficient on the interaction term ($MAX \times DTRUST$) is significantly negative at conventional levels, with the value of −0.024 ($t$-statistic = −2.33). The results suggest that the MAX effect tends to be weaker and statistically insignificant (coefficient estimate = −0.025) among firms' regions with low trust, whereas it becomes stronger among firms' regions with high trust, with the estimated coefficient of −0.024 ($t$-statistic = −2.33) for the interaction term. Thus, the results in Table 6 indicate that the impact of social trust on the MAX effect is not driven by religion, education, income, local gambling propensity, economic development, local economic conditions, urbanization, overall religiosity, or local financial development.

[23] The results are qualitatively similar when we replace MHI (median household income) with PCI (per capita personal income) as an alternative measure of income in the unreported analyses.

Overall, the results reported in Tables 5 and 6 provide further evidence that social trust conditions the MAX effect through its interaction with lottery-like return characteristics. The interaction between MAX and social trust is negative and statistically significant, indicating that the underperformance of lottery-like stocks arises primarily when firms are headquartered in high-trust regions.[24] This finding reinforces the interpretation that social trust amplifies the pricing effects associated with lottery-like stocks, rather than exerting a uniform effect on stock returns across regions. Importantly, the results in Table 6 show that this interaction effect remains robust after controlling for a broad set of regional characteristics that may be correlated with both social trust and investor preferences for lottery-like stocks. This evidence suggests that the role of social trust in shaping the MAX effect is not driven by these correlated demographic factors but reflects an independent belief-based channel through which social trust influences lottery demand and the pricing of lottery-like stocks.[25]

### 4.4. Education, wealth, and the effect of trust

Social trust reflects both institutional characteristics of the financial system, such as the perceived likelihood of fraud or opportunistic behavior, and individuals' subjective beliefs formed from prior experiences and information (Guiso et al., 2004a, 2008). When trust primarily reflects such subjective priors, its influence on financial decision-making should depend on the availability and quality of investors' information environment. More educated individuals are likely to possess greater financial knowledge and a better understanding of how markets operate, making them less reliant on social trust when forming beliefs about uncertain investment opportunities (Guiso et al., 2004a, 2008; El-Attar and

---

[24] This result is robust to the treatment of delisting returns. Our main analysis excludes firm-months with missing returns, including delisting months. When we incorporate CRSP delisting returns, the coefficient on $MAX \times DTRUST$ is −0.024 ($t$-statistic = −2.67) in model (6) of Table 5, and −0.024 ($t$-statistic = −2.28) in model (11) of Table 6. When we further set the return for performance-related delistings without a delisting payment to −55%, following Shumway and Warther (1999), the corresponding coefficients are −0.025 ($t$-statistic = −2.72) in model (6) of Table 5 and −0.024 ($t$-statistic = −2.31) in model (11) of Table 6.

[25] Following the literature on lottery-related anomalies (e.g., Bali et al., 2011), we infer investors' demand for lottery-like stocks from the return predictability of MAX. To provide more direct evidence on investor behavior, we also examine share turnover. In unreported Fama-MacBeth (1973) regressions of the natural logarithm of monthly share turnover in month $t$ on MAX, the stock characteristic controls, and the regional variables in model (11) of Table 6, all measured in month $t$ – 1, the coefficient on MAX is positive and highly significant (0.042, $t$-statistic = 15.41), indicating that lottery-like stocks attract greater trading activity. Share turnover is computed as monthly trading volume divided by shares outstanding, with Nasdaq volume adjusted for double counting following Gao and Ritter (2010). We use the natural logarithm of turnover because turnover is highly right-skewed, and we winsorize it monthly at the 1st and 99th percentiles.

Poschke, 2011; Pevzner et al., 2015). If trust primarily shapes investors' beliefs, its influence should be greatest when investors evaluate lottery-like stocks, where uncertainty surrounding extreme future payoffs leaves greater scope for trust-based priors to affect investors' beliefs. Similarly, El-Attar and Poschke (2011) show that the effect of trust on households' portfolio choices is weaker for wealthier households, who typically have access to more reliable information and greater financial sophistication. As a result, these investors are less influenced by trust-based priors when making financial decisions. Accordingly, we predict that the effect of social trust on the MAX effect will be weaker in regions with higher levels of education or income, where investors rely more on private information and less on trust when forming beliefs about extreme payoff realizations.

To further examine how education and income moderate the influence of social trust on the MAX effect, we next estimate the models separately across education and income subsamples. We first examine whether the role of social trust in shaping the MAX effect varies with regional education levels. Consistent with our belief-based framework, we expect trust to play a more important role in driving the MAX effect in regions with lower education levels, where investors are likely to rely more heavily on trust relative to private information when forming beliefs about lottery-like stocks. We follow Kumar (2009) and assume that investors residing in regions with higher (lower) education levels are more likely to be more (less) educated. In each month, we rank states by their education level and classify firms headquartered in states with a one-month lagged EDU below (above) the monthly median as low-EDU (high-EDU).[26,27] We then run the monthly Fama-MacBeth cross-sectional regression Eq. (2) separately for the low-EDU and high-EDU subsamples. Following our main analysis, we define the trust dummy (DTRUST) using the median level of TRUST for the region in which each firm is headquartered.[28] All independent variables are winsorized at the 1st and 99th percentiles on a monthly basis to mitigate the influence of outliers.

Based on our conjecture, we expect the interaction between MAX and social trust to be more negative in regions with lower education levels, reflecting a stronger role for social trust when investors rely more heavily on social trust than on private information in forming investment beliefs. Accordingly, while we expect the MAX effect to be strongest among firms headquartered in regions characterized by both high social trust and lower education levels, we do not necessarily expect the influence of social trust to disappear in higher-education regions. Rather, education is expected to moderate the strength of the trust effect. The results are reported in Table 7. The *t*-statistics, adjusted following Newey-West (1987) using six lags, are presented in parentheses. For brevity, we report results for the main specifications for each subsample.

---

[26] We use cross-sectional variation in education level (EDU) across states in the analysis because EDU varies across the 49 different US states at the same time. However, EDU does not vary much for each state over time.

[27] Education is classified using cross-sectional rankings because it represents a relatively persistent structural characteristic of a region's information environment. The same classification approach is used for income in the subsequent analysis. By contrast, DTRUST is defined relative to each region's own historical average to capture whether the prevailing level of trust is relatively high or low compared with that region's long-run norm, rather than its ranking relative to other regions. Accordingly, although the heterogeneity tests are estimated cross-sectionally each month, the trust classification is based on a region-specific historical benchmark.

[28] In Table IA10 of the Internet Appendix, we also use the mean of EDU as the threshold to split our sample by education level and mean of TRUST to define the dummy variable for TRUST (DTRUST). The results are qualitatively similar.

<Insert Table 7 about here>

Models (1) and (2) provide preliminary evidence that the pricing impact of MAX differs across education groups. When MAX and control variables are included in the regression specification, the estimated coefficient on MAX is economically larger and statistically stronger in the low-EDU subsample than in the high-EDU subsample. For example, in model (1), the coefficient on MAX for the low-EDU group is −0.039 (*t*-statistic = −2.03), compared with −0.028 (*t*-statistic = −1.79) for the high-EDU group. This pattern remains when TRUST is added to the specification in model (2), suggesting that the MAX effect is stronger in lower-education regions (Kumar, 2009).

Model (3) introduces the interaction between MAX and social trust and provides direct evidence on whether social trust continues to influence the MAX effect after conditioning on education. For the low-EDU subsample, the interaction term (MAX × DTRUST) is negative and statistically significant (−0.046, *t*-statistic = −2.92), while the coefficient on MAX becomes smaller in magnitude and statistically insignificant (−0.021, *t*-statistic = −0.95). This pattern indicates that within low-education regions, the MAX effect is weak in environments with relatively low trust but becomes economically and statistically significant in environments with relatively high trust.

In contrast, for the high-EDU subsample, the interaction term remains negative but is economically smaller and only marginally significant (−0.017, *t*-statistic = −1.68), while the coefficient on MAX remains statistically indistinguishable from zero. The difference between the coefficients on MAX × DTRUST across the low-EDU and high-EDU groups is −0.029 (*t*-statistic = −1.70). These results suggest that social trust continues to condition the MAX effect within higher-education regions, although its influence is substantially weaker than in lower-education regions. Thus, investor sophistication moderates, rather than eliminates, the influence of social trust on the pricing of lottery-like stocks.

Comparing the interaction results across education groups highlights a clear asymmetry. Among firms headquartered in high-trust regions, the MAX effect is substantially stronger in low-education regions than in high-education regions, where the interaction effect is economically smaller and only marginally significant. These findings imply that lower education amplifies the pricing impact of extreme positive returns only when combined with high levels of social trust.

Overall, the evidence in Table 7 indicates that education plays an important role in shaping how social trust affects the MAX effect. When we differentiate regions by education level, the impact of social trust on the pricing of lottery-like stocks is substantially more pronounced in lower-education regions, suggesting that education shapes the extent to which social trust influences the pricing of lottery-like stocks. In these regions, weaker private information environments lead investors to rely more heavily on trust when forming investment beliefs, strengthening the pricing effects associated with lottery-like stocks and contributing to subsequent underperformance. By contrast, in higher-education regions where investors are better informed and rely less on trust, the effect of social trust on the MAX effect is weaker but remains present, indicating that education moderates rather than eliminates the conditioning role of social trust.

In addition, while prior studies suggest that less educated investors exhibit a stronger propensity to invest in lottery-type stocks (Kumar, 2009), our results show that this preference translates into pronounced mispricing only when social trust is sufficiently high. This pattern is consistent with our belief-based framework, in which investors rely more heavily on social trust when private information is relatively limited. We next examine whether the role of social trust in shaping the MAX effect varies with regional income levels. Consistent with our belief-based framework, we expect trust to play a more important role in driving the MAX effect in lower-income regions, where investors are more likely to rely on trust rather than private information when forming investment beliefs, due to more limited access to information and financial sophistication relative to wealthier investors. Similar to the education analysis, we assume that investors residing in high-income (low-income) regions tend to have higher (lower) income levels. We use median household income (MHI) at the state level as a proxy for regional income.

Following the procedure in Table 7, we re-estimate the cross-sectional regressions using income instead of education. In each month, states are ranked by their one-month lagged MHI, and firms headquartered in states below (above) the monthly median are classified into the low-MHI (high-MHI) group.[29] We define the dummy variable for trust (DTRUST) using the median level of TRUST for the region in which each firm is headquartered.[30] Consistent with our conjecture, we expect the interaction between MAX and social trust to be more negative in low-income regions, implying that social trust exerts a stronger influence on the MAX effect where investors have more limited access to private information. Table 8 presents the results.

<Insert Table 8 about here>

Models (1) and (2) provide preliminary evidence that the pricing impact of MAX differs across income groups. When MAX and control variables are included in the regression specification, the estimated coefficient on MAX is economically larger and statistically stronger in the low-MHI subsample than in the high-MHI subsample. For example, in model (1), the coefficient on MAX for the low-MHI group is −0.045 (*t*-statistic = −2.20), compared with −0.026 (*t*-statistic = −1.67) for the high-MHI group. This pattern is consistent with prior evidence that lower-income investors exhibit stronger preferences for lottery-type stocks (Kumar, 2009), although these specifications do not yet distinguish the role of social trust.

Model (3) introduces the interaction between MAX and social trust and provides direct evidence on how trust operates within each income group. For the low-MHI subsample, the coefficient on MAX becomes economically smaller and statistically insignificant (−0.018, *t*-statistic = −0.77), while the interaction term (MAX × DTRUST) is negative and highly significant (−0.061, *t*-statistic = −3.95). This pattern indicates that within low-income regions, the MAX effect is weak in low-trust environments but becomes economically large and statistically significant in high-trust environments.

[29] Similar to EDU, there is obvious variation in MHI across the 49 different US states at the same time, whereas MHI does not change much for each state over time. Thus, we use cross-sectional variation in income level (MHI) across states in each month in the analysis.

[30] In Table IA11 of the Internet Appendix, the results remain robust when we use the mean of MHI as the threshold to split our sample by the level of income and mean of TRUST to define the dummy variable for TRUST (DTRUST).

In contrast, for the high-MHI subsample, the interaction term remains negative but is economically smaller and only marginally significant (−0.017, *t*-statistic = −1.67), while the coefficient on MAX remains statistically indistinguishable from zero (−0.021, *t*-statistic = −1.24). The difference between the coefficients on MAX × DTRUST across the low-MHI and high-MHI groups is −0.044 (*t*-statistic = −2.43). These results suggest that social trust continues to condition the MAX effect within higher-income regions, although its influence is substantially weaker than in lower-income regions. Thus, greater investor sophistication moderates, rather than eliminates, the influence of social trust on the pricing of lottery-like stocks.

Overall, the evidence in Table 8 indicates that income moderates, rather than eliminates, the influence of social trust on the MAX effect. In lower-income regions, where investors are more likely to face weaker private information environments, social trust plays a more prominent role in shaping investors' investment beliefs, strengthening the MAX effect. By contrast, in higher-income regions where investors have greater access to information and rely less on trust when making financial decisions, the influence of social trust on the MAX effect is substantially weaker. These findings are consistent with El-Attar and Poschke (2011), who show that trust plays a smaller role in the financial decisions of wealthier households due to their superior information and financial sophistication. At the same time, our results extend this insight to asset pricing by showing that income-related differences in lottery demand translate into economically meaningful mispricing only in high-trust environments. While lower-income investors exhibit a stronger propensity to invest in lottery-type stocks (Kumar, 2009), our results suggest that this preference translates into economically meaningful mispricing primarily in high-trust environments, where investors are expected to rely more heavily on social trust than on private information when forming investment beliefs.[31]

## 5. Robustness tests

We conduct a series of additional robustness analyses to verify that our main findings are not sensitive to the alternative thresholds of social trust and lottery-like characteristics, alternative model specifications, or omitted regional heterogeneity. Below, we briefly summarize these analyses, while detailed results are reported in the Internet Appendix. The relevant sections and tables of the Internet Appendix are presented in parentheses.

[31] Consistent with the credulity interpretation, we also examine whether the interaction between MAX and DTRUST is stronger among firms whose fundamentals are more difficult to verify. Each month, we split the sample by intangible intensity (intangible assets scaled by total assets, above versus below the cross-sectional median), firm age (below versus above the cross-sectional median), R&D activity (positive versus zero R&D expenditure), and earnings (non-positive versus positive income before extraordinary items, IB), and re-estimate model (6) of Table 5 in each subsample. The coefficient on MAX × DTRUST is −0.028 (*t*-statistic = −2.49) for high-intangible firms versus −0.020 (*t*-statistic = −1.83) for low-intangible firms; −0.028 (*t*-statistic = −2.21) for young firms versus −0.018 (*t*-statistic = −1.56) for older firms; −0.031 (*t*-statistic = −2.66) for firms with R&D expenditure versus −0.012 (*t*-statistic = −1.12) for firms without R&D expenditure; and −0.032 (*t*-statistic = −2.41) for firms with no earnings versus −0.020 (*t*-statistic = −2.26) for firms with positive earnings. The result generally suggests that the impact of trust on the MAX effect is stronger among firms whose fundamentals are more difficult to verify.

We show that the MAX effect disappears when the component of MAX that is orthogonal to the social trust proxy (TRUST) is used as the sorting variable in the univariate portfolio analysis (Table IA1 of Section IA-1). This finding suggests that the underperformance of lottery-like stocks is closely linked to the trust-related component of MAX, consistent with our proposed belief-based mechanism.

In our main analysis, the univariate portfolio tests classify regions into high- and low-TRUST groups by ranking TRUST over time within each region. As a robustness check, we instead use the cross-sectional variation in TRUST across regions to construct the TRUST ranking. The results remain qualitatively unchanged (Table IA2 of Section IA-2).

In our main tests of Fama-MacBeth (1973) cross-sectional regressions with an interaction term, we use the median of TRUST as the threshold to define the dummy variable for TRUST (DTRUST). As a robustness check, we use the mean of TRUST to define DTRUST and show that the results remain qualitatively unchanged (Table IA3 of Section IA-3).

In addition, our baseline interaction regressions include TRUST and the interaction term $MAX \times DTRUST$ but do not include the main effect of DTRUST. Because TRUST and DTRUST are correlated but not identical variables, we re-estimate the Fama-MacBeth regressions by simultaneously including MAX, DTRUST, and $MAX \times DTRUST$. We conduct this analysis using both the median and mean of TRUST as alternative thresholds for constructing DTRUST. The results continue to support our main findings (Table IA4 of Section IA-4).

We also show that our main results continue to hold when we replace MAX with MAX(N), defined as the average of the N (N = 2, 3, 4, and 5) highest daily returns over the previous month, in the cross-sectional regressions (Tables IA5 and IA6 of Section IA-5). Likewise, replacing MAX with idiosyncratic skewness produces similar conclusions (Table IA7 of Section IA-6).

To address the concern that the control variables in the baseline cross-sectional regressions may not fully capture persistent regional heterogeneity, we estimate pooled panel regressions with regional and monthly fixed effects (Table IA8 of Section IA-7). Our main findings remain robust after controlling for time-invariant regional differences and common time shocks.

Furthermore, to examine whether the trust effect varies with firms' institutional ownership (IO), we split the sample into high- and low-IO subsamples and re-estimate the baseline Fama-MacBeth regressions (Table IA9 of Section IA-8). We find that the interaction between MAX and social trust is stronger among firms with lower institutional ownership. This finding provides additional evidence relevant to the transmission of regional trust to stock prices, although it does not identify a local retail-investor channel.

Next, we examine the robustness of our education- and income-based analyses. In the baseline regressions, we split the sample using the median of EDU and MHI across the states in each month and construct DTRUST using the median of TRUST. As robustness checks, we alternatively use the mean of EDU (Table IA10 of Section IA-9) and MHI (Table IA11 of Section IA-10) to split the sample, together with the mean of TRUST to construct DTRUST. In addition, we replace median household income (MHI) with per capita personal income (PCI) as an alternative measure of income and obtain qualitatively similar results (Table IA12 of Section IA-11). We further show that the effect of social trust on the MAX effect remains robust after controlling for the interactions of MAX with education and income (Table IA13 of Section IA-12).

Prior studies show that social trust affects firms' financing costs (Gupta et al., 2018; Meng and Yin, 2019; Brockman et al., 2020). Because lottery-like firms tend to be younger, have greater growth opportunities, and rely more heavily on external financing, an improvement in financing conditions associated with social trust could be particularly relevant for these firms, and subsequent issuance, investment expansion, or market timing could be associated with lower future returns. To examine this alternative explanation, we compare the net share issuance (NSI), equity issuance (EQUITYISS), net external financing (XFIN), asset growth (ASSETG), capital expenditures (CAPX), and firm age (AGE) of high-MAX stocks across trust groups, and re-estimate the Fama-MacBeth (1973) regressions after including these variables and their interactions with MAX (Table IA14 of Section IA-13). High-MAX stocks in regions with relatively high trust do not exhibit greater financing activity or investment and are not younger than those in regions with relatively low trust, and the coefficient on $MAX \times DTRUST$ remains significantly negative across all specifications.

To provide further empirical support for using firms' headquarters locations to assign regional trust, we conduct a geographic permutation placebo test (Figure IA1 of Section IA-14). We randomly reassign regional trust to firms, re-estimate the baseline Fama-MacBeth regression, and repeat this procedure 5,000 times. The observed coefficient on $MAX \times DTRUST$ lies in the extreme left tail of the placebo distribution, with only 0.02% and 0.92% of the placebo coefficients being as negative as the observed coefficient when trust is reassigned across firms and across regions, respectively. These results indicate that our findings depend on the actual correspondence between firms and regional trust.

We further show that our results are robust to controlling for political ideology, racial diversity, immigrant share, and social capital (Table IA15 of Section IA-15).

Finally, we examine whether our findings reflect social transmission bias (Hirshleifer, 2020), which predicts that the MAX effect should be stronger in regions with more intensive social interaction among investors (Bali et al., 2024). We augment our baseline cross-sectional regression with three proxies for local social transmission intensity: population density, the social capital index, and the Facebook-based social connectedness measure, and their interactions with MAX. The interactions between MAX and these proxies do not show the pattern predicted by social transmission bias, whereas the interaction between MAX and DTRUST remains negative and of similar magnitude (Table IA16 of Section IA-16).

## 6. Conclusion

This paper examines whether social trust plays a role in the underperformance of lottery-type stocks. Building on prior research showing that trust influences participation in trust-intensive financial activities (Guiso et al., 2004a, 2008; El-Attar and Poschke, 2011; Barbi et al., 2023), we argue that social trust influences investors' beliefs regarding optimistic but difficult-to-verify information associated with extreme future payoffs, rather than their intrinsic risk preferences. When the level of social trust is high, investors may apply less skepticism to such information, assigning greater credibility to claims regarding extreme future payoff potential. This belief-based mechanism contributes to stronger overpricing and subsequently lower expected returns. Accordingly, we hypothesize that the negative lottery premium is stronger in high-trust regions.

Using the maximum daily return over the previous month (MAX) as a proxy for lottery-like stocks, we find strong support for this hypothesis. The MAX effect is present only in regions with relatively high trust and is absent in regions with relatively low trust, indicating that social trust conditions whether lottery-like characteristics translate into return predictability. This relation is not driven by correlated regional characteristics such as religion, education, income, local lottery participation, economic development, unemployment, population density, or local financial development. However, we show that education and income shape the extent to which trust influences the MAX effect. Specifically, the influence of social trust on the pricing of lottery-like stocks is substantially more pronounced in regions with lower education or income levels, where investors face weaker private information environments and rely more heavily on trust when forming investment beliefs. In contrast, in regions with higher education or income levels, where investors have better access to information and rely less on trust, the influence of social trust on the MAX effect is weaker but remains present, indicating that education and income moderate rather than eliminate its conditioning role. Our results are robust to alternative measures of lottery-like stocks and a range of additional robustness tests.

Overall, our findings extend the literature on lottery-related anomalies by identifying social trust as an important determinant of the MAX effect and the pricing of lottery-like stocks. More broadly, the paper contributes to the growing literature on social capital and culture in finance by showing how social trust shapes belief formation and asset prices, particularly in environments characterized by highly uncertain and positively skewed payoffs. Our findings should not be interpreted as suggesting that social trust is detrimental to financial markets. A large body of research shows that social trust promotes cooperation, facilitates economic exchange, and improves a wide range of financial and economic outcomes. Rather, our results demonstrate that the asset-pricing implications of social trust depend on the informational environment. In the specific setting of lottery-like stocks, where extreme positive outcomes are difficult to verify, higher social trust may reduce investors' skepticism toward optimistic narratives, thereby contributing to stronger belief-driven overpricing and lower subsequent returns.

## Appendix: Variable Definitions

**MAX:** The maximum daily return within a month as in Bali et al. (2011), which is calculated as the following equation:

$$\mathrm{MAX}_{i,t} = \max\left(R_{i,d}\right), \quad d = \{1,2, \ldots D_t\}, \qquad \text{(A1)}$$

where $R_{i,d}$ is the return of stock $i$ on day $d$ and $D_t$ is the number of trading days in month $t$.

**MAX(N):** the average of the N (N = 2, 3, 4, 5) highest daily returns within a month.

**BETA:** The market beta over the month. Following Scholes and Williams (1977) and Dimson (1979), we consider nonsynchronous trading and ged, current and lead market return when estimating beta. We estimate beta using daily returns within each month for each stock:

$$\mathrm{R}_{i,d} - \mathrm{r}_{f,d} = \alpha_i + \beta_{1,i}\left(R_{m,d-1} - r_{f,d-1}\right) + \beta_{2,i}\left(R_{m,d} - r_{f,d}\right) + \beta_{3,i}\left(R_{m,d+1} - r_{f,d+1}\right) + \varepsilon_{i,d}, \text{(A2)}$$

where $R_{i,d}$ is the return of stock $i$ on day $d$ and $R_{m,d}$ is the market return on day $d$, which is measured by the CRSP daily value-weighted index, and $r_{f,d}$ is the risk-free rate on day $d$, which is the one-month T-bill return obtained from Kenneth French's online data library. The market beta of stock $i$ in month $t$ is computed as $\widehat{\beta_{\iota,t}} = \widehat{\beta_{1,\iota}} + \widehat{\beta_{2,\iota}} + \widehat{\beta_{3,\iota}}$.

**MKTCAP:** The market value of equity, which is measured by a stock's price multiplied by shares outstanding in millions of dollars at the end of month $t-1$ for each stock.

**SIZE:** The natural log of market capitalization (MKTCAP).

**BTM:** Following Fama and French (1992), a firm's book-to-market ratio in month $t$ is calculated by using the market value of equity at the end of December of the previous year and the book value at the end of the previous fiscal year. The book value of equity is calculated from the balance sheet data from Compustat, which is defined as stockholder's equity plus deferred taxes and investment tax credit then subtracting the book value of preferred stock. The book value of preferred stock is taken to be the redemption value, the liquidating value or the par value, taken as available in the given order.

**BM:** The natural log of the book-to-market ratio (BTM). Observations with non-positive book equity are excluded when constructing BTM.

**MOM:** Following Jegadeesh and Titman (1993), momentum for each stock in month $t$ is defined as the cumulative return on the stock over the previous 11 months, starting two months earlier; that is, the cumulative return from month $t-12$ to month $t-2$, inclusive.

**REV:** The short-term reversal for each stock in month *t*, which is defined as the return on the stock over the past one month based on Jegadeesh (1990); that is, the return in month $t-1$.

**ILLIQ:** Following Amihud (2002), the illiquidity for each stock in month $t$ is measured by using the monthly average of the absolute daily stock return divided by its dollar trading volume:

$$\text{ILLIQ}_{\text{i,t}} = \frac{1}{D}\sum_{d=1}^{D}\frac{|R_{i,d}|}{VOLD_{i,d}}, \qquad \text{(A3)}$$

where $R_{i,d}$ is the return of stock $i$ on day $d$ and $VOLD_{i,d}$ is the daily trading volume in dollars for each stock, measured by volume times the stock price in millions of dollars. Following Lin and Liu (2018), we also require a minimum of ten non-missing daily trading volumes within a given month to compute the illiquidity.

**IO:** Institutional ownership of each stock, measured as the sum of the percentage holdings reported by institutional investors in Form 13F filings.

**NSI:** Net share issuance, following Pontiff and Woodgate (2008), measured as the change in the natural log of shares outstanding over the past 12 months, adjusted for stock splits and stock dividends using the CRSP share adjustment factor.

**EQUITYISS:** Equity issuance, measured as the sale of common and preferred stock in the most recent fiscal year, scaled by total assets at the end of the previous fiscal year.[32]

**XFIN:** Net external financing, following Bradshaw, Richardson, and Sloan (2006), measured as the sale of common and preferred stock minus the purchase of common and preferred stock, plus long-term debt issuance minus long-term debt reduction, in the most recent fiscal year, scaled by total assets at the end of the previous fiscal year.

**ASSETG:** Asset growth, following Cooper, Gulen, and Schill (2008), measured as the percentage change in total assets from the previous fiscal year to the most recent fiscal year.

**CAPX:** Capital expenditures in the most recent fiscal year, scaled by total assets at the end of the previous fiscal year.

**AGE:** Firm age, measured as the natural log of one plus the number of months since the it first appears in CRSP.

**ISKEW:** Following Bali et al. (2011) and Harvey and Siddique (2000), the idiosyncratic skewness is estimated from the following regression for each stock:

$$\text{R}_{\text{i,d}} - \text{r}_{\text{f,d}} = \alpha_{\text{i}} + \beta_{\text{i}}\left(R_{m,d} - r_{f,d}\right) + \gamma_{\text{i}}\left(R_{m,d} - r_{f,d}\right)^2 + \varepsilon_{i,d}, \qquad \text{(A4)}$$

where $R_{i,d}$ is the return of stock $i$ on day $d$, $R_{m,d}$ is the market return on day $d$, which is measured by the CRSP daily value-weighted index, $r_{f,d}$ is the risk-free rate on day $d$, which is the one-month T-bill return obtained from

[32] For EQUITYISS, XFIN, ASSETG, and CAPX, accounting data are obtained from Compustat and are used from six months after the fiscal year-end to ensure that the information is publicly available.

Kenneth French's online data library, and $\varepsilon_{i,d}$ is the idiosyncratic return on day $d$. The idiosyncratic skewness (ISKEW) of stock $i$ in month $t$ is computed as the skewness (or third moment) of daily residuals $\varepsilon_{i,d}$ in month $t$.

**TRUST:** Social trust index, which is commonly measured based on the responses to the following World Values Survey (WVS) question, "*Generally speaking, would you say that most people can be trusted or that you need to be very careful in dealing with people?*", following prior studies (Guiso et al., 2008; Aghion et al., 2010; Pevzner et al., 2015; Wei and Zhang, 2020). In this study, the region-level trust data are available and collected from the WVS from 1995 to 2020. There are five waves of the WVS surveys (waves 3, 4, 5, 6, 7) conducted in the United States in 1995, 1999, 2006, 2011, and 2017. In line with Wei and Zhang (2020), the surveys report the locations of survey respondents by 10 US geographical regions: New England (Maine, New Hampshire, Vermont, Massachusetts, Rhode Island, and Connecticut), Middle Atlantic (New York, Pennsylvania, and New Jersey), South Atlantic (Delaware, Maryland, Washington, D.C., Virginia, West Virginia, North Carolina, South Carolina, Georgia, and Florida), East South Central (Kentucky, Tennessee, Mississippi, and Alabama), West South Central (Oklahoma, Texas, Arkansas, and Louisiana), East North Central (Wisconsin, Michigan, Illinois, Indiana, and Ohio), West North Central (Missouri, North Dakota, South Dakota, Nebraska, Kansas, Minnesota, and Iowa), Rocky Mountain (Montana, Wyoming, Nevada, Utah, Colorado, Arizona, New Mexico and Idaho), Northwest (Oregon and Washington), and California. In total, there are approximately 8,479 survey respondents in the five waves of the WVS. Similar to Pevzner et al. (2015), in each region for each year, we recode the response to this question to 1 if a respondent answers that most people can be trusted and 0 otherwise, and then we calculate the mean of the responses as TRUST. Given that the index we calculated (TRUST) is available for the survey years, we use linear interpolation to estimate the value of TRUST in the years between two consecutive surveys for each region.[33] Thus, TRUST of month $t$ for each region is the value of TRUST at the corresponding year (in year $t$). Higher index values correspond to higher trust.

**DTRUST:** A dummy variable based on TRUST within each region where firms are headquartered. If a firm's region's monthly TRUST value is above (high trust) the thresholds (median and mean, respectively) of TRUST of the region over time, then DTRUST is one and zero otherwise.

$\boldsymbol{MAX \perp TRUST}$**:** The portion of MAX of stock $i$ in month $t$ ($MAX_{i,t}$) that is orthogonal to (or that cannot be explained by) the trust index in month $t$ ($TRUST_{i,t}$), which can be estimated from the following regression of MAX on TRUST for each stock $i$ over time:

[33] In the literature, interpolation is a standard approach for estimating regional characteristics for non-available data points. See, e.g., Wei and Zhang (2020); Hilary and Hui (2009); Kumar et al. (2011); and Shu et al. (2012).

$$\mathrm{MAX}_{i,t} = \alpha_i + \beta_i \mathrm{TRUST}_{i,t} + \varepsilon_{i,t}, \tag{A5}$$

where $\mathrm{MAX}_{i,t}$ is the maximum daily return of stock $i$ in month $t$, $\mathrm{TRUST}_{i,t}$ is the social trust for the region where the firm of stock $i$ is headquartered in month $t$, and $\varepsilon_{i,t}$ is the $MAX \perp TRUST$ of stock $i$ in month $t$.

**CATH:** The percentage of Catholic adherents in each state. We obtain the state-level CATH data from the Association of Religion Data Archives (ARDA). The CATH data are available for 1990, 2000, 2010, and 2020. We therefore linearly interpolate CATH for interim years following the same procedure as TRUST described above. Then, CATH for each state in month $t$ is the value of CATH at the corresponding year (in year $t$). Higher values correspond to higher percentage of Catholic adherents.

**EDU:** The percentage of the population over 25 years old in each state holding a bachelor's degree or higher. In our study, the state education level data are obtained from the Economic Research Service in the US Department of Agriculture and expressed in percent. Since data for EDU at the state level are available for 1990, 2000, 2010, and 2019, we follow the same procedure as TRUST described above to linearly interpolate EDU for interim years. Thus, EDU for each state in month $t$ is the value of EDU at the corresponding year (in year $t$). Higher values correspond to higher education level.

**MHI:** Median household income in each state. In our study, the state-level median household income data are obtained from the US Census Bureau and expressed in dollars. Since data for MHI at the state level are available for 1990, 2000 and 2010, we follow the same procedure as EDU to linearly interpolate MHI for interim years and apply its value in 2010 to the periods after 2010 until 2018.[34] Likewise, MHI for each state in month $t$ is the value of MHI at the corresponding year (in year $t$). Higher values correspond to higher income level.

**PCI:** Per capita personal income in each state. The state-level per capita personal income data are obtained from the US Census Bureau and expressed in dollars. Since data for PCI at the state level are available for 1990, 2000, and 2010, we linearly interpolate PCI for interim years, and then apply its value in 2010 to the periods after 2010 until 2018. Similarly, PCI for each state in month $t$ is the value of PCI at the corresponding year (in year $t$). Higher values correspond to higher income level.

**REL:** The percentage of the state population who are adherents of any religious congregation, obtained from the Association of Religion Data Archives (ARDA). Since the data are available for 1990, 2000, 2010 and 2020, we linearly interpolate REL for interim years following the same procedure as CATH. REL for each state in month $t$ is the value of REL at the corresponding year (in year $t$).

---

[34] We follow Hasan et al. (2017) as well as Shu et al. (2012) to backfill the data for the missing years using the available data from the most-recent preceding year.

**LOTTERY:** The natural log of one plus net state lottery revenue per capita (in dollars), where net lottery revenue is lottery ticket sales minus prizes. Lottery revenue data are from the US Census Bureau Annual Survey of State and Local Government Finances, compiled by the Urban-Brookings Tax Policy Center, and population data are from the US Bureau of Economic Analysis. Net lottery revenue is set to zero for states and years without a state lottery.[35] LOTTERY for each state in month $t$ is the value of LOTTERY at the corresponding year (in year $t$).

**GDPPC:** The natural log of state gross domestic product (GDP) per capita (in dollars). The state GDP and population are sourced from the US Bureau of Economic Analysis (BEA). The BEA state GDP data based on the North American Industry Classification System (NAICS) are available annually from 1997. For 1995 and 1996, we use the BEA state GDP data based on the Standard Industrial Classification (SIC), scaled by the ratio of NAICS-based to SIC-based GDP for each state in 1997.[36] GDPPC for each state in month $t$ is the value of GDPPC at the corresponding year (in year $t$).

**UNEMP:** The annual state unemployment rate (in percent), from the Local Area Unemployment Statistics of the US Bureau of Labor Statistics. UNEMP for each state in month $t$ is the value of UNEMP at the corresponding year (in year $t$).

**POPDEN:** The natural log of state population per square mile of land area. Population data are from the US Bureau of Economic Analysis and land area from the US Census Bureau. POPDEN for each state in month $t$ is the value of POPDEN at the corresponding year (in year $t$).

**BRANCH:** The number of bank branches per 10,000 residents in each state, which captures local financial development (Guiso et al., 2004b; Beck et al., 2007). The number of bank branches is obtained from the Summary of Deposits of the Federal Deposit Insurance Corporation (FDIC), which reports the number of branches as of June 30 of each year, and population data are from the US Bureau of Economic Analysis. BRANCH for each state in month $t$ is the value of BRANCH at the corresponding year (in year $t$). Higher values correspond to greater access to banking services.

**REP:** The Republican share of the two-party vote in presidential elections in each state. The state-level presidential election returns are obtained from the MIT Election Data and Science Lab, and REP is calculated as the number of votes for the Republican candidate divided by the total number of votes for the Republican and Democratic candidates, expressed in percent. Since the data are available only for presidential election years (1992, 1996, …2020), we linearly interpolate REP for the interim years. REP for each state in month *t* is the value of REP at the corresponding year (in year *t)*. Higher values correspond to a more conservative political ideology.

[35] Four states in our sample, Alabama, Mississippi, Nevada, and Utah, did not operate a state lottery at any point during our sample period, and LOTTERY therefore equals zero for these states in all years. In addition, eight states introduced a state lottery during our sample period, and LOTTERY equals zero for the years before the lottery's first year of operation: New Mexico (first year with lottery revenue, 1996); South Carolina, 2002; North Dakota, 2004; Tennessee, 2004; North Carolina, 2006; Oklahoma, 2006; Arkansas, 2010; and Wyoming, 2014. The remaining 37 states in our sample operated a state lottery throughout the sample period. Alaska and Hawaii, which also have no state lottery, are not represented in our sample.

[36] In 1997, the BEA changed the industry classification underlying its state GDP estimates from the SIC to the NAICS, so the two series are not directly comparable. The ratio of NAICS-based to SIC-based state GDP in 1997, the year in which both series are available, is close to one for most states.

**NONWHITE**: The percentage of the state population that is not non-Hispanic white. We obtain annual state population estimates by race and Hispanic origin from the US Census Bureau (the 1990–2000 and 2000–2010 intercensal estimates, and the 2010–2020 intercensal estimates). NONWHITE is calculated as one minus the ratio of the non-Hispanic white population to the total state population, expressed in percent. NONWHITE for each state in month *t* is the value of NONWHITE at the corresponding year (in year *t*). Higher values correspond to greater racial diversity.

**FOREIGN:** The percentage of the state population that is foreign-born. The data are obtained from the Migration Policy Institute (MPI) tabulation of the 1990 and 2000 Decennial Census and the 2010 and 2019 American Community Surveys. Since the data are available for 1990, 2000, 2010, and 2019, we linearly interpolate FOREIGN for interim years. FOREIGN for each state in month *t* is the value of FOREIGN at the corresponding year (in year *t*). Higher values correspond to a higher immigrant share of the population.

**SOCCAP**: The Social Capital Index in each state. We obtain the county-level social capital index of Rupasingha, Goetz, and Freshwater (2006) from the Northeast Regional Center for Rural Development (NERCRD) at Pennsylvania State University. The index is the first principal component of four factors: the density of associations (civic, religious, business, political, professional, labor, bowling, fitness, golf, and sports organizations), the number of nonprofit organizations, voter turnout in presidential elections, and the response rate to the census. We aggregate the county-level index to the state level using county population as weights. Since the index is available for 1997, 2005, 2009, and 2014, we linearly interpolate SOCCAP for interim years and apply the value in 1997 (2014) to the years before 1997 (after 2014). SOCCAP for each state in month *t* is the value of SOCCAP at the corresponding year (in year *t*). Higher values correspond to a higher level of social capital.

**SCIH**: The social connectedness of the state in which a firm is headquartered. We obtain the Facebook Social Connectedness Index (SCI) of Bailey et al. (2018) from the Humanitarian Data Exchange (Johnston, Kuchler, Kulkarni, and Stroebel, 2026). SCI measures the relative probability of a Facebook friendship link between the users in two locations. It is defined as the number of Facebook friendship links between locations *i* and *j* divided by the product of the numbers of Facebook users in *i* and *j*. Following Bali et al. (2024), we calculate SCIH as the natural log of the sum of the SCI between the firm's headquarters state and all US states, including the state itself. Since the SCI is available only as a cross-section based on a snapshot of Facebook friendship networks as of January 2026, SCIH is time-invariant, and we apply its value to all months in our sample period. Higher values correspond to stronger social connectedness.

## Table 1. Summary statistics

The table reports summary statistics for the 10 US geographical regions and the main variables used in our analysis. Panel A provides the details of 10 US geographical regions, the values of trust index (TRUST) for each region during the five survey years, as well as the distribution of the number of Nasdaq-listed firms across regions. In each region year, we recode the response to the World Values Survey (WVS) survey question regarding the social trust to 1 if a respondent answers that most people can be trusted and 0 otherwise and then calculate the mean of the responses as the measure of trust – TRUST. Panel B shows the summary statistics of regional variables in our sample. TRUST is the social trust index. CATH (in percent) is the state-level percentage of Catholic adherents. EDU (in percent) is the state-level percentage of the population over 25 years old with a bachelor's degree or higher. MHI (in dollars) is the state-level median household income. PCI (in dollars) is the state-level per capita personal income. REL (in percent) is the percentage of the state population who are adherents of any religious congregation. LOTTERY is the natural log of one plus net state lottery revenue (ticket sales minus prizes) per capita in dollars. GDPPC is the natural log of state GDP per capita in dollars. UNEMP (in percent) is the state unemployment rate. POPDEN is the natural log of the number of state residents per square mile of land area. BRANCH is the number of bank branches per 10,000 residents. Panel C presents the summary statistics for stock variables of interest in our study. MAX (in percent) is the maximum daily return over a month. ISKEW is the idiosyncratic skewness. BETA is the market beta. SIZE is the natural log of market capitalization in millions of dollars. BTM is the book-to-market ratio. MOM (in percent) is the momentum. REV (in percent) is the short-term reversal. ILLIQ (scaled by $10^6$) is the Amihud (2002) measure of illiquidity. Panel D reports the pairwise correlation matrix of the state-level, regional variables used in the analysis. The detailed definitions of all variables are presented in the Appendix. The sample includes Nasdaq-listed common stocks (CRSP share codes 10 and 11) headquartered in the 10 US geographical regions. No price, market capitalization, or minimum daily observation filters are imposed, except that ILLIQ requires at least ten daily observations in the month. Observations with non-positive book equity are excluded when constructing the book-to-market ratio, and firm-months with missing returns, including delisting months, are excluded. The sample period is from January 1996 to December 2018.

Panel A: Trust index by region across survey years and the number of Nasdaq-listed firms within each region

| Region | States | TRUST | | | | | Number of Nasdaq-Listed Firms |
|---|---|---|---|---|---|---|---|
| | | 1995 | 1999 | 2006 | 2011 | 2017 | |
| 1 New England | Maine, New Hampshire, Vermont, Massachusetts, Rhode Island, Connecticut | 0.331 | 0.433 | 0.425 | 0.415 | 0.487 | 848 |
| 2 Middle Atlantic | New York, Pennsylvania, and New Jersey | 0.364 | 0.405 | 0.389 | 0.362 | 0.416 | 1492 |
| 3 South Atlantic | Delaware, Maryland, Washington, D.C., Virginia, West Virginia, North Carolina, South Carolina, Georgia, Florida | 0.246 | 0.311 | 0.382 | 0.342 | 0.359 | 1452 |
| 4 East South Central | Kentucky, Tennessee, Mississippi, Alabama | 0.263 | 0.214 | 0.231 | 0.291 | 0.271 | 246 |
| 5 West South Central | Oklahoma, Texas, Arkansas, Louisiana | 0.417 | 0.313 | 0.378 | 0.299 | 0.367 | 799 |
| 6 East North Central | Wisconsin, Michigan, Illinois, Indiana, Ohio | 0.394 | 0.406 | 0.389 | 0.350 | 0.396 | 901 |
| 7 West North Central | Missouri, North Dakota, South Dakota, Nebraska, Kansas, Minnesota, Iowa | 0.315 | 0.472 | 0.407 | 0.514 | 0.504 | 467 |

*Continued on next page…*

*…Continued*

| Region | States | TRUST | | | | | Number of Nasdaq-Listed Firms |
|---|---|---|---|---|---|---|---|
| | | 1995 | 1999 | 2006 | 2011 | 2017 | |
| 8 Rocky Mountain | Montana, Wyoming, Nevada, Utah, Colorado, Arizona, New Mexico, Idaho | 0.275 | 0.367 | 0.434 | 0.394 | 0.382 | 565 |
| 9 Northwest | Oregon, Washington | 0.508 | 0.458 | 0.531 | 0.573 | 0.429 | 276 |
| 10 California | California | 0.346 | 0.333 | 0.421 | 0.421 | 0.387 | 1829 |
| Total | 49 states | - | - | - | - | - | 8875 |

Panel B: Summary Statistics of Regional Variables

| Variables | Mean | St.Dev | 25th Percentile | Median | 75th Percentile |
|---|---|---|---|---|---|
| TRUST | 0.38 | 0.06 | 0.35 | 0.38 | 0.41 |
| CATH | 23.93 | 11.62 | 15.50 | 26.38 | 30.92 |
| EDU | 27.59 | 5.15 | 24.05 | 27.17 | 30.91 |
| MHI | 48726.19 | 9263.15 | 41937.60 | 47517.00 | 55521.50 |
| PCI | 24741.23 | 4905.31 | 20940.00 | 24260.50 | 28190.00 |
| REL | 50.18 | 8.25 | 44.88 | 49.75 | 56.39 |
| LOTTERY | 3.88 | 1.11 | 3.51 | 4.08 | 4.52 |
| GDPPC | 10.68 | 0.27 | 10.48 | 10.66 | 10.87 |
| UNEMP | 5.66 | 1.91 | 4.45 | 5.27 | 6.40 |
| POPDEN | 5.31 | 0.95 | 4.63 | 5.41 | 5.83 |
| BRANCH | 2.88 | 0.74 | 2.25 | 2.92 | 3.41 |

Panel C: Summary Statistics of Stock Variables

| Variables | Mean | St.Dev | 25th Percentile | Median | 75th Percentile |
|---|---|---|---|---|---|
| MAX | 9.23 | 11.85 | 3.83 | 6.37 | 10.89 |
| ISKEW | 0.24 | 0.96 | -0.27 | 0.20 | 0.72 |
| BETA | 0.88 | 4.13 | -0.10 | 0.75 | 1.81 |
| SIZE | 5.03 | 1.80 | 3.78 | 4.95 | 6.17 |
| BTM | 0.72 | 0.86 | 0.29 | 0.54 | 0.89 |
| MOM | 14.79 | 88.34 | -25.84 | 3.32 | 33.91 |
| REV | 1.20 | 20.79 | -7.60 | 0.00 | 7.71 |
| ILLIQ | 3.65 | 14.18 | 0.01 | 0.07 | 0.85 |

*Continued on next page…*

*...Continued*

Panel D: Correlation Matrix of State-Level Regional Variables

| Variables | TRUST | CATH | EDU | MHI | PCI | REL | LOTTERY | GDPPC | UNEMP | POPDEN |
|---|---|---|---|---|---|---|---|---|---|---|
| CATH | 0.32 | | | | | | | | | |
| EDU | 0.42 | 0.38 | | | | | | | | |
| MHI | 0.41 | 0.31 | 0.86 | | | | | | | |
| PCI | 0.38 | 0.30 | 0.87 | 0.95 | | | | | | |
| REL | -0.18 | 0.23 | -0.11 | -0.17 | -0.17 | | | | | |
| LOTTERY | 0.27 | 0.41 | 0.44 | 0.36 | 0.42 | -0.23 | | | | |
| GDPPC | 0.32 | 0.21 | 0.76 | 0.79 | 0.87 | -0.01 | 0.38 | | | |
| UNEMP | -0.12 | -0.05 | 0.02 | 0.17 | 0.23 | -0.13 | 0.09 | 0.14 | | |
| POPDEN | -0.16 | 0.35 | 0.43 | 0.42 | 0.50 | 0.10 | 0.48 | 0.47 | 0.25 | |
| BRANCH | 0.19 | -0.04 | -0.14 | -0.19 | -0.11 | 0.28 | -0.04 | -0.07 | -0.25 | -0.32 |

**Table 2. Returns and alphas of portfolios sorted on MAX**

The table reports the value-weighted and equal-weighted average monthly returns as well as the CAPM alphas, Fama-French three-factor alphas (FF3 alpha), Fama-French-Carhart four-factor alphas (FFC4 alpha), Fama-French five-factor alphas (FF5 alpha), and Fama and French (2018) six-factor model, which augments FF5 with the momentum factor (FF6 alpha) on the portfolios. Decile portfolios are formed every month from January 1996 to December 2018 by sorting stocks based on the maximum daily return (MAX) over the previous month. Decile 1 (10) is the portfolio of stocks with the lowest (highest) MAX. 10−1 is the zero-cost portfolio, which is long the decile 10 portfolio (High MAX) and short the decile 1 portfolio (Low MAX). The sample includes Nasdaq-listed common stocks (CRSP share codes 10 and 11) headquartered in the 10 US geographical regions. No price, market capitalization, or minimum daily observation filters are imposed, except that ILLIQ requires at least ten daily observations in the month. Observations with non-positive book equity are excluded when constructing the book-to-market ratio, and firm-months with missing returns, including delisting months, are excluded. Average returns and alphas are given in percent. Newey-West (1987) adjusted *t*-statistics using six lags are reported in parentheses.

| | Value-Weighted Portfolios | | | | | | Equal-Weighted Portfolios | | | | | |
|---|---|---|---|---|---|---|---|---|---|---|---|---|
| Decile | Average return | CAPM alpha | FF3 alpha | FFC4 alpha | FF5 alpha | FF6 alpha | Average return | CAPM alpha | FF3 alpha | FFC4 alpha | FF5 alpha | FF6 alpha |
| 1 (Low MAX) | 1.28 | 0.71 | 0.62 | 0.62 | 0.49 | 0.50 | 1.31 | 0.75 | 0.65 | 0.67 | 0.54 | 0.56 |
| 2 | 1.33 | 0.65 | 0.55 | 0.58 | 0.43 | 0.46 | 1.38 | 0.71 | 0.60 | 0.65 | 0.51 | 0.54 |
| 3 | 1.19 | 0.42 | 0.35 | 0.39 | 0.31 | 0.34 | 1.28 | 0.53 | 0.45 | 0.51 | 0.42 | 0.46 |
| 4 | 1.04 | 0.19 | 0.12 | 0.20 | 0.22 | 0.27 | 1.12 | 0.29 | 0.22 | 0.33 | 0.31 | 0.39 |
| 5 | 1.04 | 0.12 | 0.08 | 0.17 | 0.23 | 0.29 | 1.12 | 0.22 | 0.17 | 0.29 | 0.32 | 0.40 |
| 6 | 0.93 | -0.05 | -0.06 | 0.09 | 0.22 | 0.32 | 1.08 | 0.14 | 0.11 | 0.30 | 0.37 | 0.50 |
| 7 | 0.95 | -0.10 | -0.10 | 0.07 | 0.26 | 0.37 | 1.13 | 0.13 | 0.11 | 0.31 | 0.43 | 0.57 |
| 8 | 0.79 | -0.31 | -0.30 | -0.06 | 0.14 | 0.29 | 0.99 | -0.06 | -0.06 | 0.22 | 0.36 | 0.56 |
| 9 | 0.86 | -0.27 | -0.25 | -0.03 | 0.35 | 0.49 | 1.14 | 0.06 | 0.06 | 0.33 | 0.64 | 0.82 |
| 10 (High MAX) | 0.25 | -0.86 | -0.81 | -0.54 | -0.16 | 0.01 | 0.72 | -0.36 | -0.32 | 0.03 | 0.33 | 0.56 |
| 10-1 | -1.03 | -1.57 | -1.42 | -1.16 | -0.65 | -0.49 | -0.59 | -1.11 | -0.97 | -0.64 | -0.21 | 0.00 |
| | (-1.88) | (-3.34) | (-4.70) | (-3.87) | (-2.62) | (-1.95) | (-1.05) | (-2.30) | (-2.79) | (-1.81) | (-0.62) | (0.00) |

## Table 3. Trust and MAX effect: univariate analysis

The table reports the value-weighted (Panel A) and equal-weighted (Panel B) average monthly returns as well as the CAPM alphas, Fama-French three-factor alphas (FF3 alpha), Fama-French-Carhart four-factor alphas (FFC4 alpha), Fama-French five-factor alphas (FF5 alpha), and Fama and French (2018) six-factor model, which augments FF5 with the momentum factor (FF6 alpha) on the portfolios for high-TRUST and low-TRUST subsamples (regions with high and low level of trust). We classify the sample stocks into two subsample groups (high-TRUST or low-TRUST) based on the firm's region's previous month TRUST ranking. The ranking of TRUST is created within each region where firms are headquartered and over time. The thresholds used in ranking include the median and mean of TRUST. If a firm's region's monthly TRUST is above (below) thresholds of TRUST, the firm-month is classified into the high-TRUST (low-TRUST) subsamples. For each subsample, decile portfolios are formed every month from January 1996 to December 2018 by sorting stocks based on the maximum daily return (MAX) over the previous month. Decile 1 (10) is the portfolio of stocks with the lowest (highest) MAX. 10−1 is the zero-cost portfolio, which is long the decile 10 portfolio (High MAX) and short the decile 1 portfolio (Low MAX). The sample includes Nasdaq-listed common stocks (CRSP share codes 10 and 11) headquartered in the 10 US geographical regions. No price, market capitalization, or minimum daily observation filters are imposed, except that ILLIQ requires at least ten daily observations in the month. Observations with non-positive book equity are excluded when constructing the book-to-market ratio, and firm-months with missing returns, including delisting months, are excluded. Average returns and alphas are given in percent. Newey-West (1987) adjusted *t*-statistics using six lags are reported in parentheses.

| Decile | Panel A: Value-Weighted Portfolios | | | | Panel B: Equal-Weighted Portfolios | | | |
|---|---|---|---|---|---|---|---|---|
| | Thresholds of TRUST used to define subsamples | | | | Thresholds of TRUST used to define subsamples | | | |
| | Median | | Mean | | Median | | Mean | |
| | High-TRUST | Low-TRUST | High-TRUST | Low-TRUST | High-TRUST | Low-TRUST | High-TRUST | Low-TRUST |
| 1 (Low MAX) | 1.26 | 1.33 | 1.30 | 1.30 | 1.29 | 1.35 | 1.34 | 1.32 |
| 2 | 1.27 | 1.29 | 1.24 | 1.31 | 1.32 | 1.33 | 1.29 | 1.34 |
| 3 | 1.08 | 1.23 | 1.14 | 1.15 | 1.14 | 1.32 | 1.20 | 1.25 |
| 4 | 1.12 | 1.05 | 1.09 | 0.99 | 1.19 | 1.12 | 1.15 | 1.05 |
| 5 | 0.97 | 1.15 | 0.94 | 1.21 | 1.07 | 1.25 | 1.04 | 1.33 |
| 6 | 0.88 | 0.90 | 0.87 | 0.88 | 1.01 | 1.03 | 1.00 | 1.03 |
| 7 | 0.84 | 1.09 | 0.88 | 1.00 | 1.06 | 1.20 | 1.09 | 1.14 |
| 8 | 0.67 | 0.75 | 0.67 | 0.84 | 0.85 | 1.00 | 0.83 | 1.06 |
| 9 | 0.76 | 0.94 | 0.74 | 0.96 | 1.07 | 1.23 | 1.05 | 1.27 |
| 10 (High MAX) | -0.08 | 0.46 | 0.02 | 0.30 | 0.38 | 0.91 | 0.47 | 0.79 |
| 10-1 | -1.34 | -0.87 | -1.29 | -1.00 | -0.91 | -0.45 | -0.86 | -0.53 |
| | (-2.49) | (-1.54) | (-2.42) | (-1.72) | (-1.70) | (-0.75) | (-1.64) | (-0.86) |
| CAPM alpha | -1.85 | -1.40 | -1.79 | -1.52 | -1.39 | -0.96 | -1.34 | -1.04 |
| | (-3.91) | (-2.82) | (-3.84) | (-2.93) | (-2.95) | (-1.82) | (-2.91) | (-1.88) |
| FF3 alpha | -1.72 | -1.25 | -1.65 | -1.40 | -1.26 | -0.81 | -1.21 | -0.91 |
| | (-5.21) | (-3.78) | (-5.17) | (-3.72) | (-3.51) | (-2.13) | (-3.48) | (-2.14) |
| FFC4 alpha | -1.46 | -0.97 | -1.39 | -1.11 | -0.97 | -0.45 | -0.91 | -0.54 |
| | (-4.37) | (-3.10) | (-4.29) | (-3.14) | (-2.53) | (-1.21) | (-2.45) | (-1.31) |
| FF5 alpha | -0.98 | -0.43 | -0.92 | -0.57 | -0.53 | 0.00 | -0.47 | -0.09 |
| | (-3.32) | (-1.56) | (-3.21) | (-1.72) | (-1.44) | (0.01) | (-1.31) | (-0.22) |
| FF6 alpha | -0.82 | -0.26 | -0.75 | -0.39 | -0.34 | 0.24 | -0.28 | 0.15 |
| | (-2.67) | (-0.97) | (-2.54) | (-1.24) | (-0.85) | (0.68) | (-0.71) | (0.37) |

**Table 4. Fama-MacBeth regressions in subsamples**

The table reports the time-series average of the slope coefficients from cross-sectional regressions over high-TRUST and low-TRUST subsamples (regions with high and low levels of trust). We classify the sample stocks into two subsample groups (high-TRUST or low-TRUST) based on a firm's region's previous month TRUST ranking. The ranking of TRUST is created within each region where firms are headquartered and over time. If a firm's region's monthly TRUST is above (below) thresholds of TRUST, the firm-month is classified into the high-TRUST (low-TRUST) subsamples. The thresholds used in ranking are the median and mean of TRUST. For each subsample and every month from January 1996 to December 2018 (276 months), we run the Fama-MacBeth regression of the monthly stock returns (in percent) on one-month lagged independent variables that include the maximum daily return (MAX) and a set of control variables. The control variables include market beta (BETA), natural log of market capitalization (SIZE), natural log of book-to-market ratio (BM), momentum (MOM), short-term reversal (REV) and illiquidity (ILLIQ). All variables are defined in the Appendix. The sample includes Nasdaq-listed common stocks (CRSP share codes 10 and 11) headquartered in the 10 US geographical regions. No price, market capitalization, or minimum daily observation filters are imposed, except that ILLIQ requires at least ten daily observations in the month. Observations with non-positive book equity are excluded when constructing the book-to-market ratio, and firm-months with missing returns, including delisting months, are excluded. Newey-West (1987) adjusted *t*-statistics using six lags are reported in parentheses. Ave. number of stocks is the time-series average number of stocks in the monthly cross-sectional regressions.

| Variables | Threshold of TRUST used to define high-TRUST and low-TRUST subsamples | | | | | | | |
|---|---|---|---|---|---|---|---|---|
| | Median | | | | Mean | | | |
| | High-TRUST | | Low-TRUST | | High-TRUST | | Low-TRUST | |
| MAX | -0.054 | -0.050 | -0.028 | -0.019 | -0.051 | -0.045 | -0.025 | -0.020 |
| | (-2.87) | (-3.00) | (-1.41) | (-1.10) | (-2.78) | (-2.76) | (-1.12) | (-0.97) |
| BETA | | 0.013 | | -0.051 | | -0.023 | | -0.045 |
| | | (0.26) | | (-0.99) | | (-0.49) | | (-0.78) |
| SIZE | | -0.227 | | -0.173 | | -0.221 | | -0.169 |
| | | (-2.91) | | (-2.59) | | (-2.86) | | (-2.49) |
| BM | | 0.260 | | 0.229 | | 0.245 | | 0.247 |
| | | (1.79) | | (1.87) | | (1.73) | | (1.92) |
| MOM | | 0.143 | | 0.047 | | 0.143 | | 0.090 |
| | | (0.46) | | (0.16) | | (0.47) | | (0.30) |
| REV | | -0.004 | | -0.008 | | -0.005 | | -0.006 |
| | | (-0.75) | | (-1.39) | | (-0.95) | | (-0.72) |
| ILLIQ | | 0.028 | | 0.043 | | 0.025 | | 0.059 |
| | | (2.70) | | (1.99) | | (2.57) | | (2.20) |
| Avg. number of stocks | 1,234 | 1,023 | 1,511 | 1,220 | 1,481 | 1,228 | 1,264 | 1,014 |
| Avg. adj. $R^2$ (%) | 1.25 | 4.31 | 1.10 | 4.37 | 1.22 | 4.19 | 1.26 | 5.00 |

**Table 5. Fama-MacBeth regressions with an interaction term**

The table reports the time-series average of the slope coefficients from cross-sectional regressions across the sample when the threshold to define DTRUST is median of TRUST within each region over time. Every month from January 1996 to December 2018 (276 months), we run the Fama-MacBeth regression of the monthly stock returns (in percent) on the one-month lagged independent variables that include the maximum daily return (MAX), TRUST, the interaction term of MAX and DTRUST, and control variables. DTRUST is a dummy variable based on TRUST within each region where firms are headquartered. If a firm's region's monthly TRUST is above the median of TRUST for the region over time, then DTRUST is one and zero otherwise. The control variables are market beta (BETA), natural log of market capitalization (SIZE), natural log of book-to-market ratio (BM), momentum (MOM), short-term reversal (REV) and illiquidity (ILLIQ). All variables are defined in the Appendix. The sample includes Nasdaq-listed common stocks (CRSP share codes 10 and 11) headquartered in the 10 US geographical regions. No price, market capitalization, or minimum daily observation filters are imposed, except that ILLIQ requires at least ten daily observations in the month. Observations with non-positive book equity are excluded when constructing the book-to-market ratio, and firm-months with missing returns, including delisting months, are excluded. Newey-West (1987) adjusted *t*-statistics using six lags are reported in parentheses. Ave. number of stocks is the time-series average number of stocks in the monthly cross-sectional regressions.

| Model | (1) | (2) | (3) | (4) | (5) | (6) |
|---|---|---|---|---|---|---|
| MAX | -0.038 | -0.030 | -0.038 | -0.030 | -0.030 | -0.022 |
| | (-2.08) | (-1.93) | (-2.08) | (-1.94) | (-1.51) | (-1.32) |
| TRUST | | | 0.794 | 0.379 | 1.321 | 1.070 |
| | | | (1.21) | (0.67) | (1.87) | (1.70) |
| MAX × DTRUST | | | | | -0.022 | -0.025 |
| | | | | | (-2.22) | (-2.71) |
| BETA | | -0.021 | | -0.021 | | -0.019 |
| | | (-0.49) | | (-0.49) | | (-0.45) |
| SIZE | | -0.208 | | -0.208 | | -0.208 |
| | | (-3.16) | | (-3.16) | | (-3.15) |
| BM | | 0.225 | | 0.226 | | 0.230 |
| | | (1.78) | | (1.79) | | (1.84) |
| MOM | | 0.097 | | 0.099 | | 0.099 |
| | | (0.32) | | (0.34) | | (0.34) |
| REV | | -0.006 | | -0.006 | | -0.006 |
| | | (-1.15) | | (-1.15) | | (-1.15) |
| ILLIQ | | 0.034 | | 0.034 | | 0.034 |
| | | (3.18) | | (3.20) | | (3.24) |
| Avg. number of stocks | 2,746 | 2,242 | 2,746 | 2,242 | 2,746 | 2,242 |
| Avg. adj. $R^2$ (%) | 1.09 | 4.04 | 1.12 | 4.06 | 1.22 | 4.15 |

**Table 6. Fama-MacBeth regressions, controlling for regional factors**

The table reports the time-series average of the slope coefficients from cross-sectional regressions across the sample after controlling for regional factors when the threshold to define DTRUST is median of TRUST within each region over time. Every month from January 1996 to December 2018 (276 months), we run the Fama-MacBeth regression of the monthly stock returns (in percent) on the one-month lagged independent variables that include the maximum daily return (MAX), TRUST, the interaction term of MAX and DTRUST, and control variables. DTRUST is a dummy variable based on TRUST within each region where firms are headquartered. If a firm's region's monthly TRUST is above the median of TRUST for the region over time, then DTRUST is one and zero otherwise. The control variables are market beta (BETA), natural log of market capitalization (SIZE), natural log of book-to-market ratio (BM), momentum (MOM), short-term reversal (REV), illiquidity (ILLIQ), and the regional factors which include the percentage of the Catholic population (CATH), the percentage of the population aged 25 and older who attained a bachelor's degree or higher (EDU), the median household income (MHI) measured in thousands of dollars, the natural log of one plus net state lottery revenue (ticket sales minus prizes) per capita in dollars (LOTTERY), the natural log of state GDP per capita in dollars (GDPPC), the state unemployment rate in percent (UNEMP), the natural log of the number of state residents per square mile of land area (POPDEN), the percentage of the state population who are adherents of any religious congregation (REL), and the number of bank branches per 10,000 residents (BRANCH). All variables are defined in the Appendix. The sample includes Nasdaq-listed common stocks (CRSP share codes 10 and 11) headquartered in the 10 US geographical regions. No price, market capitalization, or minimum daily observation filters are imposed, except that ILLIQ requires at least ten daily observations in the month. Observations with non-positive book equity are excluded when constructing the book-to-market ratio, and firm-months with missing returns, including delisting months, are excluded. Newey-West (1987) adjusted *t*-statistics using six lags are reported in parentheses. Ave. number of stocks is the time-series average number of stocks in the monthly cross-sectional regressions.

| Model | (1) | (2) | (3) | (4) | (5) | (6) | (7) | (8) | (9) | (10) | (11) |
|---|---|---|---|---|---|---|---|---|---|---|---|
| MAX | -0.030 | -0.024 | -0.023 | -0.024 | -0.021 | -0.024 | -0.022 | -0.022 | -0.023 | -0.024 | -0.025 |
| | (-1.51) | (-1.39) | (-1.44) | (-1.46) | (-1.27) | (-1.45) | (-1.34) | (-1.31) | (-1.37) | (-1.52) | (-1.53) |
| TRUST | 1.321 | 0.195 | 0.386 | 0.272 | 0.964 | 0.436 | 1.188 | 1.085 | 1.285 | 1.285 | 0.391 |
| | (1.87) | (0.32) | (0.70) | (0.49) | (1.55) | (0.84) | (1.81) | (1.59) | (1.85) | (1.89) | (0.57) |
| MAX × DTRUST | -0.022 | -0.025 | -0.023 | -0.023 | -0.026 | -0.022 | -0.026 | -0.026 | -0.023 | -0.023 | -0.024 |
| | (-2.22) | (-2.33) | (-2.52) | (-2.52) | (-2.74) | (-2.37) | (-2.78) | (-2.47) | (-2.46) | (-2.49) | (-2.33) |
| BETA | | -0.019 | -0.020 | -0.021 | -0.018 | -0.020 | -0.020 | -0.018 | -0.019 | -0.021 | -0.020 |
| | | (-0.45) | (-0.48) | (-0.49) | (-0.42) | (-0.47) | (-0.47) | (-0.42) | (-0.44) | (-0.51) | (-0.49) |
| SIZE | | -0.209 | -0.205 | -0.206 | -0.208 | -0.207 | -0.212 | -0.209 | -0.209 | -0.209 | -0.211 |
| | | (-3.15) | (-3.14) | (-3.14) | (-3.16) | (-3.14) | (-3.18) | (-3.16) | (-3.17) | (-3.15) | (-3.18) |
| BM | | 0.235 | 0.240 | 0.244 | 0.231 | 0.236 | 0.232 | 0.233 | 0.231 | 0.237 | 0.246 |
| | | (1.93) | (2.00) | (2.04) | (1.85) | (1.94) | (1.86) | (1.89) | (1.86) | (1.95) | (2.08) |
| MOM | | 0.102 | 0.103 | 0.104 | 0.100 | 0.108 | 0.103 | 0.103 | 0.101 | 0.103 | 0.110 |
| | | (0.35) | (0.36) | (0.36) | (0.34) | (0.38) | (0.35) | (0.35) | (0.35) | (0.36) | (0.39) |
| REV | | -0.006 | -0.006 | -0.006 | -0.006 | -0.005 | -0.006 | -0.006 | -0.005 | -0.006 | -0.006 |
| | | (-1.16) | (-1.17) | (-1.16) | (-1.15) | (-1.16) | (-1.15) | (-1.15) | (-1.16) | (-1.19) | (-1.23) |
| ILLIQ | | 0.034 | 0.034 | 0.034 | 0.034 | 0.034 | 0.034 | 0.034 | 0.034 | 0.034 | 0.034 |
| | | (3.26) | (3.26) | (3.25) | (3.25) | (3.26) | (3.23) | (3.26) | (3.24) | (3.24) | (3.26) |
| CATH | | 0.010 | | | | | | | | | 0.006 |
| | | (1.81) | | | | | | | | | (1.00) |
| EDU | | | 0.026 | | | | | | | | 0.008 |
| | | | (1.65) | | | | | | | | (0.56) |
| MHI | | | | 0.024 | | | | | | | 0.009 |
| | | | | (1.96) | | | | | | | (1.05) |
| LOTTERY | | | | | -0.003 | | | | | | -0.083 |
| | | | | | (-0.12) | | | | | | (-2.14) |
| GDPPC | | | | | | 0.741 | | | | | 0.214 |
| | | | | | | (1.74) | | | | | (0.42) |
| UNEMP | | | | | | | 0.104 | | | | 0.059 |
| | | | | | | | (2.61) | | | | (1.63) |
| POPDEN | | | | | | | | 0.086 | | | 0.028 |
| | | | | | | | | (1.66) | | | (0.46) |
| REL | | | | | | | | | -0.001 | | -0.004 |
| | | | | | | | | | (-0.15) | | (-0.88) |
| BRANCH | | | | | | | | | | -0.162 | -0.028 |
| | | | | | | | | | | (-1.95) | (-0.50) |
| Avg. number of stocks | 2,746 | 2,242 | 2,242 | 2,242 | 2,242 | 2,242 | 2,242 | 2,242 | 2,242 | 2,242 | 2,242 |
| Avg. adj. $R^2$ (%) | 1.22 | 4.23 | 4.23 | 4.24 | 4.17 | 4.22 | 4.19 | 4.21 | 4.16 | 4.25 | 4.40 |

## Table 7. Education and the effect of trust on MAX effect

The table reports the time-series average of the slope coefficients from cross-sectional regressions over two subsamples that include regions with low education level (low-EDU) and high education level (high-EDU). EDU is the state-level percentage of the population over 25 years old with a bachelor's degree or higher. We classify the sample stocks into two subsample groups (low-EDU or high-EDU) based on a firm's state's previous month EDU ranking. The threshold used in ranking is the median of EDU. The ranking of EDU is created across all states where firms are headquartered in each month. If a firm's state's monthly EDU is below (above) the median of EDU, the firm-month is classified into the low-EDU (high-EDU) subsamples. For each subsample and every month from January 1996 to December 2018 (276 months), we run the Fama-MacBeth regression of the monthly stock returns (in percent) on the one-month lagged independent variables that include the maximum daily return (MAX), TRUST, the interaction term of MAX and DTRUST, and control variables. DTRUST is a dummy variable based on TRUST within each region where firms are headquartered. If a firm's region's monthly TRUST is above the median of TRUST for the region over time, then DTRUST is one and zero otherwise. The control variables are market beta (BETA), natural log of market capitalization (SIZE), natural log of book-to-market ratio (BM), momentum (MOM), short-term reversal (REV) and illiquidity (ILLIQ). The last row (Low-High) reports the difference between the coefficients on MAX × DTRUST in model (3) for the low-EDU and high-EDU subsamples. Following the Fama-Macbeth regressions approach, we compute, for each month, the difference between the cross-sectional slope estimates on MAX × DTRUST from the low-EDU and high-EDU regressions and report the time-series average of these monthly differences. All variables are defined in the Appendix. The sample includes Nasdaq-listed common stocks (CRSP share codes 10 and 11) headquartered in the 10 US geographical regions. No price, market capitalization, or minimum daily observation filters are imposed, except that ILLIQ requires at least ten daily observations in the month. Observations with non-positive book equity are excluded when constructing the book-to-market ratio, and firm-months with missing returns, including delisting months, are excluded. Newey-West (1987) adjusted *t*-statistics using six lags are reported in parentheses. Ave. number of stocks is the time-series average number of stocks in the monthly cross-sectional regressions.

| | Low-EDU | | | | High-EDU | | |
|---|---|---|---|---|---|---|---|
| Model | (1) | (2) | (3) | | (1) | (2) | (3) |
| MAX | -0.039 | -0.039 | -0.021 | | -0.028 | -0.028 | -0.025 |
| | (-2.03) | (-2.06) | (-0.95) | | (-1.79) | (-1.81) | (-1.50) |
| TRUST | | -0.442 | 0.436 | | | -0.038 | 0.498 |
| | | (-0.61) | (0.58) | | | (-0.05) | (0.73) |
| MAX × DTRUST | | | -0.046 | | | | -0.017 |
| | | | (-2.92) | | | | (-1.68) |
| BETA | -0.018 | -0.017 | -0.019 | | -0.027 | -0.027 | -0.023 |
| | (-0.32) | (-0.30) | (-0.34) | | (-0.57) | (-0.58) | (-0.49) |
| SIZE | -0.163 | -0.164 | -0.164 | | -0.222 | -0.223 | -0.224 |
| | (-2.50) | (-2.50) | (-2.52) | | (-3.20) | (-3.19) | (-3.19) |
| BM | 0.334 | 0.334 | 0.328 | | 0.197 | 0.198 | 0.201 |
| | (2.48) | (2.48) | (2.46) | | (1.59) | (1.60) | (1.63) |
| MOM | 0.112 | 0.114 | 0.099 | | 0.099 | 0.098 | 0.101 |
| | (0.36) | (0.37) | (0.32) | | (0.33) | (0.32) | (0.34) |
| REV | 0.000 | 0.000 | 0.001 | | -0.009 | -0.009 | -0.009 |
| | (0.07) | (0.04) | (0.11) | | (-1.65) | (-1.64) | (-1.65) |
| ILLIQ | 0.023 | 0.023 | 0.023 | | 0.042 | 0.043 | 0.043 |
| | (2.16) | (2.13) | (2.15) | | (3.18) | (3.22) | (3.25) |
| Avg. number of stocks | 721 | 721 | 721 | | 1,521 | 1,521 | 1,521 |
| Avg. adj. $R^2$ (%) | 4.78 | 4.77 | 4.98 | | 4.11 | 4.12 | 4.23 |
| Low-High: MAX × DTRUST | | | | -0.029 | | | |
| | | | | (-1.70) | | | |

## Table 8. Income and the effect of trust on MAX effect

The table reports the time-series average of the slope coefficients from cross-sectional regressions over two subsamples that include regions with low income (low-MHI) and high income (high-MHI). MHI is defined as the state-level median household income. We classify the sample stocks into two subsample groups (low-MHI or high-MHI) based on a firm's state's previous month MHI ranking. The threshold used in ranking is the median of MHI. The ranking of MHI is created across all states where firms are headquartered in each month. If a firm's state's monthly MHI is below (above) the median of MHI, the firm-month is classified into the low-MHI (high-MHI) subsamples. For each subsample and every month from January 1996 to December 2018 (276 months), we run the Fama-MacBeth regression of the monthly stock returns (in percent) on the one-month lagged independent variables that include the maximum daily return (MAX), TRUST, the interaction term of MAX and DTRUST, and control variables. DTRUST is a dummy variable based on TRUST within each region where firms are headquartered. If a firm's region's monthly TRUST is above the median of TRUST for the region over time, then DTRUST is one and zero otherwise. The control variables are market beta (BETA), natural log of market capitalization (SIZE), natural log of book-to-market ratio (BM), momentum (MOM), short-term reversal (REV) and illiquidity (ILLIQ). The last row (Low-High) reports the difference between the coefficients on MAX × DTRUST in model (3) for the low-MHI and high-MHI subsamples. Following the Fama-Macbeth approach, we compute, for each month, the difference between the cross-sectional slope estimates on MAX × DTRUST from the low-MHI and high-MHI regressions, and report the time-series average of these monthly differences. All variables are defined in the Appendix. The sample includes Nasdaq-listed common stocks (CRSP share codes 10 and 11) headquartered in the 10 US geographical regions. No price, market capitalization, or minimum daily observation filters are imposed, except that ILLIQ requires at least ten daily observations in the month. Observations with non-positive book equity are excluded when constructing the book-to-market ratio, and firm-months with missing returns, including delisting months, are excluded. Newey-West (1987) adjusted *t*-statistics using six lags are reported in parentheses. Ave. number of stocks is the time-series average number of stocks in the monthly cross-sectional regressions.

| | Low-MHI | | | | High-MHI | | |
|---|---|---|---|---|---|---|---|
| Model | (1) | (2) | (3) | | (1) | (2) | (3) |
| MAX | -0.045 | -0.045 | -0.018 | | -0.026 | -0.026 | -0.021 |
| | (-2.20) | (-2.23) | (-0.77) | | (-1.67) | (-1.70) | (-1.24) |
| TRUST | | -0.436 | 0.837 | | | 0.016 | 0.729 |
| | | (-0.58) | (1.13) | | | (0.02) | (0.97) |
| MAX × DTRUST | | | -0.061 | | | | -0.017 |
| | | | (-3.95) | | | | (-1.67) |
| BETA | -0.031 | -0.030 | -0.031 | | -0.021 | -0.021 | -0.018 |
| | (-0.56) | (-0.53) | (-0.58) | | (-0.46) | (-0.46) | (-0.39) |
| SIZE | -0.178 | -0.177 | -0.178 | | -0.215 | -0.216 | -0.217 |
| | (-2.73) | (-2.70) | (-2.73) | | (-3.14) | (-3.13) | (-3.14) |
| BM | 0.286 | 0.286 | 0.269 | | 0.217 | 0.221 | 0.227 |
| | (2.03) | (2.03) | (1.91) | | (1.72) | (1.77) | (1.83) |
| MOM | 0.083 | 0.072 | 0.046 | | 0.098 | 0.100 | 0.107 |
| | (0.25) | (0.22) | (0.14) | | (0.33) | (0.34) | (0.37) |
| REV | 0.002 | 0.002 | 0.002 | | -0.008 | -0.008 | -0.008 |
| | (0.31) | (0.30) | (0.39) | | (-1.49) | (-1.50) | (-1.52) |
| ILLIQ | 0.029 | 0.029 | 0.030 | | 0.037 | 0.037 | 0.037 |
| | (1.85) | (1.85) | (1.88) | | (3.57) | (3.61) | (3.67) |
| Avg. number of stocks | 569 | 569 | 569 | | 1,673 | 1,673 | 1,673 |
| Avg. adj. $R^2$ (%) | 5.06 | 5.06 | 5.26 | | 4.06 | 4.08 | 4.19 |
| Low-High: MAX × DTRUST | | | | -0.044 | | | |
| | | | | (-2.43) | | | |

# Trust and Lottery-Related Anomalies

## Internet Appendix

This appendix shows the results of several additional empirical analyses that are further conducted to check the robustness of our main results in the paper. In short, these additional analyses demonstrate that our main finding on the role played by social trust in the underperformance of lottery-type stocks is robust. Section IA-1 reports the results of the univariate portfolio analysis by using the orthogonal component of MAX (i.e., the portion of MAX that is orthogonal to the social trust index proxy TRUST) as the sort variable. Section IA-2 presents the results of the univariate portfolio analysis that examines the relation between trust and the lottery demand when using two alternative approaches to create the ranking of TRUST and construct regions with the highest and lowest trust. Section IA-3 shows the results of the Fama-MacBeth (1973) cross-sectional regressions with an interaction term by using the mean of TRUST as the alternative threshold to define the dummy variable for TRUST (DTRUST).

Section IA-4 shows that our main findings are robust after we run the Fama-MacBeth regressions with an interaction term, which include MAX, DTRUST, and $MAX \times DTRUST$ simultaneously. Section IA-5 demonstrates that our main results are robust when instead of MAX, MAX(N), which is the average of the N (N = 2, 3, 4, 5) highest daily returns over the last month, is used as the alternative proxy for the lottery-like characteristics of stocks. Section IA-6 provides further support for the main results by using idiosyncratic skewness as the alternative measure of lottery-type characteristics instead of MAX. Section IA-7 shows that our main finding that high social trust is associated with a stronger MAX effect remains robust after accounting for time-invariant regional differences and common time shocks by estimating pooled panel regressions with regional and monthly fixed effects. Section IA-8 examines whether the relation between social trust and the MAX effect varies with firms' institutional ownership (IO) by re-estimating the cross-sectional regressions after splitting the sample into high- and low-IO subsamples using the median (mean) of IO as the threshold. The results show that the interaction between MAX and social trust is stronger among firms with lower institutional ownership. Section IA-9 confirms that our main results are robust when using the mean of EDU, as the alternative threshold, to split our sample by education level and the mean of TRUST, as the alternative threshold, to define DTRUST in the cross-sectional regression analysis. Section IA-10 shows that our main findings continue to hold when repeating the cross-sectional regression analysis, but instead using the mean of MHI, as the alternative threshold, to split the sample by the level of income and the mean of TRUST, as the alternative threshold, to define DTRUST. Section IA-11 shows that the main results continue to hold when we replace MHI (median household income) with PCI (per capita personal income) as the proxy for income in each state.

Section IA-12 shows that the effect of social trust on the MAX effect remains robust after controlling for the interactions of MAX with education and income. Section IA-13 examines whether our findings reflect a financing channel and shows that high-MAX stocks in high-trust regions do not exhibit greater financing activity or investment and are not younger, and that controlling for these characteristics and their interactions with MAX does not affect our main findings. Section IA-14 reports a geographic permutation placebo test that randomly reassigns regional trust to firms and shows that our findings depend on the actual correspondence between firms' headquarters locations and regional trust. Section IA-15 shows that our results are robust to controlling for political ideology, racial diversity, immigrant share, and social capital. Section IA-16 shows that our findings are not explained by social transmission bias, as the interaction between MAX and DTRUST remains negative and of similar magnitude after controlling for proxies for local social transmission intensity and their interactions with MAX.

*IA-1. Orthogonal test*

In this section, we test the robustness of our main results by conducting a univariate portfolio analysis using the component of MAX that is orthogonal to TRUST as the sort variable. The component of MAX that is orthogonal to the social trust index TRUST, denoted $MAX \perp TRUST$, is estimated as the residual from a time-series regression of MAX on TRUST for each stock. The information on the variable ($MAX \perp TRUST$) construction in this section is provided in the Appendix. Every month from January 1996 to December 2018, we calculate $MAX \perp TRUST$ using values of MAX and TRUST over the previous month and then use this orthogonalized MAX ($MAX \perp TRUST$) to sort all sample stocks into 10 decile portfolios to examine the cross-section of expected returns. The results are presented in Table IA1 of the Appendix. The Newey-West (1987) adjusted $t$-statistics are presented in parentheses using six lags.

The results in Table IA1 demonstrate that the MAX anomaly is not detected when only the portion of MAX that is orthogonal to TRUST ($MAX \perp TRUST$) is used as the sort variable in the univariate portfolio analysis. We find that the average raw returns of the zero-cost portfolios that are long the high $MAX \perp TRUST$ (decile 10) portfolio and short the low $MAX \perp TRUST$ (decile 1) portfolio are positive and statistically significant for both value-weighted and equal-weighted portfolios, with values of 0.57% per month ($t$-statistic = 2.71) and 0.46% per month ($t$-statistic = 2.16), respectively. This is opposite to the corresponding results in Table 2, which show the negative average raw returns of the zero-cost portfolios for both value-weighted (−1.03% per month with the $t$-statistic of −1.88) and equal-weighted (−0.59% per month with the $t$-statistic of −1.05) portfolios. In Table IA1, the CAPM, FF3, FFC4, FF5, and FF6 alphas of the zero-cost portfolio also present a similar pattern, in that they are positive and more statistically significant.

Therefore, the results in Table IA1 confirm that our main results are robust, suggesting that the negative relation between MAX and expected stock returns in Table 2 is largely driven by the relation between MAX and TRUST.

*IA-2. Trust and the lottery demand: univariate sort for regions with the highest and lowest TRUST*

In Table 3, we conduct the univariate portfolio analysis by creating a ranking (or threshold) of TRUST over time for each region to construct high- and low-trust subsamples. However, it is feasible to identify the subsamples by focusing on the cross-sectional as well as time-series variation in TRUST across all regions.

First, we consider the extreme case and define the highest and lowest trust regions using TRUST ranking in the cross-section across all regions. To do so, in each month, we select the regions with extreme values for the trust index. In other words, the subsamples include only regions with the highest and lowest TRUST in each month.

More specifically, in each month, we create a ranking of TRUST across the 10 US geographical regions where firms are headquartered and classify firms in regions where the one-month lagged value of TRUST is at the top and bottom (or the highest and lowest) ranking as the highest-TRUST and lowest-TRUST subsamples, respectively. Then, for each subsample in each month, we sort stocks into deciles based on MAX and calculate both value-weighted returns and equal-weighted average monthly raw and risk-adjusted returns for decile portfolios and for the difference between high (decile 10) and low (decile 1) MAX portfolio for the highest-TRUST and lowest-TRUST subsamples. The results are reported in Panel A of Table IA2. The numbers in parentheses are *t*-statistics, adjusted following Newey-West (1987) using six lags.

In Panel A of Table IA2, for both value-weighted and equal-weighted portfolios, the average raw returns of the zero-cost portfolios that are long the high MAX (decile 10) portfolio and short the low MAX (decile 1) portfolio tend to be more significantly negative for the highest-TRUST subsample, with values of −1.09% per month (*t*-statistic = −2.02) and −0.78% per month (*t*-statistic = −1.39), respectively. By contrast, the corresponding returns tend to be indistinguishable from zero for the lowest-TRUST subsample, with values of −0.37% per month (*t*-statistic = −0.67) and 0.01% per month (*t*-statistic = 0.01), respectively. After adjusting the risk factors, the risk-adjusted returns present a similar pattern. For instance, for value-weighted portfolios, the risk-adjusted return of the zero-cost portfolios under the Fama-French three-factor model (FF3) is −1.54% per month (*t*-statistic = −3.29) for the highest-TRUST subsample, which is more significantly negative and approximately 2.6 times larger in absolute magnitude than the corresponding return for the lowest-TRUST subsample (−0.60% per month with *t*-statistic = −1.21).

Thus, the results in Panel A of Table IA2 show a similar pattern as in Table 3. Panel A of Table IA2 indicates that the relation between MAX and expected stock returns is economically large and significantly negative for the highest-TRUST subsample (regions with the highest trust); however, it is insignificant for the lowest-TRUST subsample (regions with lowest trust). It implies that a univariate portfolio analysis using the cross-sectional variation in trust index (TRUST) across regions does not change our main findings.

Now, we turn to conducting a univariate portfolio analysis based on time-series mean variation in TRUST across regions. Unlike the monthly cross-sectional TRUST ranking used previously, in this approach, we use the alternative extreme case which only includes two regions with the highest and lowest average trust index (TRUST) over the sample period (276 months). More specifically, for each region, we calculate the averages of the one-month lagged TRUST over time and create a ranking of these 10 time-series averages. We then only select two regions of firms' headquarters where the time-series average of one-month lagged TRUST is at the top and bottom (or the highest and lowest) ranking: the highest-TRUST and lowest-TRUST regions, respectively. Next, for each region in each month, we sort stocks into deciles based on MAX. Panel B of Table IA2 reports both value-weighted returns and equal-weighted average monthly raw and risk-adjusted returns for decile portfolios and for the zero-cost portfolio for these two regions. The Newey-West (1987) adjusted $t$-statistics are presented in parentheses using six lags.

The results in Panel B of Table IA2 follow a similar pattern as in Panel A of Table IA2. The average returns and alphas of the zero-cost portfolios for both value-weighted and equal-weighted portfolios tend to be significantly negative for the highest-TRUST region, whereas they become insignificant for the lowest-TRUST region. For example, for the highest-TRUST region, the CAPM alpha of the zero-cost portfolios for equal-weighted portfolios is approximately 5.4 times larger in absolute magnitude and more significantly negative (−1.50% per month with $t$-statistic = −2.07) than the corresponding return for the lowest-TRUST region (−0.28% per month with $t$-statistic = −0.48). Consistent with Table 2, the FF5 and FF6 alphas are weaker, as these models absorb much of the unconditional MAX premium. Nevertheless, the zero-cost portfolio alpha remains more negative in the highest-TRUST region than in the lowest-TRUST region under both models. For example, for value-weighted portfolios, the FF5 alpha is −1.01% per month ($t$-statistic = −1.87) in the highest-TRUST region, compared with −0.07% per month ($t$-statistic = −0.15) in the lowest-TRUST region. The evidence in panel B of table IA2 shows that the MAX effect is economically large and, in most specifications, statistically significant in the high-trust region, whereas it is smaller and insignificant in every specification in the low-trust region.

In summary, Table IA2 shows that our main results in Table 3 are robust when separately using the cross-sectional variation and time-series mean in trust index (TRUST) across regions to form portfolios and conduct univariate portfolio analyses. It provides additional evidence for our main results that the MAX anomaly is stronger for regions with high trust than for regions with low trust.

*IA-3. Mean of TRUST as the alternative threshold to define the dummy variable for TRUST (DTRUST)*

In our main cross-sectional regression analysis in Table 5, we use the median of TRUST of the firm's headquarters region over time as the threshold to define the dummy variable for TRUST (DTRUST). In this section, we test whether our main results in Table 5 are robust when using mean of TRUST as the alternative threshold to define DTRUST.

We replace median of TRUST with mean of TRUST to define DTRUST and repeat the Fama-MacBeth (1973) cross-sectional regressions analysis in Table 5. The results are reported in Table IA3. We find the results in Table IA3 are qualitatively similar to those in Table 5. When we include the interaction term ($MAX \times DTRUST$) in the regression specification, the significant estimated coefficients on MAX become insignificant, with the values of −0.032 ($t$-statistic = −1.53) and −0.024 ($t$-statistic = −1.41) in models (5) and (6), respectively. However, the coefficients on the interaction terms ($MAX \times DTRUST$) are significantly negative except for the one in model (5), suggesting that the high level of social trust plays a significant role in the MAX effect. For example, the estimated coefficient on the interaction term ($MAX \times DTRUST$) in model (6) is $-0.018$ ($t$-statistic = $-1.85$), which implies that the MAX effect becomes weaker and statistically insignificant (coefficient estimate = $-0.024$) among firms' regions with low trust, whereas it tends to be significant and stronger among firms' regions with high trust, with the estimated coefficient of $-0.018$ (formally, $(-0.018) \times 1$). Although the coefficient on interaction term in model (5) of Table IA3 is weaker and insignificant (coefficient = $-0.016$ and $t$-statistic = $-1.45$) compared to the coefficient on interaction term in Table 5, it does not impact the interpretation of the results. Therefore, the results in Table IA3 show that our main results in Table 5 are robust when using mean of TRUST as the alternative threshold to define DTRUST in the cross-sectional regression analysis, suggesting that high social trust is associated with a stronger MAX effect.

### *IA-4. Fama-MacBeth regressions with an interaction term: using MAX, DTRUST, and $MAX \times DTRUST$ simultaneously*

In the main Fama-MacBeth regressions reported in Tables 5, 6, and IA3, we follow Walkshäusl (2014) and include TRUST and the interaction term ($MAX \times DTRUST$), while excluding the main effect of DTRUST. This specification allows us to estimate the direct relation between social trust and expected stock returns through the coefficient on TRUST. A positive coefficient on TRUST is consistent with the argument that higher social trust encourages stock market participation by reducing the perceived risk of being cheated (e.g., Guiso et al., 2008; El-Attar and Poschke, 2011).

Because TRUST and DTRUST are correlated but distinct variables, omitting the main effect of DTRUST may raise concerns regarding the specification of the interaction model. A conventional interaction specification includes the two main effects together with their interaction term. Accordingly, we re-estimate the Fama-MacBeth (1973) regressions using an alternative specification that includes MAX, DTRUST, and $MAX \times DTRUST$ simultaneously. The results are reported in Table IA4. Panel A (Panel B) presents the results when DTRUST is defined using the median (mean) value of TRUST.

The results remain qualitatively unchanged. When the median of TRUST is used to define DTRUST (Panel A), the coefficient on MAX becomes insignificant across all specifications, while the coefficient on the interaction term ($MAX \times DTRUST$) remains significantly negative. For example, the coefficients on the interaction terms ($MAX \times DTRUST$) are significantly negative, with the values of $-0.026$ ($t$-statistic = $-2.37$), $-0.030$ ($t$-statistic = $-2.64$), and $-0.030$ ($t$-statistic = $-2.60$), in models (1), (2), and (3), respectively. These findings are consistent with our main results reported in Tables 5, 6, and IA3, indicating that higher social trust is associated with a stronger MAX effect. The results also remain robust after controlling for regional characteristics, including religion, education, median household income, local lottery participation, economic development, unemployment, population density, overall religiosity, and local financial development. Similar conclusions are obtained when the mean of TRUST is used to define DTRUST (Panel B).

Overall, Table IA4 confirms that our main findings are robust to an alternative interaction specification that includes the main effects of both MAX and DTRUST. The significantly negative coefficient on the interaction term continues to indicate that higher social trust is associated with a stronger MAX effect, and that this relation is not explained by the regional factors.

*IA-5. Alternative constructions of MAX*

In the main tests, we use a single day of maximum return from the last month (MAX) as the proxy for the lottery-like characteristics of stocks. In this section, we examine whether our main results are robust when replacing MAX with MAX(N) (N = 2, 3, 4, 5). MAX(N) is defined as the average of the N (N = 2, 3, 4, 5) highest daily returns over the previous month. Given the concerns about the seemingly arbitrary measure of a single day of maximum return (MAX) from the previous month, Bali et al. (2011) also introduce MAX(N) as alternative constructions of MAX to check the robustness of their main results. We replace MAX with MAX(N) (N = 2, 3, 4, 5) to repeat the Fama-MacBeth (1973) cross-sectional regression analysis in Table 4.

The results are presented in Table IA5. Panel A and Panel B separately report the results using the median and mean of TRUST as thresholds to define high-TRUST and low-TRUST groups. The patterns obtained in Table IA5 are similar to the results in Table 4, showing that the MAX effect is associated with social trust and is more negative and statistically significant among regions with a high level of social trust than among regions with a low level of social trust. For instance, in Panel A of Table IA5, for the high-TRUST group, the average slope coefficients on MAX(2) without and with control variables are −0.080 ($t$-statistic = −2.90) and −0.080 ($t$-statistic = −3.18), respectively. However, for the low-TRUST group, the coefficients on MAX(2) without and with control variables are −0.047 ($t$-statistic = −1.62) and −0.039 ($t$-statistic = −1.55), respectively. We find that the slope coefficients of MAX(2) for the high-TRUST group are approximately 1.7 to 2.1 times larger in absolute magnitude than those for the low-TRUST group. In comparison to Table 4, Table IA5 also shows that the MAX effect tends to be more pronounced as the number of days for calculating MAX(N) increases, regardless of the level of trust.

We next use MAX(N) (N = 2, 3, 4, 5) instead of MAX to repeat Fama-MacBeth (1973) cross-sectional regressions with interaction terms in Table 5 and Table IA3, which use different thresholds, median and mean of TRUST, respectively, to define the dummy variables for TRUST (DTRUST). For brevity, we only repeat analysis for the main specifications of Table 5 and Table IA3. The results of these analyses are presented in Table IA6. We report the results with the median and mean of TRUST used as thresholds to define DTRUST in Panel A and Panel B, respectively.

We find that the results of Panel A and Panel B in Table IA6 are consistent with those shown in Table 5 and Table IA3. Particularly, the significant slope coefficients on MAX(N) become weaker and less significant after introducing the interaction term between MAX(N) and the trust dummy variable (DTRUST). The coefficients on $MAX(N) \times DTRUST$ are significantly negative across all specifications in Panel A, and negative in Panel B, where they are significant at the 10% level for all specifications except for N = 5, suggesting that a high level of social trust contributes to the MAX effect. Similarly, compared to Table 5 and Table IA3, we find that the MAX effect becomes more significant, both economically and statistically, and the coefficients on the interaction terms $MAX(N) \times DTRUST$ are more negative as we average more days to calculate MAX(N).

Overall, the results in Tables IA5 and IA6 show that our main findings do not change after replacing MAX with MAX(N), confirming that social trust plays a significant role in the MAX effect, and that this effect is stronger among the regions with a high level of social trust in comparison to regions with a low level of social trust.

*IA-6. Alternative proxy for lottery-type characteristics: idiosyncratic skewness*

In this section, we examine whether our main results still hold when we use idiosyncratic skewness instead of MAX as the proxy for lottery-type characteristics.[37]

The literature on lottery-related anomalies suggests that skewness measures can also be used as proxies for lottery-type characteristics, since similar to MAX, they capture one dimension of a return distribution, that is, the small probability of a large payoff. We follow Bali et al. (2011) to construct idiosyncratic skewness (ISKEW) at the end of each month. We measure ISKEW as the skewness (or third moment) of daily residuals by running a two-factor time-series regression of daily stock returns within the month. Detailed information on the variable is provided in the variable definitions of the Appendix.

We next replace MAX with idiosyncratic skewness (ISKEW) to repeat the analysis in Table 5 and Table IA3; that is, the Fama-MacBeth (1973) cross-sectional regressions with an interaction term using different thresholds, median and mean of TRUST to define the dummy variables for TRUST (DTRUST).[38] Table IA7 presents the results when ISKEW is used as the measure for lottery-like characteristics of stocks. We only report the results of key models for brevity. Panel A (Panel B) shows the results when median (mean) of TRUST is used as the threshold to define DTRUST.

The results of Panel A and Panel B in Table IA7 present a similar pattern to those in Table 5 and Table IA3. When we include the interaction term of the idiosyncratic skewness and the dummy variable for TRUST, $(ISKEW \times DTRUST)$, in the regression specification, the significant slope coefficients of the idiosyncratic skewness become weaker and less significant. The slope coefficients for the interaction terms are significantly negative at the 10% significance level, indicating that lottery-related anomalies are influenced by the high level of social trust. Thus, the results in Table IA7 provide additional support that our main results are robust when replacing MAX with a skewness measure as the proxy for lottery-type characteristics.

*IA-7. Pooled panel regressions with regional and monthly fixed effects*

[37] In unreported analyses, we replace MAX with total skewness and repeat the analysis in Tables 5 and IA3. While the coefficients for the interaction term of the total skewness and the dummy variable for trust (DTRUST) are less significant compared to the results using idiosyncratic skewness, our main findings continue to hold.

[38] In unreported analyses, we replace MAX with skewness measures (idiosyncratic skewness and total skewness, respectively) and repeat the Fama-MacBeth (1973) cross-sectional regression analysis in Table 4 for high-TRUST and low-TRUST groups, respectively. The results are robust. Although for the low-TRUST group, the average slope coefficients on skewness measures tend to be significantly negative after including control variables, they are less significant, both economically and statistically, in comparison to those coefficients for the high-TRUST group, and our main findings continue to hold.

Although the baseline Fama-MacBeth (1973) regressions control for standard firm characteristics, they may not fully account for unobserved regional characteristics that are persistent over time. To address this concern, we estimate pooled panel regressions with regional and monthly fixed effects using the following specification:

$$R_{i,t} = \lambda_{0,t-1} + \lambda_{1,t-1}MAX_{i,t-1} + \lambda_{2,t-1}TRUST_{i,t-1} + \lambda_{3,t-1}MAX_{i,t-1} \times DTRUST_{i,t-1} + \boldsymbol{\Lambda}_{t-1}\boldsymbol{Z}_{i,t-1} + \delta_r + \lambda_t + \varepsilon_{i,t} \quad \text{(IA1)}$$

where $R_{i,t}$ is the monthly stock return (in percent), $MAX_{i,t-1}$ is the maximum daily return during the previous month, $TRUST_{i,t-1}$ is the regional social trust index, and $DTRUST_{i,t-1}$ is a dummy variable indicating whether regional trust is above the median (or mean) threshold. $Z_{i,t-1}$ denotes the vector of control variables. $\delta_r$ represents regional fixed effects, and $\lambda_t$ represents monthly fixed effects. Variable definitions are provided in the Appendix. Following the main analysis, all independent variables are winsorized monthly at the 1st and 99th percentiles. Heteroskedasticity-robust standard errors are two-way clustered by region and month, and *t*-statistics are reported in parentheses.

Table IA8 reports the estimation results. After controlling for both time-invariant regional differences and common time shocks, the coefficient on the interaction term, $MAX \times DTRUST$, remains significantly negative across the specifications. For example, when the median of TRUST is used to define DTRUST, the coefficient on MAX becomes statistically insignificant after including the stock control variables, whereas the coefficient on $MAX \times DTRUST$ remains significantly negative (−0.021, *t*-statistic = −2.70). Similarly, when the mean of TRUST is used to define DTRUST, the coefficient on $MAX \times DTRUST$ is −0.018 (*t*-statistic = −2.63). These results indicate that the MAX effect is concentrated among firms headquartered in high-trust regions, while it largely disappears in low-trust regions. Overall, the findings confirm that our main results are robust to controlling for unobserved regional heterogeneity and common monthly shocks through regional and monthly fixed effects. This evidence suggests that the stronger MAX effect observed in high-trust regions reflects trust-related variation rather than persistent differences across regions.

*IA-8. Institutional ownership and the impact of trust on the MAX effect*

One potential concern regarding our findings is that regional social trust may not be reflected in stock prices if local investors have only a limited influence on price formation. Prior research documents that individual investors exhibit a preference for geographically proximate firms (Ivković and Weisbenner, 2005; Seasholes and Zhu, 2010). Motivated by this evidence, we examine whether the relation between social trust and the MAX effect varies with firms' institutional ownership (IO), since firms with lower institutional ownership are generally expected to have greater participation by retail investors. However, because institutional ownership does not identify the geographical location of non-institutional investors, this analysis provides an indirect test and does not directly identify a local retail-

investor channel.[39]

Following the approach in Tables 7 and 8, we re-estimate the Fama-MacBeth (1973) cross-sectional regressions after splitting the sample into high- and low-IO subsamples. In each month, firms are ranked according to their one-month lagged institutional ownership and classified using either the monthly mean or median as the threshold. Within each subsample, DTRUST is defined using the corresponding median of TRUST. Panel A (Panel B) of Table IA9 reports the results based on the mean (median) thresholds for IO. All independent variables are winsorized monthly at the 1st and 99th percentiles, and Newey-West (1987) adjusted $t$-statistics based on six lags are reported in parentheses. Consistent with our expectations, the results in Panel A show that the interaction between MAX and DTRUST is significantly negative for firms with lower institutional ownership but becomes weaker and insignificant among firms with higher institutional ownership. The interaction in model (3) is −0.042 ($t$-statistic = −2.81) for low-IO firms and −0.000 ($t$-statistic = −0.01) for high-IO firms. The difference between the coefficients on $MAX \times DTRUST$ across the low-IO and high-IO subsamples is −0.042 ($t$-statistic = −2.14). To examine whether the stronger effect of trust among firms with low institutional ownership reflects differences in arbitrage costs, model (4) of Table IA9 additionally includes the interactions of MAX with illiquidity ($MAX \times ILLIQ$) and firm size ($MAX \times SIZE$) within each subsample. In Panel A, the coefficient on $MAX \times DTRUST$ for low-IO firms remains significantly negative (−0.044, $t$-statistic = −2.94), whereas that for high-IO firms remains insignificant (−0.002, $t$-statistic = −0.13). The difference between the two subsamples remains statistically significant (−0.043, $t$-statistic = −2.23). The results suggest that the stronger effect of trust among low-IO firms is not explained by differences in arbitrage costs. The results are qualitatively similar in Panel B, although the difference between subsamples is weaker.[40] Overall, the findings in Table IA9 indicate that the relation between social trust and the MAX effect is more pronounced when institutional ownership is lower. We interpret these results as supportive evidence relevant to the transmission of regional social trust to stock prices.

*IA-9. Mean of EDU as the alternative threshold to split sample by education level*

In the main tests, we introduce the education level measured by EDU and find that the effect of trust on the MAX effect varies across regions with different education levels. In our main analysis in Table 7, we use the median of EDU as the threshold to divide our sample stocks into two groups (low-EDU and high-EDU) by education level. For each group, we run the monthly Fama and MacBeth (1973) cross-sectional regressions using the median of TRUST of the firm's headquarters region over time to define the dummy variable for TRUST (DTRUST). In this section, as a robustness test, we use alternative thresholds of EDU and TRUST to examine whether the main results in Table 7 persist. We repeat the Fama and MacBeth (1973) cross-sectional regression analysis in Table 7 using mean of EDU as the alternative threshold to split our sample by education level as well as mean of TRUST as the alternative threshold to define the dummy variable for TRUST (DTRUST).

[39] A direct test of a local-investor transmission channel requires investor-level data on the geographical location of investors and their holdings or trading activity. We do not have access to these data. Accordingly, the institutional ownership analysis in this section provides an indirect test of whether the relation between trust and the MAX effect varies with the composition of firm ownership, rather than a direct test of investor locality.

[40] In untabulated results, we also examine the role of the supply of lendable shares and short-sale constraints, proxied by IO (Nagel, 2005). Because institutional ownership is the sorting variable in Table IA9, we instead add IO and its interaction with MAX to the regression in model (6) of Table 5, estimated on the subsample of firm-months with available institutional ownership data. The coefficient on $MAX \times DTRUST$ remains significantly negative (−0.032, $t$-statistic = −2.96). These results are available upon request.

Table IA10 presents the results. The results in Table IA10 are qualitatively similar to those in Table 7. The coefficient on interaction term ($MAX \times DTRUST$) in model (3) for the low-EDU group is significantly negative (coefficient = −0.032 and $t$-statistic = −2.26); however, for the high-EDU group, it is insignificant (coefficient = −0.012 and $t$-statistic = −0.92). This does not change the interpretation of the main results, and Table IA10 provides further support for our main findings, indicating that a higher level of social trust is associated with a stronger MAX effect, depending on the education level. Table IA10 suggests that when considering the different levels of education, the effect of trust on the MAX effect is stronger when the region's education level is lower, indicating that education serves as a crucial channel through which social trust impacts the MAX effect. In Table IA10, the results support Kumar (2009) who shows that less educated investors exhibit a greater propensity to invest in stocks with lottery features, whereas we show that less educated investors invest disproportionately more in lottery-type stocks only in high-trust regions.

*IA-10.Mean of MHI as the alternative threshold to split sample by income level*

In our main analysis in Table 8, we introduce the income level proxied by MHI to investigate how the effect of trust on the MAX effect varies across regions with different income levels. We use the median of MHI as the threshold to divide our sample into low-MHI and high-MHI groups by the level of income. Then we run the monthly Fama and MacBeth (1973) cross-sectional regressions using the median of TRUST to define the dummy variable for TRUST (DTRUST) for each group. In this section, we use alternative thresholds of MHI and TRUST to examine whether our main findings in Table 8 continue to hold. We repeat the Fama and MacBeth (1973) cross-sectional regression analysis in Table 8, but using mean of MHI as the alternative threshold to split the sample by the level of income as well as mean of TRUST as the alternative threshold to define the dummy variable for TRUST (DTRUST). The results are reported in Table IA11.

In Table IA11, the results show a similar pattern to those presented in Table 8. The coefficient on the interaction term ($MAX \times DTRUST$) in model (3) for the high-MHI group is insignificant (coefficient = −0.013 and $t$-statistic = −0.88), whereas that of the low-MHI group is significantly negative (coefficient = −0.034 and $t$-statistic = −2.21). The results in Table 8 remain robust when we use the mean of MHI as the threshold to split our sample by the income level and mean of TRUST to define DTRUST. In summary, Table IA11 shows that a high level of social trust contributes to the MAX effect, which depends on the income level. Again, Table IA11 shows that the effect of trust on the MAX effect is more pronounced in regions with lower income, suggesting that income acts as a key channel through which trust influences the underperformance of the lottery-type stocks. Also, the results in Table IA11 support Kumar (2009) who shows that low-income investors exhibit a greater propensity to invest in lottery-type stocks. However, we show that less wealthy investors invest disproportionately more in stocks with lottery features only in high-trust regions.

*IA-11.Alternative proxy for income: per capita personal income, and the effect of trust on MAX effect*

In our analyses related to the relation between income level and the effect of trust on the MAX effect in Table 8 and Table IA11, we use MHI (median household income) as the proxy for income level in each state and run the Fama and MacBeth (1973) cross-sectional regressions using different thresholds: median (mean) of MHI as the threshold to split the sample by the level of income as well as median (mean) of TRUST as the threshold to define the dummy variable for TRUST (DTRUST) in Table 8 (Table IA11). We show the effect of income levels on the relation between the MAX effect and social trust. In this section, as a robustness test, we use PCI (per capita personal income) as the proxy for income in each state instead of MHI (median household income) to repeat the analyses in Tables 8 and IA11. Similarly, we conduct the cross-sectional regression analyses with different thresholds: median (mean) of PCI as the threshold to split the sample by the level of income as well as median (mean) of TRUST as the threshold to define DTRUST. The results are presented in Table IA12. For brevity, Panel A and Panel B show the results for median of PCI with median of TRUST for DTRUST and mean of PCI with mean of TRUST for DTRUST, respectively. The results of Panels A and B in Table IA12 are substantially the same as those presented in Tables 8 and IA11, indicating that replacing MHI with PCI does not change our main finding that the effect of social trust on the MAX effect is stronger in low-income regions than in high-income regions.

*IA-12.Fama-MacBeth regressions with an interaction term, controlling for interactions of MAX with education and income*

In our main analysis in Tables 7 and 8, we split the sample by education (EDU) and income (MHI). To provide a robustness test, we augment the Fama-MacBeth (1973) regressions in Table 5 with EDU and its interaction with MAX ( $MAX \times EDU$ ) and with MHI and its interaction with MAX ($MAX \times MHI$). The results are reported in Table IA13.

The results in Table IA13 show that our main findings remain robust. The coefficient on $MAX \times DTRUST$ remains significantly negative and very similar in magnitude to that in model (6) of Table 5, with values of −0.026 ($t$-statistic = −2.73) in model (7) and −0.026 ($t$-statistic = −2.68) in model (8). The coefficient on $MAX \times EDU$ is insignificant (0.002, $t$-statistic = 1.22). The coefficient on $MAX \times MHI$ is positive and significant (0.003, $t$-statistic = 2.86), indicating that the MAX effect is weaker in regions with higher income. This is consistent with the evidence that lower-income investors exhibit stronger preferences for lottery-type stocks (Kumar, 2009). Importantly, the interaction between MAX and trust is not explained by these effects.

*IA-13.Financing conditions, investment, and the effect of trust on MAX effect*

Prior studies show that social trust affects firms' financing costs and financing decisions (Gupta et al., 2018; Meng and Yin, 2019; Brockman et al., 2020). This raises the possibility that our findings reflect a financing channel rather than investors' beliefs. Lottery-like firms tend to be younger, have greater growth opportunities, and rely more heavily on external financing. Any improvement in financing conditions associated with social trust could therefore be particularly relevant for these firms, and subsequent equity issuance, investment expansion, or market timing could be associated with lower future returns, generating a similar interaction pattern between MAX and social trust. In this section, we examine this alternative explanation.

We consider six firm-level variables that capture firms' financing activity, investment and age. Net share issuance (NSI) is the change in the natural log of split-adjusted shares outstanding over the past 12 months (Pontiff and Woodgate, 2008) and captures net equity issuance. Equity issuance (EQUITYISS) is the sale of common and preferred stock scaled by lagged total assets and captures seasoned equity offerings. Net external financing (XFIN) is the net cash raised from equity and debt, scaled by lagged total assets (Bradshaw, Richardson, and Sloan, 2006), and captures firms' dependence on external financing. Asset growth (ASSETG) is the annual growth rate of total assets (Cooper, Gulen, and Schill, 2008), and CAPX is the capital expenditures scaled by lagged total assets; both capture investment expansion. Firm age (AGE) is the natural log of one plus the number of months since the firm first appears in CRSP and captures how recently the firm went public. Detailed definitions of these variables are provided in the Appendix.

We first compare these variables for high-MAX stocks across high- and low-TRUST groups. Each month, we sort stocks into quintiles by MAX in the previous month and classify stocks in the top quintile as high-MAX stocks. We then compute the average of each variable of high-MAX stocks in the low-TRUST group (DTRUST = 0) and the high-TRUST group (DTRUST = 1), and the difference between the two groups. Panel A of Table IA14 reports the time-series average of these monthly values. None of the differences is statistically significant. For example, the average NSI of high-MAX stocks is 0.104 in the high-TRUST group and 0.105 in the low-TRUST group (difference = −0.001, $t$-statistic = −0.14). Moreover, the average values of equity issuance, net external financing, asset growth, and capital expenditures are slightly lower in the high-TRUST group, which is opposite to the prediction of the financing channel, and high-MAX stocks in the high-TRUST group are not younger than those in the low-TRUST group.

We next re-estimate the Fama-MacBeth (1973) regressions in Table 5 after including each variable and its interaction with MAX. Each variable is winsorized monthly at the 1$^{st}$ and 99$^{th}$ percentiles and standardized each month to have a mean of zero and a standard deviation of one.[41] Models (1) to (6) of Panel B in Table IA14 add the variables one at a time, model (7) includes all six variables together, and model (8) additionally includes the regional controls in Table 6. The coefficient on $MAX \times DTRUST$ remains significantly negative and stable across all specifications, ranging from −0.022 ($t$-statistic = −2.43) to −0.027 ($t$-statistic = −3.00). When all six variables and their interactions with MAX are included together with the regional controls in model (8), the coefficient on $MAX \times DTRUST$ is −0.023 ($t$-statistic = −2.12), which is close to the corresponding estimate of −0.025 ($t$-statistic = −2.71) in model

[41] The results are qualitatively similar without standardization of these variables.

(6) of Table 5. These results indicate that the relation between social trust and the MAX effect is not explained by differences in financing activity, investment, or firm age, and is unlikely to be driven by a financing channel.

*IA-14. Geographic permutation placebo test*

In our analysis, we assign regional social trust to firms based on the locations of their headquarters. To provide empirical support for this choice, we conduct a geographic permutation placebo test. If our findings reflect the actual correspondence between firms and the trust of their headquarters regions, the interaction between MAX and DTRUST should be substantially weaker when regional trust is randomly reassigned to firms. We conduct the placebo test in two ways. First, in the firm-level reassignment, we randomly reassign the headquarters regions of each firm across all firms in the sample, while keeping the number of firms in each region unchanged. Each firm is then assigned the time series of TRUST and DTRUST of its randomly assigned region. Second, in the region-level reassignment, we randomly permute the time series of TRUST and DTRUST across the 10 US geographical regions, so that all firms headquartered in the same region receive the trust series of another, randomly chosen region. The first approach breaks the link between individual firms and their regional trust, whereas the second approach preserves the grouping of firms within regions but breaks the link between regions and their trust. For each reassignment, we re-estimate the baseline Fama-MacBeth (1973) regression in model (6) of Table 5 and record the coefficient on $MAX \times DTRUST$. We repeat each procedure 5,000 times. Figure IA1 plots the distributions of the placebo coefficients on $MAX \times DTRUST$, with the observed coefficient of $-0.025$ ($t$-statistic = $-2.71$) indicated by the dashed vertical line. Under the firm-level reassignment (Panel A), the placebo coefficients are centered around zero, with a mean of $-0.0001$, and only 0.02% of the placebo coefficients are as negative as the observed coefficient. Under the region-level reassignment (Panel B), the placebo coefficients are also centered around zero, with a mean of $-0.001$, and only 0.92% of the placebo coefficients are as negative as the observed coefficient.

In summary, the observed coefficient of $MAX \times DTRUST$ lies in the left tail of both placebo distributions. These results indicate that our findings depend on the actual correspondence between firms' headquarters locations and regional trust and provide further empirical support for using firms' headquarters locations to assign regional trust.

*IA-15. Additional regional characteristics: political ideology, racial diversity, immigrant share, and social capital*

One potential concern regarding our findings is that social trust may be correlated with other regional characteristics that are not captured by the regional controls in Table 6, such as political ideology, racial diversity, the share of immigrants, and other dimensions of social capital. If these characteristics are also related to investors' preferences for lottery-like stocks, the relationship between social trust and the MAX effect could reflect these unmeasured regional factors rather than social trust itself. To address this concern, we control for additional state-level variables and determine whether the relationship between social trust and the MAX effect remains robust after controlling for them.

We measure these four characteristics as follows. Political ideology (REP) is the Republican share of the two-party vote in presidential elections in each state, in percent. We obtain state-level presidential election returns from the MIT Election Data and Science Lab. Since the data are available only for presidential election years, we linearly interpolate REP for interim years. Racial diversity (NONWHITE) is the percentage of the state population that is not non-Hispanic white. We construct it from the annual state population estimates by race and Hispanic origin of the US Census Bureau. Immigrant share (FOREIGN) is the percentage of the state population that is foreign born. We obtain it from the Migration Policy Institute tabulation of the 1990 and 2000 Decennial Census and the 2010 and 2019 American Community Surveys, and we linearly interpolate FOREIGN for interim years. Social capital (SOCCAP) is the county-level social capital index of Rupasingha, Goetz, and Freshwater (2006), obtained from the Northeast Regional Center for Rural Development at Pennsylvania State University. This index is widely used in the literature (e.g., Hasan, Hoi, Wu, and Zhang, 2017). It is constructed from the density of civic, religious, business, political, professional, labor, and recreational associations, voter turnout, and the response rate to the Census. We aggregate the county-level index to the state level by weighting each county by its population. Since the index is available for 1997, 2005, 2009, and 2014, we linearly interpolate SOCCAP for interim years and apply the values in 1997 (2014) to the years before 1997 (after 2014).

Similar to the other regional variables, for each state in month $t$, these variables take their values in the corresponding year $t$. We then repeat the Fama-MacBeth (1973) cross-sectional regression analysis in Table 6. Models (1) to (4) add each of the four variables separately to the specification with the stock characteristic controls. Model (5) includes all four variables simultaneously together with all regional controls in Table 6 (CATH, EDU, MHI, LOTTERY, GDPPC, UNEMP, POPDEN, REL, and BRANCH). All independent variables are winsorized monthly at the 1$^{st}$ and 99$^{th}$ percentiles, and Newey-West (1987) adjusted $t$-statistics using six lags are reported in parentheses. The results are reported in Table IA15.

The results in Table IA15 are consistent with those in Table 6. In all specifications, the coefficient on MAX is statistically insignificant, whereas the coefficient on the interaction term $MAX \times DTRUST$ remains significant and negative. For example, in model (5), which simultaneously controls for all 13 regional characteristics, the coefficient on $MAX \times DTRUST$ is −0.024 (t-statistic = −2.26), very close to that in model (11), Table 6. In summary, Table IA15 confirms that the effect of social trust on the MAX effect is not driven by political ideology, racial diversity, immigrant share, or other dimensions of social capital.

*IA-16.Social transmission intensity and the effect of trust on the MAX effect*

One potential concern is that our results may reflect social transmission bias from Hirshleifer (2020), rather than social trust. The social transmission bias framework predicts that the MAX effect should be stronger in regions where social interaction among investors is more intensive (Bali, Hirshleifer, Peng, Tang, and Wang, 2024). If high-trust regions are also regions with more intensive social interaction, the stronger MAX effect we document in high-trust regions could reflect social transmission rather than trust. To distinguish the two channels, we augment the Fama-MacBeth (1973) regressions in Table 6 with a proxy for local social transmission intensity (PROXY), and its interaction with MAX, $MAX \times PROXY$. We use three proxies: population density (POPDEN), the social capital index (SOCCAP) of Rupasingha, Goetz, and Freshwater (2006), and social connectedness (SCIH). SCIH, following Bali et al. (2024), is the natural log of the sum of the Facebook Social Connectedness Index (Bailey et al., 2018) between the state where a firm is headquartered and all US states, including the state itself.

Each proxy is standardized to have a mean of zero and a standard deviation of one.[42] For each proxy, we estimate one specification with the stock characteristic controls and another that additionally includes the regional controls in Table 6 (excluding POPDEN when it is used as the PROXY). The results are reported in Table IA16. The results indicate that the effect of social trust on the MAX effect is distinct from the social transmission channel. The coefficient on $MAX \times PROXY$ is not significantly negative for any of the three proxies. It is insignificant for SOCCAP and SCIH and significantly positive for POPDEN. The coefficients on $MAX \times DTRUST$ remain negative across all specifications, ranging from −0.018 to −0.029, similar in magnitude to that in Table 6. It is significant at the 5% level in all specifications with the stock characteristic controls, and at the 10% level or better when the regional controls are included. These findings indicate that the relation between social trust and the MAX effect is not explained by differences in the intensity of local social transmission.

[42] The results are qualitatively similar without standardization of these variables.

## Table IA1. Univariate portfolios sorted on orthogonalized MAX ($MAX \perp TRUST$)

The table reports the value-weighted and equal-weighted average monthly returns as well as the CAPM alphas, Fama-French three-factor alphas (FF3 alpha), Fama-French-Carhart four-factor alphas (FFC4 alpha), Fama-French five-factor alphas (FF5 alpha), and Fama and French (2018) six-factor model, which augments FF5 with the momentum factor (FF6 alpha) on the portfolios. Decile portfolios are formed every month from January 1996 to December 2018 by sorting stocks based on the portion of the maximum daily return (MAX) that is orthogonal to the social trust index TRUST ($MAX \perp TRUST$) over the previous month. Decile 1 (10) refers to the portfolio of stocks with the lowest (highest) $MAX \perp TRUST$. 10−1 is the zero-cost portfolio, which is long the decile 10 portfolio (High $MAX \perp TRUST$) and short the decile 1 portfolio (Low $MAX \perp TRUST$). The sample includes Nasdaq-listed common stocks (CRSP share codes 10 and 11) headquartered in the 10 US geographical regions. No price, market capitalization, or minimum daily observation filters are imposed, except that ILLIQ requires at least ten daily observations in the month. Observations with non-positive book equity are excluded when constructing the book-to-market ratio, and firm-months with missing returns, including delisting months, are excluded. Average returns and alphas are in percent. Newey-West (1987) adjusted *t*-statistics using six lags are reported in parentheses.

| | Value-Weighted Portfolios | | | | | | Equal-Weighted Portfolios | | | | | |
|---|---|---|---|---|---|---|---|---|---|---|---|---|
| Decile | Average return | CAPM alpha | FF3 alpha | FFC4 alpha | FF5 alpha | FF6 alpha | Average return | CAPM alpha | FF3 alpha | FFC4 alpha | FF5 alpha | FF6 alpha |
| 1 (Low $MAX \perp TRUST$) | 0.17 | -0.86 | -0.83 | -0.66 | -0.40 | -0.29 | 0.54 | -0.47 | -0.46 | -0.23 | -0.03 | 0.12 |
| 2 | 0.88 | -0.09 | -0.08 | 0.01 | 0.15 | 0.20 | 1.08 | 0.15 | 0.14 | 0.27 | 0.36 | 0.44 |
| 3 | 1.11 | 0.24 | 0.21 | 0.28 | 0.38 | 0.43 | 1.26 | 0.42 | 0.37 | 0.47 | 0.54 | 0.61 |
| 4 | 1.16 | 0.31 | 0.26 | 0.33 | 0.30 | 0.34 | 1.25 | 0.43 | 0.36 | 0.46 | 0.39 | 0.46 |
| 5 | 1.19 | 0.37 | 0.31 | 0.38 | 0.31 | 0.36 | 1.28 | 0.49 | 0.42 | 0.51 | 0.43 | 0.50 |
| 6 | 1.10 | 0.27 | 0.21 | 0.31 | 0.34 | 0.41 | 1.21 | 0.41 | 0.34 | 0.47 | 0.46 | 0.55 |
| 7 | 1.21 | 0.35 | 0.29 | 0.38 | 0.44 | 0.51 | 1.33 | 0.49 | 0.41 | 0.55 | 0.56 | 0.66 |
| 8 | 1.06 | 0.15 | 0.10 | 0.25 | 0.33 | 0.44 | 1.18 | 0.29 | 0.22 | 0.41 | 0.47 | 0.62 |
| 9 | 0.97 | 0.00 | -0.01 | 0.18 | 0.36 | 0.49 | 1.14 | 0.21 | 0.18 | 0.43 | 0.52 | 0.69 |
| 10 (High $MAX \perp TRUST$) | 0.75 | -0.27 | -0.25 | -0.02 | 0.27 | 0.41 | 0.99 | -0.01 | 0.01 | 0.29 | 0.52 | 0.71 |
| 10-1 | 0.57 | 0.59 | 0.58 | 0.64 | 0.66 | 0.70 | 0.46 | 0.46 | 0.46 | 0.52 | 0.55 | 0.59 |
| | (2.71) | (2.80) | (2.76) | (3.08) | (2.77) | (2.94) | (2.16) | (2.21) | (2.17) | (2.42) | (2.36) | (2.51) |

**Table IA2. Trust and MAX effect: univariate analysis using different approach to define highest and lowest trust regions**

The table reports the average monthly returns, CAPM alphas, Fama-French three-factor alphas (FF3 alpha), Fama-French-Carhart four-factor alphas (FFC4 alpha), Fama-French five-factor alphas (FF5 alpha), and Fama and French (2018) six-factor model, which augments FF5 with the momentum factor (FF6 alpha) on the portfolios for highest-TRUST and lowest-TRUST subsamples defined using two approaches. In Panel A, we classify the sample stocks into subsamples based on a firm's region's previous month TRUST ranking created across all regions where firms are headquartered in each month. If a firm's region's monthly TRUST is at the highest (lowest) ranking, the firm-month is classified into the highest-TRUST (lowest-TRUST) subsample. In Panel B, we calculate the average of the one-month lagged TRUST over time for each region where firms are headquartered and create a ranking of the regions' time-series averages. If the time-series average of one-month lagged TRUST for a region of a firm's headquarters is at the highest (lowest) ranking, the firm is classified into the highest-TRUST (lowest-TRUST) subsample. For each subsample, decile portfolios are formed every month from January 1996 to December 2018 by sorting stocks based on the maximum daily return (MAX) over the previous month. Decile 1 (10) is the portfolio of stocks with the lowest (highest) MAX. 10−1 is the zero-cost portfolio, which is long the decile 10 portfolio (High MAX) and short the decile 1 portfolio (Low MAX). The sample includes Nasdaq-listed common stocks (CRSP share codes 10 and 11) headquartered in the 10 US geographical regions. No price, market capitalization, or minimum daily observation filters are imposed, except that ILLIQ requires at least ten daily observations in the month. Observations with non-positive book equity are excluded when constructing the book-to-market ratio, and firm-months with missing returns, including delisting months, are excluded. Returns and alphas are in percent. Newey-West (1987) adjusted *t*-statistics using six lags are reported in parentheses.

| | Panel A: Regions sorted based on cross-sectional ranking of TRUST | | | | Panel B: Regions sorted based on time-series mean of TRUST | | | |
|---|---|---|---|---|---|---|---|---|
| | Value-Weighted Portfolios | | Equal-Weighted Portfolios | | Value-Weighted Portfolios | | Equal-Weighted Portfolios | |
| Decile | Highest-TRUST | Lowest-TRUST | Highest-TRUST | Lowest-TRUST | Highest-TRUST | Lowest-TRUST | Highest-TRUST | Lowest-TRUST |
| 1 (Low MAX) | 1.32 | 1.15 | 1.31 | 1.11 | 1.42 | 1.18 | 1.38 | 1.13 |
| 2 | 1.43 | 1.05 | 1.52 | 1.13 | 1.56 | 1.09 | 1.63 | 1.15 |
| 3 | 1.23 | 1.64 | 1.23 | 1.72 | 1.15 | 1.57 | 1.07 | 1.66 |
| 4 | 1.02 | 1.12 | 1.08 | 1.13 | 1.02 | 1.11 | 1.06 | 1.14 |
| 5 | 1.04 | 1.04 | 1.16 | 1.25 | 0.92 | 1.02 | 1.16 | 1.23 |
| 6 | 1.42 | 0.63 | 1.41 | 0.73 | 1.37 | 0.67 | 1.36 | 0.76 |
| 7 | 1.09 | 0.84 | 1.10 | 0.90 | 1.27 | 0.87 | 1.24 | 0.90 |
| 8 | 1.11 | 0.43 | 1.30 | 0.59 | 0.91 | 0.46 | 1.23 | 0.64 |
| 9 | 1.45 | 0.56 | 1.54 | 0.73 | 1.18 | 0.47 | 1.44 | 0.61 |
| 10 (High MAX) | 0.24 | 0.78 | 0.52 | 1.12 | -0.07 | 0.85 | 0.32 | 1.13 |
| 10-1 | -1.09 | -0.37 | -0.78 | 0.01 | -1.49 | -0.33 | -1.06 | 0.00 |
| | (-2.02) | (-0.67) | (-1.39) | (0.01) | (-2.47) | (-0.60) | (-1.54) | (-0.01) |
| CAPM alpha | -1.61 | -0.61 | -1.30 | -0.26 | -1.97 | -0.58 | -1.50 | -0.28 |
| | (-3.22) | (-1.11) | (-2.49) | (-0.45) | (-3.31) | (-1.04) | (-2.07) | (-0.48) |
| FF3 alpha | -1.54 | -0.60 | -1.22 | -0.27 | -1.90 | -0.56 | -1.42 | -0.28 |
| | (-3.29) | (-1.21) | (-2.45) | (-0.50) | (-3.46) | (-1.13) | (-2.08) | (-0.54) |
| FFC4 alpha | -1.18 | -0.44 | -0.79 | -0.04 | -1.59 | -0.41 | -1.00 | -0.06 |
| | (-2.41) | (-0.89) | (-1.34) | (-0.07) | (-2.68) | (-0.82) | (-1.25) | (-0.11) |
| FF5 alpha | -0.83 | -0.09 | -0.50 | 0.34 | -1.01 | -0.07 | -0.48 | 0.32 |
| | (-1.81) | (-0.19) | (-0.98) | (0.62) | (-1.87) | (-0.15) | (-0.63) | (0.59) |
| FF6 alpha | -0.60 | 0.00 | -0.21 | 0.49 | -0.83 | 0.02 | -0.21 | 0.46 |
| | (-1.33) | (0.01) | (-0.38) | (0.87) | (-1.50) | (0.03) | (-0.26) | (0.83) |

**Table IA3. Fama-MacBeth regressions with an interaction term: using mean of TRUST as the threshold to define DTRUST**

The table reports the time-series average of the slope coefficients from cross-sectional regressions across the sample when the threshold to define DTRUST is mean of TRUST within each region over time. Every month from January 1996 to December 2018 (276 months), we run the Fama-MacBeth regression of the monthly stock returns (in percent) on the one-month lagged independent variables that include the maximum daily return (MAX), TRUST, the interaction term of MAX and DTRUST, and control variables. DTRUST is a dummy variable based on TRUST within each region where firms are headquartered. If a firm's region's monthly TRUST is above the mean of TRUST for the region over time, then DTRUST is one and zero otherwise. The control variables are market beta (BETA), natural log of market capitalization (SIZE), natural log of book-to-market ratio (BM), momentum (MOM), short-term reversal (REV) and illiquidity (ILLIQ). All variables are defined in the Appendix. The sample includes Nasdaq-listed common stocks (CRSP share codes 10 and 11) headquartered in the 10 US geographical regions. No price, market capitalization, or minimum daily observation filters are imposed, except that ILLIQ requires at least ten daily observations in the month. Observations with non-positive book equity are excluded when constructing the book-to-market ratio, and firm-months with missing returns, including delisting months, are excluded. Newey-West (1987) adjusted *t*-statistics using six lags are reported in parentheses. Ave. number of stocks is the time-series average number of stocks in the monthly cross-sectional regressions.

| Model | (1) | (2) | (3) | (4) | (5) | (6) |
|---|---|---|---|---|---|---|
| MAX | -0.038 | -0.030 | -0.038 | -0.030 | -0.032 | -0.024 |
| | (-2.08) | (-1.93) | (-2.08) | (-1.94) | (-1.53) | (-1.41) |
| TRUST | | | 0.794 | 0.379 | 1.219 | 0.939 |
| | | | (1.21) | (0.67) | (1.65) | (1.49) |
| MAX × DTRUST | | | | | -0.016 | -0.018 |
| | | | | | (-1.45) | (-1.85) |
| BETA | | -0.021 | | -0.021 | | -0.020 |
| | | (-0.49) | | (-0.49) | | (-0.46) |
| SIZE | | -0.208 | | -0.208 | | -0.208 |
| | | (-3.16) | | (-3.16) | | (-3.15) |
| BM | | 0.225 | | 0.226 | | 0.228 |
| | | (1.78) | | (1.79) | | (1.83) |
| MOM | | 0.097 | | 0.099 | | 0.099 |
| | | (0.32) | | (0.34) | | (0.34) |
| REV | | -0.006 | | -0.006 | | -0.006 |
| | | (-1.15) | | (-1.15) | | (-1.17) |
| ILLIQ | | 0.034 | | 0.034 | | 0.034 |
| | | (3.18) | | (3.20) | | (3.20) |
| Avg. number of stocks | 2,746 | 2,242 | 2,746 | 2,242 | 2,746 | 2,242 |
| Avg. adj. $R^2$ (%) | 1.09 | 4.04 | 1.12 | 4.06 | 1.24 | 4.17 |

**Table IA4. Fama-MacBeth regressions with an interaction term: using MAX, DTRUST, and $MAX \times DTRUST$**

The table reports the time-series average of the slope coefficients from the Fama-MacBeth (1973) cross-sectional regressions in which MAX, DTRUST, and $MAX \times DTRUST$ are included simultaneously. Every month from January 1996 to December 2018 (276 months), we run the Fama-MacBeth regression of the monthly stock returns (in percent) on the one-month lagged independent variables that include the maximum daily return (MAX), DTRUST, the interaction term of MAX and DTRUST, and control variables as well as the regional factors. DTRUST is a dummy variable based on TRUST within each region where firms are headquartered. If a firm's region's monthly TRUST is above the threshold of TRUST for the region over time, then DTRUST is one and zero otherwise. The thresholds are the median and mean of TRUST within each region over time. The control variables include market beta (BETA), natural log of market capitalization (SIZE), natural log of book-to-market ratio (BM), momentum (MOM), short-term reversal (REV) and illiquidity (ILLIQ). The regional factors include the percentage of the Catholic population (CATH), the percentage of the population aged 25 and older who attained a bachelor's degree or higher (EDU), the median household income (MHI) measured in thousands of dollars, the natural log of one plus net state lottery revenue (ticket sales minus prizes) per capita in dollars (LOTTERY), the natural log of state GDP per capita in dollars (GDPPC), the state unemployment rate in percent (UNEMP), the natural log of the number of state residents per square mile of land area (POPDEN), the percentage of the state population who are adherents of any religious congregation (REL), and the number of bank branches per 10,000 residents (BRANCH). All variables are defined in the Appendix. Panel A (Panel B) reports the results when median (mean) of TRUST is used as the threshold to define DTRUST. The sample includes Nasdaq-listed common stocks (CRSP share codes 10 and 11) headquartered in the 10 US geographical regions. No price, market capitalization, or minimum daily observation filters are imposed, except that ILLIQ requires at least ten daily observations in the month. Observations with non-positive book equity are excluded when constructing the book-to-market ratio, and firm-months with missing returns, including delisting months, are excluded. Newey-West (1987) adjusted *t*-statistics using six lags are reported in parentheses. Ave. number of stocks is the time-series average number of stocks in the monthly cross-sectional regressions.

| | Panel A: Median of TRUST as the threshold to define DTRUST | | | Panel B: Mean of TRUST as the threshold to define DTRUST | | |
|---|---|---|---|---|---|---|
| Model | (1) | (2) | (3) | (1) | (2) | (3) |
| MAX | -0.028 | -0.020 | -0.022 | -0.025 | -0.019 | -0.022 |
| | (-1.41) | (-1.21) | (-1.39) | (-1.12) | (-1.01) | (-1.18) |
| DTRUST | 0.061 | 0.057 | 0.038 | 0.097 | 0.085 | 0.107 |
| | (0.61) | (0.59) | (0.32) | (0.92) | (0.81) | (0.79) |
| $MAX \times DTRUST$ | -0.026 | -0.030 | -0.030 | -0.026 | -0.027 | -0.027 |
| | (-2.37) | (-2.64) | (-2.60) | (-1.92) | (-1.83) | (-1.82) |
| BETA | | -0.020 | -0.022 | | -0.020 | -0.022 |
| | | (-0.46) | (-0.53) | | (-0.47) | (-0.54) |
| SIZE | | -0.209 | -0.212 | | -0.208 | -0.212 |
| | | (-3.16) | (-3.20) | | (-3.15) | (-3.19) |
| BM | | 0.227 | 0.245 | | 0.227 | 0.245 |
| | | (1.82) | (2.08) | | (1.81) | (2.08) |
| MOM | | 0.096 | 0.110 | | 0.096 | 0.109 |
| | | (0.32) | (0.39) | | (0.32) | (0.39) |
| REV | | -0.006 | -0.006 | | -0.006 | -0.006 |
| | | (-1.13) | (-1.21) | | (-1.14) | (-1.22) |
| ILLIQ | | 0.034 | 0.034 | | 0.033 | 0.033 |
| | | (3.23) | (3.27) | | (3.19) | (3.24) |
| CATH | | | 0.010 | | | 0.008 |
| | | | (1.89) | | | (1.50) |

*Continued on next page…*

*...Continued*

| | | | | | | |
|---|---|---|---|---|---|---|
| EDU | | | 0.004 | | | 0.002 |
| | | | (0.29) | | | (0.18) |
| MHI | | | 0.009 | | | 0.009 |
| | | | (1.03) | | | (1.01) |
| LOTTERY | | | -0.080 | | | -0.076 |
| | | | (-2.13) | | | (-2.01) |
| GDPPC | | | 0.303 | | | 0.334 |
| | | | (0.59) | | | (0.66) |
| UNEMP | | | 0.057 | | | 0.045 |
| | | | (1.53) | | | (1.21) |
| POPDEN | | | 0.020 | | | 0.032 |
| | | | (0.37) | | | (0.59) |
| REL | | | -0.005 | | | -0.005 |
| | | | (-1.05) | | | (-1.08) |
| BRANCH | | | 0.046 | | | 0.028 |
| | | | (0.78) | | | (0.48) |
| Avg. number of stocks | 2,746 | 2,242 | 2,242 | 2,746 | 2,242 | 2,242 |
| Avg. adj. $R^2$ (%) | 1.23 | 4.15 | 4.41 | 1.24 | 4.18 | 4.44 |

## Table IA5. Fama-MacBeth regressions using MAX(N)

The table reports the time-series average of the slope coefficients from cross-sectional regressions of four alternative MAX measures over high-TRUST and low-TRUST subsamples (regions with high and low level of trust). We classify the sample stocks into two subsamples based on the firm's region's previous month TRUST ranking. The ranking of TRUST is created within each region where firms are headquartered and over time. The thresholds used in ranking include the median and mean of TRUST. If a firm's region's monthly TRUST is above (below) thresholds of TRUST, the firm-month is classified into the high-TRUST (low-TRUST) subsamples. For each subsample and every month from January 1996 to December 2018 (276 months), we run the Fama-MacBeth regression of the monthly stock returns (in percent) on one-month lagged independent variables that include the alternative MAX measures (MAX(N)) and a set of control variables. MAX(N) is the average of the N (N=2, 3, 4, and 5) highest daily returns over a month. The control variables include market beta (BETA), natural log of market capitalization (SIZE), natural log of book-to-market ratio (BM), momentum (MOM), short-term reversal (REV) and illiquidity (ILLIQ). All variables are defined in the Appendix. Panel A (Panel B) reports the results when the threshold used to define high-TRUST and low-TRUST subsamples is median (mean) of TRUST within each region over time. The sample includes Nasdaq-listed common stocks (CRSP share codes 10 and 11) headquartered in the 10 US geographical regions. No price, market capitalization, or minimum daily observation filters are imposed, except that ILLIQ requires at least ten daily observations in the month. Observations with non-positive book equity are excluded when constructing the book-to-market ratio, and firm-months with missing returns, including delisting months, are excluded. Newey-West (1987) adjusted *t*-statistics using six lags are reported in parentheses. Ave. number of stocks is the time-series average number of stocks in the monthly cross-sectional regressions.

Panel A: Median of TRUST as the threshold to define high-TRUST and low-TRUST subsamples

| Variables | N=2 | | | | N=3 | | | | N=4 | | | | N=5 | | | |
|---|---|---|---|---|---|---|---|---|---|---|---|---|---|---|---|---|
| | High-TRUST | | Low-TRUST | | High-TRUST | | Low-TRUST | | High-TRUST | | Low-TRUST | | High-TRUST | | Low-TRUST | |
| MAX(N) | -0.080 | -0.080 | -0.047 | -0.039 | -0.102 | -0.108 | -0.065 | -0.061 | -0.122 | -0.134 | -0.082 | -0.079 | -0.141 | -0.160 | -0.098 | -0.097 |
| | (-2.90) | (-3.18) | (-1.62) | (-1.55) | (-2.94) | (-3.40) | (-1.77) | (-1.90) | (-2.96) | (-3.53) | (-1.85) | (-2.03) | (-2.98) | (-3.62) | (-1.93) | (-2.15) |
| BETA | | 0.029 | | -0.038 | | 0.039 | | -0.030 | | 0.044 | | -0.024 | | 0.047 | | -0.020 |
| | | (0.63) | | (-0.75) | | (0.87) | | (-0.60) | | (0.98) | | (-0.50) | | (1.05) | | (-0.41) |
| SIZE | | -0.238 | | -0.181 | | -0.245 | | -0.187 | | -0.248 | | -0.189 | | -0.250 | | -0.190 |
| | | (-3.12) | | (-2.78) | | (-3.23) | | (-2.91) | | (-3.30) | | (-2.95) | | (-3.31) | | (-2.98) |
| BM | | 0.254 | | 0.221 | | 0.249 | | 0.215 | | 0.246 | | 0.212 | | 0.244 | | 0.210 |
| | | (1.80) | | (1.85) | | (1.79) | | (1.82) | | (1.78) | | (1.80) | | (1.77) | | (1.79) |
| MOM | | 0.131 | | 0.042 | | 0.122 | | 0.037 | | 0.119 | | 0.036 | | 0.117 | | 0.033 |
| | | (0.42) | | (0.14) | | (0.38) | | (0.13) | | (0.37) | | (0.12) | | (0.37) | | (0.11) |
| REV | | -0.005 | | -0.008 | | -0.005 | | -0.009 | | -0.006 | | -0.009 | | -0.006 | | -0.009 |
| | | (-0.84) | | (-1.49) | | (-0.91) | | (-1.54) | | (-0.96) | | (-1.60) | | (-0.99) | | (-1.65) |
| ILLIQ | | 0.030 | | 0.044 | | 0.031 | | 0.045 | | 0.032 | | 0.046 | | 0.032 | | 0.046 |
| | | (2.81) | | (2.02) | | (2.90) | | (2.05) | | (2.94) | | (2.07) | | (2.98) | | (2.08) |
| Avg. number of stocks | 1,234 | 1,023 | 1,511 | 1,220 | 1,234 | 1,023 | 1,511 | 1,220 | 1,234 | 1,023 | 1,511 | 1,220 | 1,234 | 1,023 | 1,511 | 1,220 |
| Avg. adj. $R^2$ (%) | 1.47 | 4.43 | 1.30 | 4.48 | 1.56 | 4.49 | 1.38 | 4.53 | 1.61 | 4.52 | 1.42 | 4.55 | 1.62 | 4.54 | 1.43 | 4.56 |

*Continued on next page…*

*...Continued*

Panel B: Mean of TRUST as the threshold to define high-TRUST and low-TRUST subsamples

| Variables | N=2 | | | | N=3 | | | | N=4 | | | | N=5 | | | |
|---|---|---|---|---|---|---|---|---|---|---|---|---|---|---|---|---|
| | High-TRUST | | Low-TRUST | | High-TRUST | | Low-TRUST | | High-TRUST | | Low-TRUST | | High-TRUST | | Low-TRUST | |
| MAX(N) | -0.077 | -0.074 | -0.042 | -0.039 | -0.099 | -0.102 | -0.059 | -0.060 | -0.118 | -0.127 | -0.076 | -0.078 | -0.138 | -0.150 | -0.095 | -0.101 |
| | (-2.85) | (-3.00) | (-1.36) | (-1.35) | (-2.91) | (-3.25) | (-1.52) | (-1.65) | (-2.94) | (-3.37) | (-1.65) | (-1.82) | (-2.96) | (-3.46) | (-1.77) | (-2.04) |
| BETA | | -0.006 | | -0.033 | | 0.004 | | -0.026 | | 0.010 | | -0.021 | | 0.013 | | -0.017 |
| | | (-0.14) | | (-0.58) | | (0.09) | | (-0.46) | | (0.22) | | (-0.37) | | (0.29) | | (-0.30) |
| SIZE | | -0.232 | | -0.178 | | -0.240 | | -0.184 | | -0.243 | | -0.186 | | -0.244 | | -0.189 |
| | | (-3.07) | | (-2.72) | | (-3.19) | | (-2.84) | | (-3.25) | | (-2.89) | | (-3.27) | | (-2.95) |
| BM | | 0.238 | | 0.237 | | 0.233 | | 0.231 | | 0.230 | | 0.229 | | 0.228 | | 0.226 |
| | | (1.73) | | (1.89) | | (1.72) | | (1.86) | | (1.71) | | (1.84) | | (1.70) | | (1.81) |
| MOM | | 0.133 | | 0.089 | | 0.124 | | 0.090 | | 0.121 | | 0.090 | | 0.120 | | 0.088 |
| | | (0.43) | | (0.29) | | (0.40) | | (0.30) | | (0.39) | | (0.30) | | (0.38) | | (0.29) |
| REV | | -0.006 | | -0.006 | | -0.006 | | -0.006 | | -0.007 | | -0.007 | | -0.007 | | -0.007 |
| | | (-1.04) | | (-0.73) | | (-1.11) | | (-0.78) | | (-1.15) | | (-0.83) | | (-1.18) | | (-0.89) |
| ILLIQ | | 0.026 | | 0.062 | | 0.028 | | 0.063 | | 0.029 | | 0.064 | | 0.029 | | 0.065 |
| | | (2.67) | | (2.23) | | (2.75) | | (2.25) | | (2.79) | | (2.28) | | (2.83) | | (2.31) |
| Avg. number of stocks | 1,481 | 1,228 | 1,264 | 1,014 | 1,481 | 1,228 | 1,264 | 1,014 | 1,481 | 1,228 | 1,264 | 1,014 | 1,481 | 1,228 | 1,264 | 1,014 |
| Avg. adj. $R^2$ (%) | 1.43 | 4.31 | 1.43 | 5.08 | 1.53 | 4.38 | 1.49 | 5.11 | 1.57 | 4.41 | 1.53 | 5.14 | 1.58 | 4.42 | 1.54 | 5.15 |

**Table IA6. Fama-MacBeth regressions with an interaction term and MAX(N)**

The table reports the time-series average of the slope coefficients from cross-sectional regressions of four alternative MAX measures across the sample. Every month from January 1996 to December 2018 (276 months), we run the Fama-MacBeth regression of the monthly stock returns (in percent) on the one-month lagged independent variables that include the alternative MAX measures (MAX(N)), TRUST, the interaction term of MAX(N) and DTRUST, and control variables. MAX(N) is the average of the N (N=2, 3, 4, and 5) highest daily returns over a month. DTRUST is a dummy variable based on TRUST within each region where firms are headquartered. If a firm's region's monthly TRUST is above the threshold of TRUST for the region over time, then DTRUST is one and zero otherwise. The control variables are market beta (BETA), natural log of market capitalization (SIZE), natural log of book-to-market ratio (BM), momentum (MOM), short-term reversal (REV) and illiquidity (ILLIQ). All variables are defined in the Appendix. Panel A (Panel B) reports the results when the threshold used to define DTRUST is median (mean) of TRUST within each region over time. The sample includes Nasdaq-listed common stocks (CRSP share codes 10 and 11) headquartered in the 10 US geographical regions. No price, market capitalization, or minimum daily observation filters are imposed, except that ILLIQ requires at least ten daily observations in the month. Observations with non-positive book equity are excluded when constructing the book-to-market ratio, and firm-months with missing returns, including delisting months, are excluded. Newey-West (1987) adjusted *t*-statistics using six lags are reported in parentheses. Ave. number of stocks is the time-series average number of stocks in the monthly cross-sectional regressions.

Panel A: Median of TRUST as the threshold to define DTRUST

| | N=2 | | | N=3 | | | N=4 | | | N=5 | | |
|---|---|---|---|---|---|---|---|---|---|---|---|---|
| Model | (1) | (2) | (3) | (1) | (2) | (3) | (1) | (2) | (3) | (1) | (2) | (3) |
| MAX(N) | -0.053 | -0.053 | -0.044 | -0.076 | -0.077 | -0.066 | -0.097 | -0.098 | -0.085 | -0.117 | -0.118 | -0.104 |
| | (-2.21) | (-2.23) | (-1.73) | (-2.49) | (-2.51) | (-2.06) | (-2.63) | (-2.65) | (-2.21) | (-2.74) | (-2.76) | (-2.33) |
| TRUST | | 0.377 | 1.140 | | 0.379 | 1.178 | | 0.379 | 1.200 | | 0.378 | 1.206 |
| | | (0.66) | (1.79) | | (0.67) | (1.85) | | (0.67) | (1.88) | | (0.66) | (1.88) |
| MAX(N) × DTRUST | | | -0.030 | | | -0.035 | | | -0.040 | | | -0.045 |
| | | | (-2.54) | | | (-2.45) | | | (-2.46) | | | (-2.45) |
| BETA | -0.007 | -0.007 | -0.005 | 0.002 | 0.002 | 0.004 | 0.007 | 0.007 | 0.009 | 0.011 | 0.011 | 0.013 |
| | (-0.17) | (-0.17) | (-0.12) | (0.04) | (0.04) | (0.09) | (0.17) | (0.18) | (0.22) | (0.27) | (0.27) | (0.32) |
| SIZE | -0.217 | -0.217 | -0.217 | -0.223 | -0.223 | -0.223 | -0.226 | -0.226 | -0.226 | -0.227 | -0.227 | -0.227 |
| | (-3.39) | (-3.38) | (-3.37) | (-3.52) | (-3.52) | (-3.51) | (-3.58) | (-3.57) | (-3.57) | (-3.61) | (-3.60) | (-3.59) |
| BM | 0.218 | 0.219 | 0.214 | 0.213 | 0.214 | 0.219 | 0.210 | 0.211 | 0.216 | 0.207 | 0.209 | 0.213 |
| | (1.78) | (1.80) | (1.85) | (1.76) | (1.77) | (1.82) | (1.74) | (1.76) | (1.81) | (1.72) | (1.74) | (1.80) |
| MOM | 0.088 | 0.090 | 0.090 | 0.080 | 0.082 | 0.082 | 0.076 | 0.078 | 0.078 | 0.073 | 0.074 | 0.075 |
| | (0.29) | (0.30) | (0.31) | (0.26) | (0.27) | (0.28) | (0.25) | (0.26) | (0.26) | (0.24) | (0.25) | (0.25) |
| REV | -0.007 | -0.007 | -0.007 | -0.007 | -0.007 | -0.007 | -0.007 | -0.007 | -0.007 | -0.007 | -0.008 | -0.008 |
| | (-1.25) | (-1.26) | (-1.26) | (-1.31) | (-1.32) | (-1.33) | (-1.36) | (-1.37) | (-1.38) | (-1.40) | (-1.41) | (-1.42) |
| ILLIQ | 0.035 | 0.035 | 0.035 | 0.036 | 0.036 | 0.036 | 0.036 | 0.037 | 0.037 | 0.037 | 0.037 | 0.038 |
| | (3.23) | (3.25) | (3.29) | (3.27) | (3.29) | (3.34) | (3.29) | (3.31) | (3.36) | (3.30) | (3.32) | (3.38) |
| Avg. number of stocks | 2,242 | 2,242 | 2,242 | 2,242 | 2,242 | 2,242 | 2,242 | 2,242 | 2,242 | 2,242 | 2,242 | 2,242 |
| Avg. adj. $R^2$ (%) | 4.15 | 4.17 | 4.26 | 4.20 | 4.22 | 4.31 | 4.24 | 4.25 | 4.34 | 4.25 | 4.27 | 4.36 |

*Continued on next page…*

*...Continued*

Panel B: Mean of TRUST as the threshold to define DTRUST

| | N=2 | | | N=3 | | | N=4 | | | N=5 | | |
|---|---|---|---|---|---|---|---|---|---|---|---|---|
| Model | (1) | (2) | (3) | (1) | (2) | (3) | (1) | (2) | (3) | (1) | (2) | (3) |
| MAX(N) | -0.053 | -0.053 | -0.047 | -0.076 | -0.077 | -0.071 | -0.097 | -0.098 | -0.091 | -0.117 | -0.118 | -0.112 |
| | (-2.21) | (-2.23) | (-1.82) | (-2.49) | (-2.51) | (-2.15) | (-2.63) | (-2.65) | (-2.30) | (-2.74) | (-2.76) | (-2.45) |
| TRUST | | 0.377 | 1.012 | | 0.379 | 1.052 | | 0.379 | 1.069 | | 0.378 | 1.075 |
| | | (0.66) | (1.57) | | (0.67) | (1.63) | | (0.67) | (1.65) | | (0.66) | (1.65) |
| MAX(N) × DTRUST | | | -0.022 | | | -0.025 | | | -0.029 | | | -0.031 |
| | | | (-1.73) | | | (-1.68) | | | (-1.69) | | | (-1.63) |
| BETA | -0.007 | -0.007 | -0.005 | 0.002 | 0.002 | 0.003 | 0.007 | 0.007 | 0.009 | 0.011 | 0.011 | 0.012 |
| | (-0.17) | (-0.17) | (-0.13) | (0.04) | (0.04) | (0.08) | (0.17) | (0.18) | (0.22) | (0.27) | (0.27) | (0.31) |
| SIZE | -0.217 | -0.217 | -0.217 | -0.223 | -0.223 | -0.224 | -0.226 | -0.226 | -0.226 | -0.227 | -0.227 | -0.228 |
| | (-3.39) | (-3.38) | (-3.37) | (-3.52) | (-3.52) | (-3.50) | (-3.58) | (-3.57) | (-3.56) | (-3.61) | (-3.60) | (-3.58) |
| BM | 0.218 | 0.219 | 0.222 | 0.213 | 0.214 | 0.217 | 0.210 | 0.211 | 0.214 | 0.207 | 0.209 | 0.212 |
| | (1.78) | (1.80) | (1.83) | (1.76) | (1.77) | (1.81) | (1.74) | (1.76) | (1.80) | (1.72) | (1.74) | (1.78) |
| MOM | 0.088 | 0.090 | 0.091 | 0.080 | 0.082 | 0.083 | 0.076 | 0.078 | 0.079 | 0.073 | 0.074 | 0.076 |
| | (0.29) | (0.30) | (0.31) | (0.26) | (0.27) | (0.28) | (0.25) | (0.26) | (0.26) | (0.24) | (0.25) | (0.25) |
| REV | -0.007 | -0.007 | -0.007 | -0.007 | -0.007 | -0.007 | -0.007 | -0.007 | -0.007 | -0.007 | -0.008 | -0.008 |
| | (-1.25) | (-1.26) | (-1.27) | (-1.31) | (-1.32) | (-1.33) | (-1.36) | (-1.37) | (-1.38) | (-1.40) | (-1.41) | (-1.43) |
| ILLIQ | 0.035 | 0.035 | 0.035 | 0.036 | 0.036 | 0.036 | 0.036 | 0.037 | 0.037 | 0.037 | 0.037 | 0.037 |
| | (3.23) | (3.25) | (3.25) | (3.27) | (3.29) | (3.29) | (3.29) | (3.31) | (3.31) | (3.30) | (3.32) | (3.33) |
| Avg. number of stocks | 2,242 | 2,242 | 2,242 | 2,242 | 2,242 | 2,242 | 2,242 | 2,242 | 2,242 | 2,242 | 2,242 | 2,242 |
| Avg. adj. $R^2$ (%) | 4.15 | 4.17 | 4.28 | 4.20 | 4.22 | 4.33 | 4.24 | 4.25 | 4.36 | 4.25 | 4.27 | 4.37 |

**Table IA7. Fama-MacBeth regressions with an interaction term and ISKEW as a measure for lottery-type characteristics**

The table reports the time-series average of the slope coefficients from cross-sectional regressions across the sample when idiosyncratic skewness is used as the measure for lottery-type characteristics. Every month from January 1996 to December 2018 (276 months), we run the Fama-MacBeth regression of the monthly stock returns (in percent) on the one-month lagged independent variables that include ISKEW, TRUST, the interaction term of ISKEW and DTRUST, and control variables. ISKEW is the idiosyncratic skewness over a month. DTRUST is a dummy variable based on TRUST within each region where firms are headquartered. If a firm's region's monthly TRUST is above the threshold of TRUST for the region over time, then DTRUST is one and zero otherwise. The thresholds are the median and mean of TRUST within each region over time. The control variables include market beta (BETA), natural log of market capitalization (SIZE), natural log of book-to-market ratio (BM), momentum (MOM), short-term reversal (REV) and illiquidity (ILLIQ). All variables are defined in the Appendix. Panel A (Panel B) reports the results when the median (mean) of TRUST is used as the threshold to define DTRUST. The sample includes Nasdaq-listed common stocks (CRSP share codes 10 and 11) headquartered in the 10 US geographical regions. No price, market capitalization, or minimum daily observation filters are imposed, except that ILLIQ requires at least ten daily observations in the month. Observations with non-positive book equity are excluded when constructing the book-to-market ratio, and firm-months with missing returns, including delisting months, are excluded. Newey-West (1987) adjusted *t*-statistics using six lags are reported in parentheses. Ave. number of stocks is the time-series average number of stocks in the monthly cross-sectional regressions.

| | Panel A: Median of TRUST as the threshold to define DTRUST | | | Panel B: Mean of TRUST as the threshold to define DTRUST | | |
|---|---|---|---|---|---|---|
| Model | (1) | (2) | (3) | (1) | (2) | (3) |
| ISKEW | -0.147 | -0.148 | -0.117 | -0.147 | -0.148 | -0.109 |
| | (-2.37) | (-2.37) | (-1.84) | (-2.37) | (-2.37) | (-1.51) |
| TRUST | | 0.386 | 0.473 | | 0.386 | 0.463 |
| | | (0.68) | (0.83) | | (0.68) | (0.82) |
| ISKEW × DTRUST | | | -0.106 | | | -0.125 |
| | | | (-1.79) | | | (-1.83) |
| BETA | -0.051 | -0.051 | -0.050 | -0.051 | -0.051 | -0.051 |
| | (-1.03) | (-1.03) | (-1.02) | (-1.03) | (-1.03) | (-1.03) |
| SIZE | -0.189 | -0.189 | -0.188 | -0.189 | -0.189 | -0.188 |
| | (-2.47) | (-2.46) | (-2.46) | (-2.47) | (-2.46) | (-2.45) |
| BM | 0.235 | 0.237 | 0.239 | 0.235 | 0.237 | 0.239 |
| | (1.64) | (1.66) | (1.68) | (1.64) | (1.66) | (1.67) |
| MOM | 0.113 | 0.115 | 0.114 | 0.113 | 0.115 | 0.113 |
| | (0.37) | (0.39) | (0.38) | (0.37) | (0.39) | (0.38) |
| REV | -0.005 | -0.005 | -0.005 | -0.005 | -0.005 | -0.005 |
| | (-0.88) | (-0.89) | (-0.90) | (-0.88) | (-0.89) | (-0.90) |
| ILLIQ | 0.032 | 0.032 | 0.032 | 0.032 | 0.032 | 0.032 |
| | (3.00) | (3.02) | (3.02) | (3.00) | (3.02) | (3.02) |
| Avg. number of stocks | 2,242 | 2,242 | 2,242 | 2,242 | 2,242 | 2,242 |
| Avg. adj. $R^2$ (%) | 3.61 | 3.62 | 3.64 | 3.61 | 3.62 | 3.64 |

**Table IA8. Pooled panel regressions with regional and monthly fixed effects**

The table reports the estimated coefficients from pooled panel regressions with regional and monthly fixed effects. We run the pooled panel regressions of the monthly stock returns (in percent) on the one-month lagged independent variables that include the maximum daily return (MAX), TRUST, the interaction term of MAX and DTRUST, the stock characteristic controls, and the regional and monthly fixed effects. DTRUST is a dummy variable based on TRUST within each region where firms are headquartered. If a firm's region's monthly TRUST is above the threshold of TRUST for the region over time, then DTRUST is one and zero otherwise. The thresholds are the median and mean of TRUST within each region over time. The control variables are market beta (BETA), natural log of market capitalization (SIZE), natural log of book-to-market ratio (BM), momentum (MOM), short-term reversal (REV), and illiquidity (ILLIQ). All variables are defined in the Appendix. The sample is from January 1996 to December 2018. Panel A (Panel B) reports the results when median (mean) of TRUST is used as the threshold to define DTRUST. The sample includes Nasdaq-listed common stocks (CRSP share codes 10 and 11) headquartered in the 10 US geographical regions. No price, market capitalization, or minimum daily observation filters are imposed, except that ILLIQ requires at least ten daily observations in the month. Observations with non-positive book equity are excluded when constructing the book-to-market ratio, and firm-months with missing returns, including delisting months, are excluded. A constant is included in each specification but not reported in the table. The inclusion of fixed effects is indicated at the bottom of the table. Heteroskedasticity-robust standard errors are two-way clustered by region and month. *t*-statistics are reported in parentheses.

| Variables | Threshold of TRUST is used to define DTRUST | | | |
|---|---|---|---|---|
| | Panel A: Median | | Panel B: Mean | |
| MAX | -0.002 | 0.011 | -0.001 | 0.011 |
| | (-0.32) | (0.36) | (-0.16) | (0.34) |
| TRUST | 0.592 | 0.549 | 0.594 | 0.295 |
| | (0.57) | (0.35) | (0.57) | (0.21) |
| MAX × DTRUST | -0.019 | -0.021 | -0.019 | -0.018 |
| | (-2.10) | (-2.70) | (-2.09) | (-2.63) |
| BETA | | -0.082 | | -0.082 |
| | | (-2.25) | | (-2.25) |
| SIZE | | -0.223 | | -0.223 |
| | | (-3.34) | | (-3.34) |
| BM | | 0.466 | | 0.465 |
| | | (3.20) | | (3.19) |
| MOM | | 0.510 | | 0.510 |
| | | (1.73) | | (1.73) |
| REV | | -0.011 | | -0.011 |
| | | (-0.96) | | (-0.96) |
| ILLIQ | | 0.000 | | 0.000 |
| | | (0.11) | | (0.10) |
| Region FE | Yes | Yes | Yes | Yes |
| Month FE | Yes | Yes | Yes | Yes |
| Number of observations | 757,808 | 618,897 | 757,808 | 618,897 |
| Adj. $R^2$ | 0.102 | 0.108 | 0.102 | 0.108 |

**Table IA9. Institutional ownership and the impact of trust on the MAX effect**

The table reports the time-series averages of the slope coefficients from the monthly Fama-MacBeth (1973) cross-sectional regressions estimated separately for firms with low and high institutional ownership (IO). IO is the aggregate percentage ownership held by institutional investors, based on firms with available 13F filings. We classify the sample stocks into two subsample groups (low-IO or high-IO) based on the firm's previous month IO ranking. The ranking of IO is created across all firms in each month. The thresholds used in ranking include the mean and median of IO. If a firm's monthly IO is below (above) the threshold of IO, the firm-month is classified into the low-IO (high-IO) subsamples. For each subsample and every month from January 1996 to December 2018 (276 months), we run the Fama-MacBeth regression of the monthly stock returns (in percent) on the one-month lagged independent variables that include the maximum daily return (MAX), TRUST, the interaction term of MAX and DTRUST, and control variables. DTRUST is a dummy variable based on TRUST within each region where firms are headquartered. If a firm's region's monthly TRUST is above the median of TRUST for the region over time, then DTRUST is one and zero otherwise. The control variables are market beta (BETA), natural log of market capitalization (SIZE), natural log of book-to-market ratio (BM), momentum (MOM), short-term reversal (REV) and illiquidity (ILLIQ). Model (4) additionally includes the interaction term of MAX and ILLIQ and the interaction term of MAX and SIZE. The last row (Low-High) reports the difference between the coefficients on MAX × DTRUST in models (3) and (4) for the low-IO and high-IO subsamples. Following the Fama-MacBeth approach, we compute, for each month, the difference between the cross-sectional slope estimates on MAX × DTRUST from the low-IO and high-IO regressions, and report the time-series average of these monthly differences. All variables are defined in the Appendix. Panel A (Panel B) reports the results when the mean (median) of IO is used to define the low- and high-IO subsamples. The sample includes Nasdaq-listed common stocks (CRSP share codes 10 and 11) headquartered in the 10 US geographical regions. No price, market capitalization, or minimum daily observation filters are imposed, except that ILLIQ requires at least ten daily observations in the month. Observations with non-positive book equity are excluded when constructing the book-to-market ratio, and firm-months with missing returns, including delisting months, are excluded. Newey-West (1987) adjusted *t*-statistics using six lags are reported in parentheses. Ave. number of stocks is the time-series average number of stocks in the monthly cross-sectional regressions.

Panel A: Mean of IO as the threshold to define low-IO and high-IO subsamples

| | Low-IO | | | | High-IO | | | |
|---|---|---|---|---|---|---|---|---|
| Model | (1) | (2) | (3) | (4) | (1) | (2) | (3) | (4) |
| MAX | -0.039 | -0.039 | -0.019 | -0.047 | -0.005 | -0.005 | -0.016 | -0.025 |
| | (-2.06) | (-2.08) | (-0.89) | (-2.17) | (-0.30) | (-0.31) | (-0.94) | (-1.38) |
| TRUST | | 0.557 | 1.735 | 1.943 | | -0.528 | -0.174 | -0.187 |
| | | (0.65) | (1.97) | (2.21) | | (-0.83) | (-0.24) | (-0.26) |
| MAX × DTRUST | | | -0.042 | -0.044 | | | 0.000 | -0.002 |
| | | | (-2.81) | (-2.94) | | | (-0.01) | (-0.13) |
| MAX × ILLIQ | | | | 0.008 | | | | 0.023 |
| | | | | (0.80) | | | | (0.27) |
| MAX × SIZE | | | | -0.065 | | | | -0.035 |
| | | | | (-5.38) | | | | (-3.01) |
| BETA | -0.029 | -0.029 | -0.025 | -0.008 | 0.022 | 0.022 | 0.023 | 0.037 |
| | (-0.65) | (-0.64) | (-0.57) | (-0.17) | (0.38) | (0.39) | (0.40) | (0.68) |
| SIZE | -0.497 | -0.496 | -0.495 | -0.157 | -0.223 | -0.224 | -0.223 | -0.059 |
| | (-5.34) | (-5.36) | (-5.37) | (-1.61) | (-3.21) | (-3.22) | (-3.18) | (-0.81) |
| BM | 0.252 | 0.253 | 0.257 | 0.210 | 0.045 | 0.047 | 0.056 | 0.047 |
| | (1.86) | (1.88) | (1.92) | (1.58) | (0.35) | (0.36) | (0.43) | (0.36) |
| MOM | 0.013 | 0.013 | 0.019 | 0.034 | 0.081 | 0.080 | 0.075 | 0.081 |
| | (0.05) | (0.05) | (0.07) | (0.13) | (0.25) | (0.25) | (0.24) | (0.26) |

*Continued on next page…*

*…Continued*

| | | | | | | | | |
|---|---|---|---|---|---|---|---|---|
| REV | -0.007 | -0.007 | -0.007 | -0.007 | -0.001 | -0.001 | -0.001 | -0.001 |
| | (-1.13) | (-1.13) | (-1.16) | (-1.10) | (-0.21) | (-0.23) | (-0.24) | (-0.25) |
| ILLIQ | 0.015 | 0.015 | 0.015 | -0.003 | 0.127 | 0.122 | 0.101 | 0.864 |
| | (1.64) | (1.66) | (1.66) | (-0.22) | (0.57) | (0.55) | (0.47) | (1.57) |
| Avg. number of stocks | 894 | 894 | 894 | 894 | 881 | 881 | 881 | 881 |
| Avg. adj. $R^2$ (%) | 4.46 | 4.46 | 4.67 | 5.35 | 4.63 | 4.66 | 4.78 | 5.08 |
| Low-High: MAX × DTRUST | | | -0.042 | -0.043 | | | | |
| | | | (-2.14) | (-2.23) | | | | |

Panel B: Median of IO as the threshold to define low-IO and high-IO subsamples

| | Low-IO | | | | High-IO | | | |
|---|---|---|---|---|---|---|---|---|
| Model | (1) | (2) | (3) | (4) | (1) | (2) | (3) | (4) |
| MAX | -0.038 | -0.038 | -0.022 | -0.047 | -0.006 | -0.006 | 0.000 | -0.016 |
| | (-2.05) | (-2.06) | (-1.10) | (-2.36) | (-0.36) | (-0.37) | (0.02) | (-0.85) |
| TRUST | | 0.416 | 1.678 | 1.959 | | -0.372 | 0.224 | 0.092 |
| | | (0.48) | (1.89) | (2.21) | | (-0.60) | (0.32) | (0.13) |
| MAX × DTRUST | | | -0.042 | -0.047 | | | -0.022 | -0.022 |
| | | | (-3.22) | (-3.56) | | | (-1.66) | (-1.65) |
| MAX × ILLIQ | | | | 0.010 | | | | 0.010 |
| | | | | (0.98) | | | | (0.14) |
| MAX × SIZE | | | | -0.063 | | | | -0.037 |
| | | | | (-5.23) | | | | (-3.14) |
| BETA | -0.036 | -0.036 | -0.033 | -0.015 | 0.019 | 0.019 | 0.018 | 0.033 |
| | (-0.82) | (-0.81) | (-0.75) | (-0.34) | (0.33) | (0.34) | (0.32) | (0.60) |
| SIZE | -0.495 | -0.493 | -0.490 | -0.159 | -0.238 | -0.240 | -0.238 | -0.065 |
| | (-5.09) | (-5.11) | (-5.09) | (-1.59) | (-3.47) | (-3.49) | (-3.44) | (-0.92) |
| BM | 0.258 | 0.259 | 0.264 | 0.221 | 0.050 | 0.050 | 0.062 | 0.045 |
| | (1.85) | (1.87) | (1.91) | (1.61) | (0.39) | (0.40) | (0.50) | (0.36) |
| MOM | 0.032 | 0.034 | 0.037 | 0.054 | 0.051 | 0.052 | 0.047 | 0.049 |
| | (0.12) | (0.14) | (0.15) | (0.21) | (0.16) | (0.16) | (0.15) | (0.16) |
| REV | -0.008 | -0.007 | -0.007 | -0.007 | -0.001 | -0.001 | -0.001 | -0.001 |
| | (-1.25) | (-1.25) | (-1.24) | (-1.16) | (-0.21) | (-0.22) | (-0.21) | (-0.20) |
| ILLIQ | 0.015 | 0.016 | 0.016 | -0.003 | -0.292 | -0.307 | -0.287 | 0.772 |
| | (1.73) | (1.76) | (1.78) | (-0.18) | (-1.13) | (-1.18) | (-1.12) | (1.16) |
| Avg. number of stocks | 843 | 843 | 843 | 843 | 932 | 932 | 932 | 932 |
| Avg. adj. $R^2$ (%) | 4.43 | 4.44 | 4.63 | 5.30 | 4.61 | 4.64 | 4.76 | 5.08 |
| Low-High: MAX × DTRUST | | | -0.020 | -0.025 | | | | |
| | | | (-1.34) | (-1.66) | | | | |

**Table IA10. Education and the effect of trust on MAX effect: using mean of EDU as the threshold to define subsamples**

The table reports the time-series average of the slope coefficients from cross-sectional regressions over two subsamples that include regions with low education level (low-EDU) and high education level (high-EDU). EDU is the state-level percentage of the population over 25 years old with a bachelor's degree or higher. We classify the sample stocks into two subsample groups (low-EDU or high-EDU) based on the firm's state's previous month EDU ranking. The threshold used in the ranking of EDU is its mean across all states where firms are headquartered in each month. If a firm's state's monthly EDU is below (above) the mean of EDU, the firm-month is classified into the low-EDU (high-EDU) subsamples. For each subsample and every month from January 1996 to December 2018 (276 months), we run the Fama-MacBeth regression of the monthly stock returns (in percent) on the one-month lagged independent variables that include the maximum daily return (MAX), TRUST, the interaction term of MAX and DTRUST, and control variables. DTRUST is a dummy variable based on TRUST within each region where firms are headquartered. If a firm's region's monthly TRUST is above the mean of TRUST for the region over time, then DTRUST is one and zero otherwise. The control variables are market beta (BETA), natural log of market capitalization (SIZE), natural log of book-to-market ratio (BM), momentum (MOM), short-term reversal (REV) and illiquidity (ILLIQ). All variables are defined in the Appendix. The sample includes Nasdaq-listed common stocks (CRSP share codes 10 and 11) headquartered in the 10 US geographical regions. No price, market capitalization, or minimum daily observation filters are imposed, except that ILLIQ requires at least ten daily observations in the month. Observations with non-positive book equity are excluded when constructing the book-to-market ratio, and firm-months with missing returns, including delisting months, are excluded. Newey-West (1987) adjusted *t*-statistics using six lags are reported in parentheses. Ave. number of stocks is the time-series average number of stocks in the monthly cross-sectional regressions.

| | Low-EDU | | | High-EDU | | |
|---|---|---|---|---|---|---|
| Model | (1) | (2) | (3) | (1) | (2) | (3) |
| MAX | -0.040 | -0.040 | -0.023 | -0.028 | -0.028 | -0.026 |
| | (-2.19) | (-2.20) | (-1.04) | (-1.82) | (-1.83) | (-1.42) |
| TRUST | | -0.389 | 0.520 | | -0.415 | -0.137 |
| | | (-0.54) | (0.71) | | (-0.59) | (-0.20) |
| MAX × DTRUST | | | -0.032 | | | -0.012 |
| | | | (-2.26) | | | (-0.92) |
| BETA | -0.010 | -0.010 | -0.010 | -0.026 | -0.026 | -0.021 |
| | (-0.19) | (-0.18) | (-0.19) | (-0.57) | (-0.57) | (-0.45) |
| SIZE | -0.163 | -0.163 | -0.168 | -0.228 | -0.228 | -0.230 |
| | (-2.63) | (-2.63) | (-2.72) | (-3.16) | (-3.15) | (-3.16) |
| BM | 0.342 | 0.342 | 0.331 | 0.184 | 0.185 | 0.186 |
| | (2.65) | (2.65) | (2.59) | (1.45) | (1.46) | (1.48) |
| MOM | 0.134 | 0.135 | 0.128 | 0.083 | 0.081 | 0.086 |
| | (0.42) | (0.42) | (0.40) | (0.28) | (0.27) | (0.30) |
| REV | -0.002 | -0.002 | -0.002 | -0.009 | -0.009 | -0.009 |
| | (-0.29) | (-0.31) | (-0.27) | (-1.64) | (-1.62) | (-1.65) |
| ILLIQ | 0.024 | 0.024 | 0.023 | 0.043 | 0.043 | 0.044 |
| | (2.31) | (2.29) | (2.27) | (3.20) | (3.24) | (3.28) |
| Avg. number of stocks | 851 | 851 | 851 | 1,392 | 1,392 | 1,392 |
| Avg. adj. $R^2$ (%) | 4.55 | 4.54 | 4.72 | 4.14 | 4.14 | 4.25 |

**Table IA11. Income and the effect of trust on MAX effect: using mean of MHI as the threshold to define subsamples**

The table reports the time-series average of the slope coefficients from cross-sectional regressions over two subsamples that include regions with low income (low-MHI) and high income (high-MHI). MHI is defined as the state-level median household income. We classify the sample stocks into two subsample groups (low-MHI or high-MHI) based on the firm's state's previous month MHI ranking. The threshold used in the ranking of MHI is its mean across all states where firms are headquartered in each month. If a firm's state's monthly MHI is below (above) the mean of MHI, the firm-month is classified into the low-MHI (high-MHI) subsamples. For each subsample and every month from January 1996 to December 2018 (276 months), we run the Fama-MacBeth regression of the monthly stock returns (in percent) on the one-month lagged independent variables that include the maximum daily return (MAX), TRUST, the interaction term of MAX and DTRUST, and control variables. DTRUST is a dummy variable based on TRUST within each region where firms are headquartered. If a firm's region's monthly TRUST is above the mean of TRUST for the region over time, then DTRUST is one and zero otherwise. The control variables are market beta (BETA), natural log of market capitalization (SIZE), natural log of book-to-market ratio (BM), momentum (MOM), short-term reversal (REV) and illiquidity (ILLIQ). All variables are defined in the Appendix. The sample includes Nasdaq-listed common stocks (CRSP share codes 10 and 11) headquartered in the 10 US geographical regions. No price, market capitalization, or minimum daily observation filters are imposed, except that ILLIQ requires at least ten daily observations in the month. Observations with non-positive book equity are excluded when constructing the book-to-market ratio, and firm-months with missing returns, including delisting months, are excluded. Newey-West (1987) adjusted *t*-statistics using six lags are reported in parentheses. Ave. number of stocks is the time-series average number of stocks in the monthly cross-sectional regressions.

| | Low-MHI | | | High-MHI | | |
|---|---|---|---|---|---|---|
| Model | (1) | (2) | (3) | (1) | (2) | (3) |
| MAX | -0.038 | -0.039 | -0.021 | -0.029 | -0.029 | -0.026 |
| | (-2.06) | (-2.08) | (-0.93) | (-1.87) | (-1.88) | (-1.28) |
| TRUST | | -0.415 | 0.071 | | 0.040 | 0.636 |
| | | (-0.68) | (0.11) | | (0.05) | (0.90) |
| MAX × DTRUST | | | -0.034 | | | -0.013 |
| | | | (-2.21) | | | (-0.88) |
| BETA | -0.033 | -0.032 | -0.033 | -0.018 | -0.019 | -0.013 |
| | (-0.62) | (-0.61) | (-0.65) | (-0.39) | (-0.40) | (-0.29) |
| SIZE | -0.172 | -0.172 | -0.174 | -0.220 | -0.220 | -0.221 |
| | (-2.78) | (-2.78) | (-2.84) | (-3.08) | (-3.08) | (-3.07) |
| BM | 0.337 | 0.334 | 0.325 | 0.187 | 0.189 | 0.194 |
| | (2.60) | (2.58) | (2.52) | (1.44) | (1.47) | (1.52) |
| MOM | 0.141 | 0.139 | 0.132 | 0.082 | 0.082 | 0.088 |
| | (0.44) | (0.43) | (0.41) | (0.28) | (0.28) | (0.31) |
| REV | -0.001 | -0.001 | -0.001 | -0.009 | -0.009 | -0.009 |
| | (-0.15) | (-0.16) | (-0.14) | (-1.63) | (-1.62) | (-1.66) |
| ILLIQ | 0.027 | 0.027 | 0.027 | 0.040 | 0.040 | 0.041 |
| | (2.14) | (2.12) | (2.12) | (3.64) | (3.69) | (3.76) |
| Avg. number of stocks | 797 | 797 | 797 | 1,445 | 1,445 | 1,445 |
| Avg. adj. $R^2$ (%) | 4.61 | 4.61 | 4.81 | 4.12 | 4.13 | 4.26 |

**Table IA12. Income and the effect of trust on MAX effect: using per capita personal income as the proxy for income**

The table reports the time-series average of the slope coefficients from cross-sectional regressions over two subsamples that include regions with low income (low-PCI) and high income (high-PCI). PCI is defined as the state-level per capita personal income. We classify the sample stocks into two subsample groups (low-PCI or high-PCI) based on the firm's state's previous month PCI ranking. The ranking of PCI is created across all states where firms are headquartered in each month. The thresholds used in ranking include the median and mean of PCI. If a firm's state's monthly PCI is below (above) the threshold of PCI, the firm-month is classified into the low-PCI (high-PCI) subsamples. For each subsample and every month from January 1996 to December 2018 (276 months), we run the Fama-MacBeth regression of the monthly stock returns (in percent) on the one-month lagged independent variables that include the maximum daily return (MAX), TRUST, the interaction term of MAX and DTRUST, and control variables. DTRUST is a dummy variable based on TRUST within each region where firms are headquartered. If a firm's region's monthly TRUST is above the threshold of TRUST for the region over time, then DTRUST is one and zero otherwise. The thresholds to define DTRUST are median and mean of TRUST within each region over time. The control variables are market beta (BETA), natural log of market capitalization (SIZE), natural log of book-to-market ratio (BM), momentum (MOM), short-term reversal (REV) and illiquidity (ILLIQ). All variables are defined in the Appendix. Panel A (Panel B) reports the results when the threshold used to define low-PCI and high-PCI subsamples is median (mean) of PCI and the threshold to define DTRUST is median (mean) of TRUST within each region over time. The sample includes Nasdaq-listed common stocks (CRSP share codes 10 and 11) headquartered in the 10 US geographical regions. No price, market capitalization, or minimum daily observation filters are imposed, except that ILLIQ requires at least ten daily observations in the month. Observations with non-positive book equity are excluded when constructing the book-to-market ratio, and firm-months with missing returns, including delisting months, are excluded. Newey-West (1987) adjusted *t*-statistics using six lags are reported in parentheses. Ave. number of stocks is the time-series average number of stocks in the monthly cross-sectional regressions.

Panel A: Median of PCI as the threshold to define low-PCI and high-PCI subsamples

| | Low-PCI | | | High-PCI | | |
|---|---|---|---|---|---|---|
| Model | (1) | (2) | (3) | (1) | (2) | (3) |
| MAX | -0.038 | -0.038 | -0.021 | -0.030 | -0.030 | -0.023 |
| | (-1.98) | (-1.99) | (-1.04) | (-1.88) | (-1.91) | (-1.37) |
| TRUST | | 0.457 | 1.206 | | -0.223 | 0.620 |
| | | (0.62) | (1.60) | | (-0.31) | (0.94) |
| MAX × DTRUST | | | -0.047 | | | -0.020 |
| | | | (-2.52) | | | (-1.97) |
| BETA | -0.065 | -0.063 | -0.064 | -0.011 | -0.011 | -0.008 |
| | (-1.12) | (-1.10) | (-1.12) | (-0.23) | (-0.24) | (-0.18) |
| SIZE | -0.163 | -0.164 | -0.162 | -0.214 | -0.214 | -0.213 |
| | (-2.94) | (-2.96) | (-2.98) | (-2.97) | (-2.97) | (-2.95) |
| BM | 0.342 | 0.341 | 0.336 | 0.188 | 0.190 | 0.194 |
| | (2.93) | (2.93) | (2.88) | (1.45) | (1.47) | (1.51) |
| MOM | 0.252 | 0.253 | 0.242 | 0.053 | 0.052 | 0.056 |
| | (0.80) | (0.80) | (0.78) | (0.18) | (0.17) | (0.19) |
| REV | 0.002 | 0.002 | 0.002 | -0.009 | -0.009 | -0.009 |
| | (0.34) | (0.32) | (0.35) | (-1.62) | (-1.62) | (-1.66) |
| ILLIQ | 0.035 | 0.035 | 0.035 | 0.037 | 0.038 | 0.038 |
| | (1.66) | (1.65) | (1.67) | (3.67) | (3.73) | (3.78) |
| Avg. number of stocks | 584 | 584 | 584 | 1,658 | 1,658 | 1,658 |
| Avg. adj. $R^2$ (%) | 4.74 | 4.72 | 4.95 | 4.17 | 4.17 | 4.27 |

*Continued on next page…*

*...Continued*

Panel B: Mean of PCI as the threshold to define low-PCI and high-PCI subsamples

| | Low-PCI | | | High-PCI | | |
|---|---|---|---|---|---|---|
| Model | (1) | (2) | (3) | (1) | (2) | (3) |
| MAX | -0.040 | -0.040 | -0.021 | -0.028 | -0.028 | -0.024 |
| | (-2.20) | (-2.22) | (-1.02) | (-1.77) | (-1.78) | (-1.19) |
| TRUST | | -0.052 | 0.334 | | 0.054 | 0.688 |
| | | (-0.08) | (0.47) | | (0.07) | (0.97) |
| MAX × DTRUST | | | -0.039 | | | -0.017 |
| | | | (-2.29) | | | (-1.10) |
| BETA | -0.058 | -0.057 | -0.056 | -0.013 | -0.013 | -0.008 |
| | (-1.04) | (-1.02) | (-1.01) | (-0.28) | (-0.29) | (-0.19) |
| SIZE | -0.155 | -0.155 | -0.158 | -0.215 | -0.214 | -0.215 |
| | (-2.79) | (-2.80) | (-2.86) | (-2.95) | (-2.94) | (-2.92) |
| BM | 0.331 | 0.328 | 0.318 | 0.204 | 0.206 | 0.211 |
| | (2.83) | (2.81) | (2.75) | (1.58) | (1.61) | (1.66) |
| MOM | 0.117 | 0.116 | 0.102 | 0.104 | 0.105 | 0.112 |
| | (0.36) | (0.35) | (0.31) | (0.35) | (0.36) | (0.39) |
| REV | 0.001 | 0.001 | 0.001 | -0.010 | -0.010 | -0.010 |
| | (0.11) | (0.10) | (0.14) | (-1.71) | (-1.72) | (-1.76) |
| ILLIQ | 0.032 | 0.032 | 0.032 | 0.038 | 0.038 | 0.039 |
| | (2.28) | (2.25) | (2.26) | (3.53) | (3.58) | (3.66) |
| Avg. number of stocks | 689 | 689 | 689 | 1,554 | 1,554 | 1,554 |
| Avg. adj. $R^2$ (%) | 4.50 | 4.50 | 4.71 | 4.18 | 4.19 | 4.31 |

**Table IA13. Fama-MacBeth regressions with an interaction term, controlling for interactions of MAX with education and income**

The table reports the time-series average of the slope coefficients from cross-sectional regressions that control for education (EDU), and income (MHI), and their interactions with MAX. Every month from January 1996 to December 2018 (276 months), we run the Fama-MacBeth regression of the monthly stock returns (in percent) on the one-month lagged independent variables that include the maximum daily return (MAX), TRUST, the interaction term of MAX and DTRUST, the interaction term of MAX and EDU, the interaction term of MAX and MHI, and control variables. DTRUST is a dummy variable based on TRUST within each region where firms are headquartered. If a firm's region's monthly TRUST is above the median of TRUST for the region over time, then DTRUST is one and zero otherwise. MHI is defined as the state-level median household income, and EDU is the state-level percentage of the population over 25 years old with a bachelor's degree or higher. The control variables are market beta (BETA), natural log of market capitalization (SIZE), natural log of book-to-market ratio (BM), momentum (MOM), short-term reversal (REV) and illiquidity (ILLIQ). All variables are defined in the Appendix. The sample includes Nasdaq-listed common stocks (CRSP share codes 10 and 11) headquartered in the 10 US geographical regions. No price, market capitalization, or minimum daily observation filters are imposed, except that ILLIQ requires at least ten daily observations in the month. Observations with non-positive book equity are excluded when constructing the book-to-market ratio, and firm-months with missing returns, including delisting months, are excluded. Newey-West (1987) adjusted *t*-statistics using six lags are reported in parentheses. Ave. number of stocks is the time-series average number of stocks in the monthly cross-sectional regressions.

| Model | (1) | (2) | (3) | (4) | (5) | (6) | (7) | (8) |
|---|---|---|---|---|---|---|---|---|
| MAX | -0.038 | -0.030 | -0.038 | -0.030 | -0.030 | -0.022 | -0.075 | -0.159 |
| | (-2.08) | (-1.93) | (-2.08) | (-1.94) | (-1.51) | (-1.32) | (-1.74) | (-3.27) |
| TRUST | | | 0.794 | 0.379 | 1.321 | 1.070 | 0.485 | 0.394 |
| | | | (1.21) | (0.67) | (1.87) | (1.70) | (0.91) | (0.72) |
| MAX × DTRUST | | | | | -0.022 | -0.025 | -0.026 | -0.026 |
| | | | | | (-2.22) | (-2.71) | (-2.73) | (-2.68) |
| EDU | | | | | | | 0.008 | |
| | | | | | | | (0.52) | |
| MAX × EDU | | | | | | | 0.002 | |
| | | | | | | | (1.22) | |
| MHI | | | | | | | | -0.000 |
| | | | | | | | | (-0.04) |
| MAX × MHI | | | | | | | | 0.003 |
| | | | | | | | | (2.86) |
| BETA | | -0.021 | | -0.021 | | -0.019 | -0.021 | -0.021 |
| | | (-0.49) | | (-0.49) | | (-0.45) | (-0.50) | (-0.51) |
| SIZE | | -0.208 | | -0.208 | | -0.208 | -0.206 | -0.207 |
| | | (-3.16) | | (-3.16) | | (-3.15) | (-3.16) | (-3.17) |
| BM | | 0.225 | | 0.226 | | 0.230 | 0.239 | 0.243 |
| | | (1.78) | | (1.79) | | (1.84) | (2.00) | (2.04) |
| MOM | | 0.097 | | 0.099 | | 0.099 | 0.104 | 0.107 |
| | | (0.32) | | (0.34) | | (0.34) | (0.36) | (0.37) |
| REV | | -0.006 | | -0.006 | | -0.006 | -0.006 | -0.006 |
| | | (-1.15) | | (-1.15) | | (-1.15) | (-1.19) | (-1.17) |
| ILLIQ | | 0.034 | | 0.034 | | 0.034 | 0.034 | 0.034 |
| | | (3.18) | | (3.20) | | (3.24) | (3.27) | (3.28) |
| Avg. number of stocks | 2,746 | 2,242 | 2,746 | 2,242 | 2,746 | 2,242 | 2,242 | 2,242 |
| Avg. adj. $R^2$ (%) | 1.09 | 4.04 | 1.12 | 4.06 | 1.22 | 4.15 | 4.32 | 4.34 |

**Table IA14. Trust and the MAX effect: financing, investment and firm age**

This table presents the time-series averages of firms' financing, investment and age for high-MAX stocks in the high- and low-TRUST groups (Panel A), and the time-series average of the slope coefficients from cross-sectional regressions that control for these variables and their interactions with MAX (Panel B). In Panel A, each month from January 1996 to December 2018 (276 months), stocks are sorted into quintiles by the maximum daily return (MAX) in the previous month, and stocks in the top quintile are classified as high-MAX stocks. DTRUST is a dummy variable based on TRUST within each region where firms are headquartered. If a firm's region's monthly TRUST is above the median of TRUST for the region over time, then DTRUST is one and zero otherwise. We compute the average of each variable for high-MAX stocks in the low-TRUST group (DTRUST = 0) and the high-TRUST group (DTRUST = 1), and the difference between the two groups, and report the time-series averages of these monthly values. In Panel B, we run the Fama-MacBeth regression of the monthly stock returns (in percent), from January 1996 to December 2018 (276 months), on the one-month lagged independent variables that include MAX, TRUST, the interaction term of MAX and DTRUST, the firm-level variable, its interaction with MAX, and stock characteristic controls. NSI is net share issuance, the change in the natural log of shares outstanding over the past 12 months, adjusted for stock splits and stock dividends. EQUITYISS is the sale of common and preferred stock scaled by lagged total assets. XFIN is net external financing, the sale of stock minus the purchase of stock plus long-term debt issuance minus long-term debt reduction, scaled by lagged total assets. ASSETG is the annual growth rate of total assets. CAPX is capital expenditure scaled by lagged total assets. AGE is the natural log of one plus the number of months since the firm first appears in CRSP. Accounting variables are used from six months after the fiscal year-end. In Panel B, each model is estimated on the firm-months with available data on the added variables, and the stock characteristic and regional controls are the same as those in Table 6 (CATH, EDU, MHI, LOTTERY, GDPPC, UNEMP, POPDEN, REL and BRANCH). All variables are defined in the Appendix. The sample includes Nasdaq-listed common stocks (CRSP share codes 10 and 11) headquartered in the 10 US geographical regions. No price, market capitalization, or minimum daily observation filters are imposed, except that ILLIQ requires at least ten daily observations in the month. Observations with non-positive book equity are excluded when constructing the book-to-market ratio, and firm-months with missing returns, including delisting months, are excluded. Newey-West (1987) adjusted *t*-statistics using six lags are reported in parentheses. Ave. number of stocks is the time-series average number of stocks in the monthly cross-sectional regressions.

Panel A: High-MAX stocks across trust groups and financing, investment and firm age

| Variable | Low trust | High trust | High-Low |
|---|---|---|---|
| NSI | 0.105 | 0.104 | -0.001 |
| | | | (-0.14) |
| EQUITYISS | 0.538 | 0.501 | -0.037 |
| | | | (-1.14) |
| XFIN | 0.568 | 0.543 | -0.024 |
| | | | (-0.72) |
| ASSETG | 0.525 | 0.480 | -0.044 |
| | | | (-1.32) |
| CAPX | 0.070 | 0.070 | -0.001 |
| | | | (-0.24) |
| AGE | 4.307 | 4.324 | 0.017 |
| | | | (0.68) |

Panel B: Fama-MacBeth regressions with financing, investment and firm age

| Model | (1) | (2) | (3) | (4) | (5) | (6) | (7) | (8) |
|---|---|---|---|---|---|---|---|---|
| MAX | -0.013 | -0.021 | -0.018 | -0.021 | -0.021 | -0.021 | -0.017 | -0.021 |
| | (-0.83) | (-1.37) | (-1.22) | (-1.28) | (-1.44) | (-1.31) | (-1.25) | (-1.46) |
| TRUST | 1.052 | 1.131 | 1.155 | 1.052 | 0.997 | 0.961 | 0.881 | 0.005 |
| | (1.69) | (1.63) | (1.66) | (1.70) | (1.38) | (1.51) | (1.26) | (0.01) |
| MAX × DTRUST | -0.024 | -0.026 | -0.026 | -0.027 | -0.026 | -0.022 | -0.025 | -0.023 |
| | (-2.72) | (-2.82) | (-2.83) | (-3.00) | (-2.80) | (-2.43) | (-2.77) | (-2.12) |
| NSI | -0.177 | | | | | | -0.148 | -0.151 |
| | (-2.12) | | | | | | (-1.63) | (-1.68) |
| MAX × NSI | -0.009 | | | | | | -0.011 | -0.011 |
| | (-1.53) | | | | | | (-1.50) | (-1.44) |
| EQUITYISS | | -0.270 | | | | | 0.460 | 0.443 |
| | | (-2.45) | | | | | (1.40) | (1.32) |
| MAX × EQUITYISS | | 0.006 | | | | | 0.004 | 0.005 |
| | | (0.95) | | | | | (0.15) | (0.18) |
| XFIN | | | -0.327 | | | | -0.905 | -0.888 |
| | | | (-2.99) | | | | (-2.03) | (-1.99) |

*Continued on next page…*

*...Continued*

| | | | | | | | | |
|---|---|---|---|---|---|---|---|---|
| MAX × XFIN | | | 0.008 | | | | 0.047 | 0.045 |
| | | | (1.19) | | | | (1.18) | (1.11) |
| ASSETG | | | | -0.266 | | | 0.255 | 0.256 |
| | | | | (-2.75) | | | (1.10) | (1.12) |
| MAX × ASSETG | | | | -0.003 | | | -0.057 | -0.056 |
| | | | | (-0.42) | | | (-2.08) | (-2.07) |
| CAPX | | | | | -0.111 | | -0.005 | 0.013 |
| | | | | | (-1.32) | | (-0.05) | (0.15) |
| MAX × CAPX | | | | | -0.004 | | 0.000 | 0.000 |
| | | | | | (-0.62) | | (0.06) | (0.04) |
| AGE | | | | | | 0.216 | 0.093 | 0.103 |
| | | | | | | (3.02) | (1.12) | (1.30) |
| MAX × AGE | | | | | | -0.011 | -0.014 | -0.013 |
| | | | | | | (-1.28) | (-1.46) | (-1.39) |
| Stock characteristic controls | Yes | Yes | Yes | Yes | Yes | Yes | Yes | Yes |
| Regional controls in Table 6 | No | No | No | No | No | No | No | Yes |
| Avg. number of stocks | 2,242 | 1,979 | 2,000 | 2,185 | 1,988 | 2,242 | 1,967 | 1,967 |
| Avg. adj. $R^2$ (%) | 4.53 | 4.22 | 4.21 | 4.46 | 4.10 | 4.36 | 5.31 | 5.51 |

**Figure IA1. Geographic permutation placebo test**

The figure plots the distributions of the coefficients on $MAX \times DTRUST$ from 5,000 placebo Fama-MacBeth (1973) regressions in which regional trust is randomly reassigned to firms. In each placebo regression, we re-estimate the baseline regression in model (6) of Table 5. In Panel A, the headquarters region of each firm is randomly reassigned across firms, keeping the number of firms in each region unchanged. In Panel B, the time series of TRUST and DTRUST are randomly permuted across the 10 US geographical regions. The dashed vertical line indicates the observed coefficient on $MAX \times DTRUST$. The p-value is the share of placebo coefficients that are less than or equal to the observed coefficient. The sample includes Nasdaq-listed common stocks (CRSP share codes 10 and 11) headquartered in the 10 US geographical regions from January 1996 to December 2018 (276 months). No price, market capitalization, or minimum daily observation filters are imposed, except that ILLIQ requires at least ten daily observations in the month. Observations with non-positive book equity are excluded when constructing the book-to-market ratio, and firm-months with missing returns, including delisting months, are excluded.

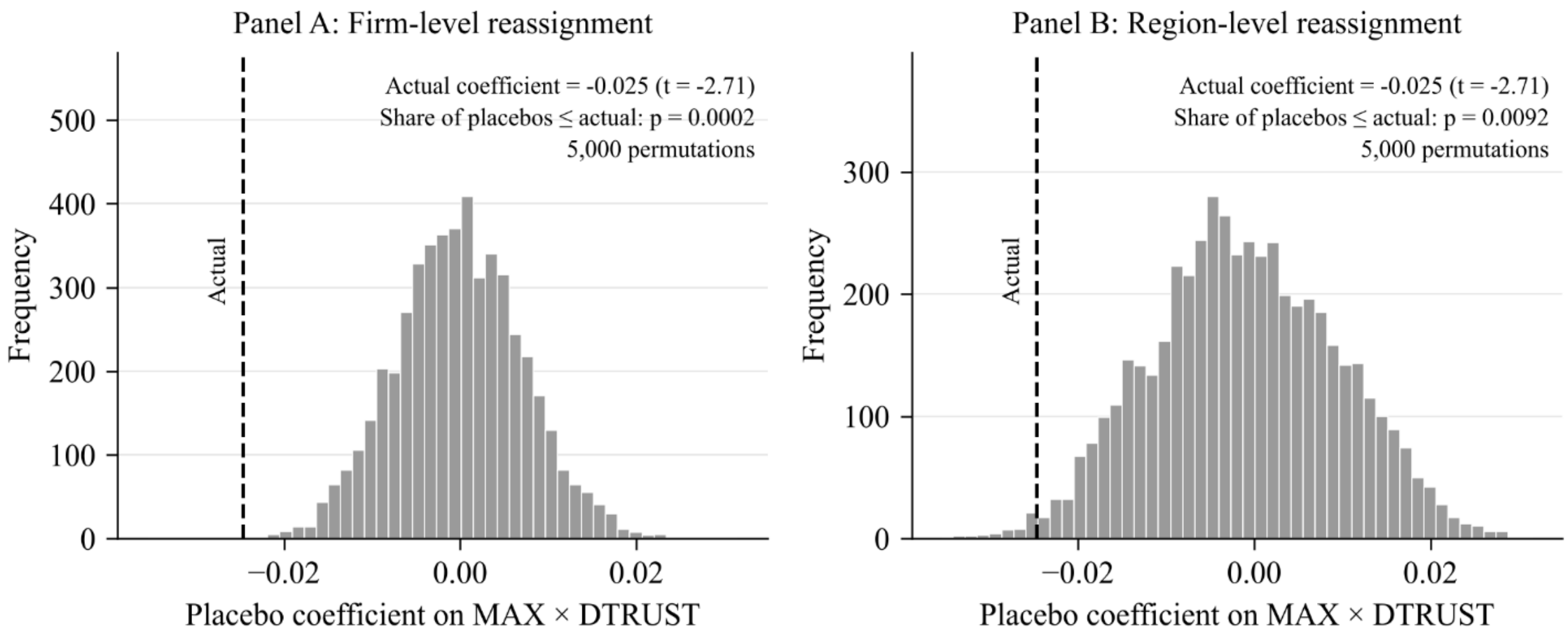

**Table IA15. Fama-MacBeth regressions, controlling for political ideology, racial diversity, immigrant share and social capital**

The table reports the time-series average of the slope coefficients from cross-sectional regressions across the sample after controlling for additional regional characteristics when the threshold to define DTRUST is median of TRUST within each region over time. Every month from January 1996 to December 2018 (276 months), we run the Fama-MacBeth regression of the monthly stock returns (in percent) on the one-month lagged independent variables that include the maximum daily return (MAX), TRUST, the interaction term of MAX and DTRUST, stock characteristic controls and regional control variables. DTRUST is a dummy variable based on TRUST within each region where firms are headquartered. If a firm's region's monthly TRUST is above the median of TRUST for the region over time, then DTRUST is one and zero otherwise. The stock characteristic controls are market beta (BETA), natural log of market capitalization (SIZE), natural log of book-to-market ratio (BM), momentum (MOM), short-term reversal (REV), and illiquidity (ILLIQ). The regional variables include political ideology as the Republican share of the two-party vote in presidential elections in each state, in percent (REP); racial diversity as the percentage of the state population that is not non-Hispanic white (NONWHITE); immigrant share as the percentage of the state population that is foreign-born (FOREIGN); and social capital as the population-weighted average of the county-level social capital index (SOCCAP) of Rupasingha, Goetz, and Freshwater (2006). All models include stock characteristic controls and model (5), additionally, includes regional control factors of Table 6 (CATH, EDU, MHI, LOTTERY, GDPPC, UNEMP, POPDEN, REL and BRANCH). All regional variables are measured for the state in which a firm is headquartered. All variables are defined in the Appendix. The sample includes Nasdaq-listed common stocks (CRSP share codes 10 and 11) headquartered in the 10 US geographical regions. No price, market capitalization, or minimum daily observation filters are imposed, except that ILLIQ requires at least ten daily observations in the month. Observations with non-positive book equity are excluded when constructing the book-to-market ratio, and firm-months with missing returns, including delisting months, are excluded. Newey-West (1987) adjusted *t*-statistics using six lags are reported in parentheses. Ave. number of stocks is the time-series average number of stocks in the monthly cross-sectional regressions.

| Model | (1) | (2) | (3) | (4) | (5) |
|---|---|---|---|---|---|
| MAX | -0.024 | -0.023 | -0.025 | -0.023 | -0.025 |
| | (-1.43) | (-1.47) | (-1.63) | (-1.39) | (-1.55) |
| TRUST | 0.349 | 1.882 | 1.087 | 1.804 | -0.075 |
| | (0.65) | (2.04) | (1.68) | (1.98) | (-0.10) |
| MAX × DTRUST | -0.024 | -0.023 | -0.021 | -0.025 | -0.024 |
| | (-2.40) | (-2.45) | (-2.30) | (-2.63) | (-2.26) |
| REP | -0.014 | | | | 0.006 |
| | (-1.77) | | | | (0.71) |
| NONWHITE | | 0.010 | | | 0.001 |
| | | (2.14) | | | (0.28) |
| FOREIGN | | | 0.019 | | 0.013 |
| | | | (2.11) | | (1.17) |
| SOCCAP | | | | -0.117 | 0.182 |
| | | | | (-1.55) | (1.83) |
| Stock characteristic controls | Yes | Yes | Yes | Yes | Yes |
| Regional controls in Table 6 | No | No | No | No | Yes |
| Avg. number of stocks | 2,242 | 2,242 | 2,242 | 2,242 | 2,242 |
| Avg. adj. $R^2$ (%) | 4.23 | 4.26 | 4.29 | 4.22 | 4.42 |

**Table IA16. Trust, social transmission intensity, and the MAX effect**

The table reports the time-series average of the slope coefficients from cross-sectional regressions that include proxies for the intensity of local social transmission and their interactions with MAX. Every month from January 1996 to December 2018 (276 months), we run the Fama-MacBeth regression of the monthly stock returns (in percent) on the one-month lagged independent variables that include the maximum daily return (MAX), TRUST, the interaction term of MAX and DTRUST, a proxy for local social transmission intensity (PROXY), the interaction term of MAX and PROXY, and control variables. DTRUST is a dummy variable based on TRUST within each region where firms are headquartered. If a firm's region's monthly TRUST is above the median of TRUST for the region over time, then DTRUST is one and zero otherwise. In panel A, PROXY is population density (POPDEN) which is the natural log of the number of state residents per square mile of land area. In panel B, PROXY is the social capital index (SOCCAP) which is the population-weighted average of the county-level social capital index of Rupasingha, Goetz, and Freshwater (2006). In panel C, PROXY is social connectedness (SCIH), measured following Bali, Hirshleifer, Peng, Tang, and Wang (2024), as the natural log of the sum of the Facebook Social Connectedness index (Bailey et al., 2018) between the state where a firm is headquartered and all US states, including the state itself. The stock characteristic controls are market beta (BETA), natural log of market capitalization (SIZE), natural log of book-to-market ratio (BM), momentum (MOM), short-term reversal (REV) and illiquidity (ILLIQ). All models include stock characteristic controls and model (2) in each panel, additionally, includes regional control factors of Table 6 (CATH, EDU, MHI, LOTTERY, GDPPC, UNEMP, POPDEN, REL and BRANCH), excluding POPDEN in Panel A, where it is the PROXY. All variables are defined in the Appendix. The sample includes Nasdaq-listed common stocks (CRSP share codes 10 and 11) headquartered in the 10 US geographical regions. No price, market capitalization, or minimum daily observation filters are imposed, except that ILLIQ requires at least ten daily observations in the month. Observations with non-positive book equity are excluded when constructing the book-to-market ratio, and firm-months with missing returns, including delisting months, are excluded. Newey-West (1987) adjusted *t*-statistics using six lags are reported in parentheses. Ave. number of stocks is the time-series average number of stocks in the monthly cross-sectional regressions.

| | Panel A: POPDEN | | Panel B: SOCCAP | | Panel C: SCIH | |
|---|---|---|---|---|---|---|
| Model | (1) | (2) | (1) | (2) | (1) | (2) |
| MAX | -0.023 | -0.025 | -0.023 | -0.026 | -0.026 | -0.028 |
| | (-1.32) | (-1.55) | (-1.43) | (-1.64) | (-1.58) | (-1.71) |
| TRUST | 1.190 | 0.438 | 1.728 | -0.069 | 1.246 | 0.256 |
| | (1.64) | (0.63) | (1.92) | (-0.10) | (1.79) | (0.37) |
| MAX × DTRUST | -0.029 | -0.027 | -0.022 | -0.018 | -0.020 | -0.020 |
| | (-2.41) | (-2.31) | (-2.27) | (-1.68) | (-2.00) | (-1.75) |
| PROXY | -0.052 | -0.117 | -0.063 | 0.070 | -0.044 | 0.078 |
| | (-1.26) | (-1.77) | (-1.08) | (1.37) | (-0.77) | (1.58) |
| MAX × PROXY | 0.015 | 0.016 | -0.003 | -0.003 | -0.004 | -0.003 |
| | (2.14) | (2.23) | (-0.44) | (-0.45) | (-0.62) | (-0.55) |
| Stock characteristic controls | Yes | Yes | Yes | Yes | Yes | Yes |
| Regional controls in Table 6 | No | Yes | No | Yes | No | Yes |
| Avg. number of stocks | 2,242 | 2,242 | 2,242 | 2,242 | 2,242 | 2,242 |
| Avg. adj. $R^2$ (%) | 4.33 | 4.51 | 4.32 | 4.50 | 4.31 | 4.48 |